\documentclass{emulateapj}
\usepackage{enumerate,multirow}
\usepackage{amsmath,amssymb,graphicx,natbib,subfigure,color} 
\usepackage[normalem]{ulem}
\usepackage[percent]{overpic}

\usepackage{amsmath}
\usepackage{amssymb}

\shorttitle{formation of galactic nuclei}
\shortauthors{Antonini, Barausse and Silk}
\begin{document}
\def\gap{\;\rlap{\lower 2.5pt
\hbox{$\sim$}}\raise 1.5pt\hbox{$>$}\;}
\def\lap{\;\rlap{\lower 2.5pt
 \hbox{$\sim$}}\raise 1.5pt\hbox{$<$}\;}

\newcommand \sbh{\rm MBH}
\newcommand\NSC{NSC}
\title{co-evolution of nuclear star clusters, massive black holes and their host galaxies}

\author{Fabio Antonini$^{1}$, Enrico Barausse$^{2,3}$ \& Joseph Silk$^{2,3,4,5}$}
\affil{(1) Center for Interdisciplinary Exploration and Research in Astrophysics (CIERA)
and Department of Physics and Astrophysics, 
Northwestern University,  Evanston, IL 60208\\
(2) Sorbonne Universit\'{e}s, UPMC Univ Paris 06, UMR 7095, Institut d'Astrophysique de Paris, F-75014, Paris, France\\
(3) CNRS, UMR 7095, Institut d'Astrophysique de Paris, F-75014, Paris, France \\
(4) Laboratoire AIM-Paris-Saclay, CEA/DSM/IRFU, CNRS, Universite Paris Diderot,  F-91191 Gif-sur-Yvette, France   \\
(5) Department of Physics and Astronomy, Johns Hopkins University,  Baltimore MD 21218, USA
}

\begin{abstract}
Studying how nuclear star clusters~(NSCs) form and how they are related to the growth of the central massive black holes (MBHs) and their host galaxies is fundamental for our understanding of the evolution of galaxies and the processes that have shaped their central structures. We present the results of a semi-analytical galaxy formation model that follows the evolution of dark matter halos along merger trees, as well as that of the baryonic components. This model allows us to study the evolution of NSCs in a cosmological context, by taking into account the growth of NSCs due to both dynamical friction-driven migration of stellar clusters and star formation triggered by infalling gas, while also accounting for dynamical heating from (binary) MBHs. We find that in-situ star formation contributes a significant fraction (up to 
$\sim 80\%$ ) of the total mass of NSCs in our model. Both \NSC \ growth through in-situ star formation and  through  
star cluster migration  are found to generate
\NSC \ --  host galaxy scaling correlations that are  shallower than the same correlations for  \sbh s.
We explore the role of galaxy mergers on the evolution of NSCs, and
show that observational data on NSC -- host galaxy scaling relations provide evidence
of partial erosion of \NSC s by MBH binaries  in luminous galaxies. We show that this observational
feature is reproduced by our models, and we make predictions about the \NSC \ and  \sbh \ occupation 
fraction in galaxies. We conclude by discussing several implications for theories of \NSC\ formation.
\end{abstract}


\section{introduction}
Over the past two decades, high-resolution observations
with the \emph{Hubble space telescope} have shown that massive stellar
clusters reside at the photometric and dynamical  centers of
most intermediate and low luminosity galaxies of
all Hubble types~\citep{1998AJ....116...68C,1999AJ....118..208M,2002AJ....123.1389B,2003ApJ...582L..79B,2003AJ....125.2936G,Cote:2006}. 
With sizes in the range $2$ to $10~$pc and masses in
the range $10^5$ to $10^8~M_{\odot}$, these nuclear clusters (NCs)
have central densities
up to $ \sim 10^{6-7}~M_{\odot}~{\rm pc^{-3}}$, making them
the densest stellar systems observed~\citep[e.g.,][]{1996AJ....111.1566P,2005ApJ...618..237W}.

\NSC s are  observed to be larger
 and brighter, and to follow different 
 structural scaling relations than  globular clusters.  
 The \NSC \ half light radii  scale with their total mass roughly as $r_{\rm h}\sim M_{\rm \NSC}^{0.5}$,
 while globular clusters have $r_{\rm h}\approx 3~{\rm pc}$ irrespective 
 of their luminosity, albeit with a large scatter\ \citep{Harris}.  
However, the \NSC \ mass distribution overlaps  with that of globular clusters at its low 
 mass end, suggesting a possible connection between
 the two types of stellar clusters~\citep[e.g.,][]{Cote:2006,2012ApJS..203....5T}.

 \NSC s have been studied in detail for large samples of galaxies in different environments. 
Observational work has revealed
that the nucleation fraction of galaxies is similar for  galaxies belonging to 
very different environments, including 
the Virgo, Fornax, and Coma clusters as well as galaxies in the 
field~\citep{Cote:2006,2012ApJS..203....5T,F,2014ApJ...791..133B}.
 On average  $80\%$ of all galaxies of all Hubble types with 
magnitude in the range $-20 \lesssim M_{B} \lesssim -12$ contain 
 a well defined central cluster. 
  Hence \NSC s appear to be ubiquitous structures in galaxies 
although they tend to be missing in galaxies brighter  than  magnitude $M_{B}\sim-20$
and fainter than $M_{B}\sim -12$~\citep[e.g.,][]{VdB,Cote:2006}.
The  real nucleated fraction could be however
larger than the value constrained from observations given that the
morphological complexity and high surface brightness often characterizing
the center of galaxies  can make  \NSC s  difficult to identify, especially
in massive spheroids.

High resolution spectroscopic surveys have been used to place constraints  on \NSC \
 ages and star formation histories~\citep{2006ApJ...649..692W,2006AJ....132.1074R,2005ApJ...631..280B}. The common finding emerging  from these studies is
 that \NSC s are characterized by  a complex star formation history with a mixture of
 morphological components  and different stellar populations spanning a wide range of characteristic ages
  from $10~$Myr to $10~$Gyr. Cluster ages and masses are also found 
 to depend on the host galaxy Hubble type, with \NSC s in early-type spirals
being older and more massive than those of late-type spirals. 
The luminosity-weighted ages of \NSC s are typically 
 smaller than the ages of host galactic disks, indicating that the growth of the  nuclei
 is a continuous and ongoing process occurring during and after most of the host galaxy was formed~\citep{2006AJ....132.1074R}.

 Age estimates for most \NSC s remain however susceptible to large uncertainties due to the fact  that the 
 light is often dominated by  young stellar populations while at the same time the mass is dominated by the old stars.
 The inability to infer spectroscopically the age of old cluster components can  introduce a bias toward 
 younger ages. 
The bulk of the stellar population is typically found in an older and
spatially more extended stellar component~\citep{2015arXiv150105586C}.
 This appears to be the case for the Milky Way in which more than $\gtrsim 80~\%$
 of the stars inside the inner parsec formed more than $5~{\rm Gyr}$ ago, while luminous 
 massive stars dominate the central cluster light within $\sim 0.5$pc~\citep{2005ApJ...631..280B,2011ApJ...741..108P,2012ApJ...744...24Y,2012ApJ...745..121L}.

\NSC \ old and young stellar populations also differ morphologically.
\citet{2006AJ....132.2539S} showed that the three edge-on late-type galaxies IC 5052, NGC 4206, and NGC 4244 have nuclei
 that are strongly flattened along the plane of their host galaxy disks. Such \NSC s show evidence for young 
stellar components distributed onto a disk-like or ring structure superimposed on an older more spherical component.  
In the edge-on spiral galaxy NGC 4244, old and young stellar components  both show clear signs of rotation~\citep{2008ApJ...687..997S}.
\citet{2006AJ....132.2539S}  suggested that the presence   
of such multiple  rotating morphological components and the presence of a flattened young cluster aligned
 with the major axis of the host galaxy
point toward  an in situ build up of nuclear clusters, whereby stars form episodically in compact 
nuclear disks and then lose angular momentum or heat vertically to form an older spheroidal structure. 
It has been noted however that  both rotation and the presence of young structural components 
does not exclude other formation mechanisms such as for example episodic accretion 
of young star clusters  in the central part of the galaxy due to dynamical friction~\citep{2014ApJ...794..106A}. 
In the case of NGC 4244 for example, it is likely that 
 accreation of star clusters has contributed at least in part to the growth of its 
\NSC~\citep{2011MNRAS.418.2697H,2013MNRAS.429.2974D}.

Since the early studies, it has been realized that \NSC\ masses correlate fairly well with galaxy properties
such as bulge velocity dispersion, bulge and galaxy total 
luminosity~\citep{2003ApJ...582L..79B,2004AJ....127..105B,2006AJ....132.1074R}.
The existence of such correlations suggests 
that the formation of \NSC s is intimately connected to the 
formation and evolution of the host galaxy\ \citep[e.g.,][]{2015MNRAS.451.5378L}.
Given that \NSC s and  \sbh s are found to coexist in some galaxies, that in these 
systems they have comparable masses, and that both follow tight correlations
with galaxy properties, it is also natural to ask  whether  \NSC s and  \sbh s are somehow connected to each other. 

\NSC s and  \sbh s are known to co-exist in galaxies with masses $\sim 10^{10}~M_{\odot}$~\citep{2008ApJ...687..997S,GD08};
galaxies with masses lower than this value show clear evidence for nucleation but 
little evidence for a  \sbh .
Conversely   galaxies with  masses above $\sim 10^{11}~M_{\odot}$
are dominated by   \sbh s but show no evidence for nucleation~\citep[e.g.,][]{2006ApJ...644L..17W,Neumayer:2012}.
\citet{Ferrarese:2006} found that the nuclei and  \sbh s obey a similar scaling relation linking
their mass to the virial mass of the host galaxy.
More recently, \citet{G12} obtained  \NSC \  -- host galaxy correlations
using a sample in which he excluded what were likely to be nuclear stellar
discs from the galaxy sample of Ferrarese et al., while including  an additional 13 
\NSC s in galaxies with velocity dispersions  out to about $200~{\rm km/s}$.
Graham found that  \NSC \  and  \sbh \ masses follow different scaling correlations with host galaxy properties.
Other authors  confirmed this result,
 showing that the mass of \NSC s scales with the host-galaxy spheroid's velocity dispersion as $M_{\rm NSC} \sim \sigma^2$,
while the mass of  \sbh s follows the much steeper relation 
$M_{\rm  \sbh} \sim \sigma^5$~\citep{G12,Erwin:2012,Leigh:2012,2013ApJ...763...76S,2013ARAA..51..511K,F}.
Although this might suggest that  \sbh s and \NSC s did not form from the same mechanism~\citep{2013ApJ...763...62A},
whether the formation of  \sbh s is connected to the evolution of \NSC s and whether the two types of
central objects grow together or in competition 
from the same physical process remains unclear. 

The observational findings mentioned above provided motivation for theoretical work 
aimed to understand  how \NSC s form and how their evolution is linked to the evolution of their host galaxy.
Two models have been suggested for the formation of \NSC s:
(i) the cluster infall scenario, in which stellar clusters are driven to the galactic nucleus by dynamical friction, merge,
and build up a \NSC \ \citep{1975ApJ...196..407T}; (ii) The nuclear star formation scenario, in which gas falls into the nucleus and
 forms stars~\citep[e.g.,][]{2004ApJ...605L..13M,2006ApJ...650L..37M}.  
Due to the inherent complexity of gas dynamics in star formation, which makes the second of these
two processes difficult to model, theoretical work has been mostly directed toward the cluster infall 
scenario~\citep[but see][]{2015ApJ...799..185A}. However, both dissipative and dissipationless processes are likely to 
play an important role in \NSC\ formation\ \citep{2015ApJ...806L...8A}.

Theoretical studies have employed two different methodologies: $N$-body simulations and semi-analytical modeling.
$N$-body techniques are used to simulate the last stage of inspiral and merger of stellar clusters in the inner  region
of galaxies. These models have shown that a star cluster merger scenario can explain without obvious difficulties the
observed properties of \NSC s, including their density and  velocity dispersion 
profiles~\citep[e.g.,][]{2004ApJ...610L..13B,2008MNRAS.388L..69C,AM12,2014ApJ...794..106A,2014ApJ...784L..44P}.
While these studies make reliable predictions about the aspect of
\NSC s, they suffer from the fact that the adopted initial conditions are often not well motivated, hence  the need to rely on semi-analytical models.
These are used to make
predictions for the appearance of star clusters, of known mass and radius, in the center of galaxies, and to compute the accumulated mass
at the center through dynamical friction migration over a Hubble time.
In semi-analytical models,  the rate at which the \NSC \ grows by accreting young clusters is
 estimated from empirical cluster formation rates, dynamical friction timescales and dissolution 
 times~\citep[e.g.,][]{2011ApJ...729...35A,2014ApJ...785...71G,2014MNRAS.444.3738A}.
Studies based on semi-analytical approaches have demonstrated that the \NSC -- host galaxy
property scaling relations and their half-mass radius --  mass relation
are both consistent with formation by star cluster accretion~\citep{2013ApJ...763...62A}.

All previous calculations assumed \NSC\ formation to take place in isolated galactic spheroids,
thus neglecting the role of galaxy evolution, mergers, and the role of in-situ star formation.
Also, these former idealized attempts could not explore the details of the interplay between  \sbh\ and \NSC \ evolution. 
In this paper, we present a semi-analytical galaxy-formation model that allows us to
shed light on exactly these points, i.e. it allows us to assess  the role of galaxy mergers,  \sbh \ mergers and 
nuclear star formation on the growth of \NSC s.
We follow the formation and evolution of galaxies, 
 \sbh s and \NSC s along cosmic history, including the growth of \NSC s due to both central migration of stellar clusters and 
in-situ star formation, while also accounting  for dynamical heating from (binary) MBHs.

The paper is organized as follows.
In Section~2 we introduce and discuss the numerical methods employed in our study. Section~3
describes the sample data  to which  our numerical results are compared.
In Sections~4  and 5 we describe the main results of our calculations and discuss some of their implications
 in Section~6. We summarize in Section~7.

\begin{figure*}
\centering 
\includegraphics[width=14cm,angle=0.]{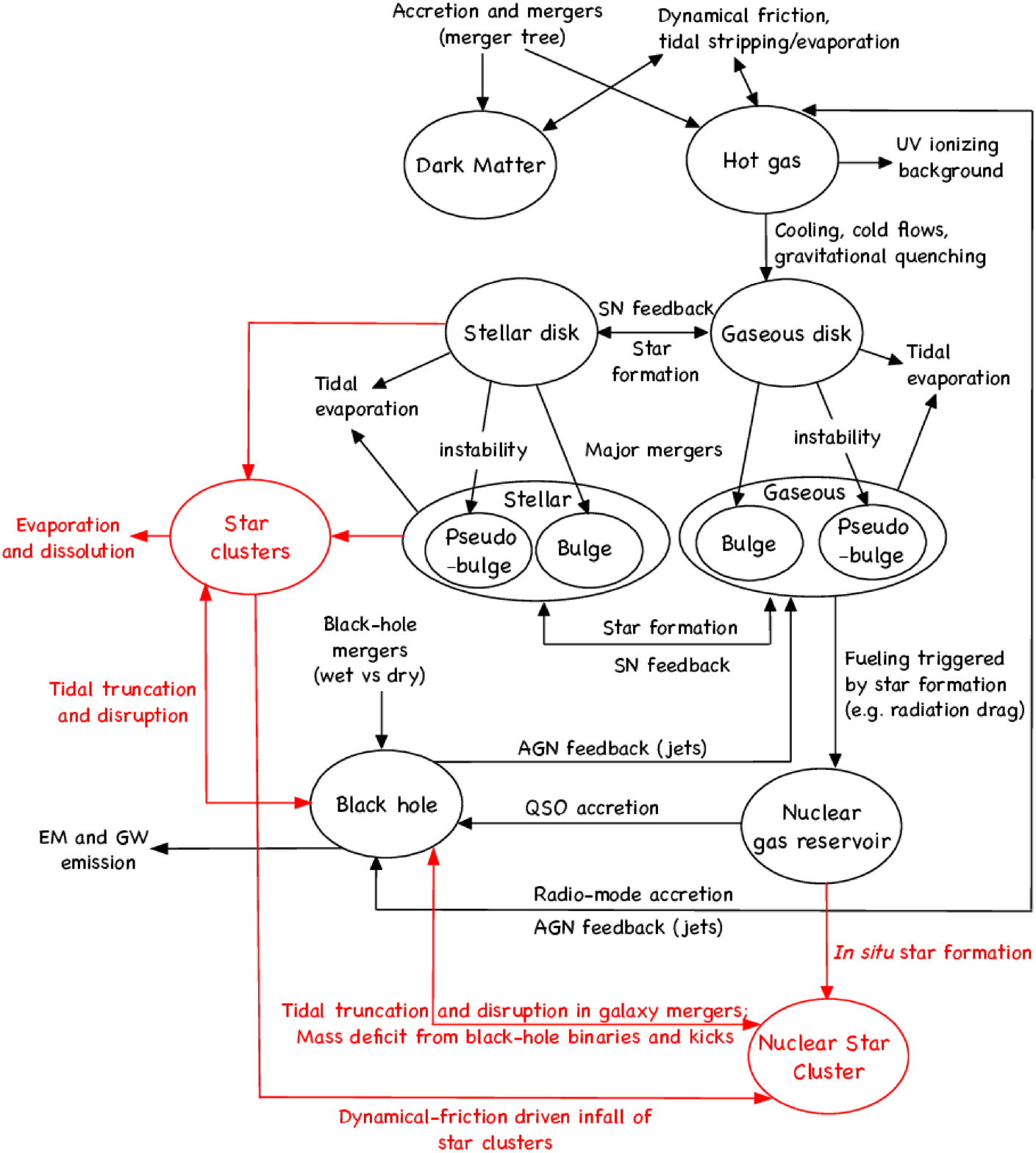}
\caption{Schematic representation of the model of~\citet{2012MNRAS.423.2533B} (in black; including the improvements described in \citet{2014ApJ...794..104S}) with the additions described in this paper (in red) to model the formation
and evolution of NSCs.
\label{Fig_model}}
\end{figure*}

 \section{Semi-analytical models}
 We study the formation of \NSC s and their co-evolution with 
  \sbh s along the merger history of their host galaxies by semianalytical techniques. 

 First, we consider a purely dissipationless formation  model in which  stellar clusters
 form  and migrate to the center of a galaxy through dynamical friction. 
 We also consider  a second semi-analytical model,
which unlike the first follows the formation and merger history of galaxies (both in their Dark-Matter and
baryonic components). 
This allows us to study the formation, evolution and disruption of NSCs in a self-consistent way,
accounting not only for the dynamical friction-driven inspiral of stellar clusters to the nucleus, but also for 
star formation in the nuclear gas, and for the effect of galaxy and MBH mergers.

We begin in this section by describing our methodology,
highlighting in particular the simplifying approximations that are made in our models.

 \subsection{Cluster inspiral model~(CliN)}\label{Mod1}
 In our semi-analytical model $CliN$ a galactic nucleus forms 
 through the orbital decay (via dynamical friction) and merger of  star clusters in the
 central region of a galaxy. This code adopts  a rather idealized model of an isolated galactic spheroid
and  follows the consecutive inspiral of  stellar clusters  along 
their orbits until they decay  into the galactic center or they are disrupted in the process. 
This relatively simple approach has the advantage that
the details of the tidal interaction of the clusters with the background galaxy and central  \sbh\
are easily included, and in a relatively precise manner. 
On the other hand, as previously stated, this model has the important limitation that galaxy evolution
 as well as dissipative processes that can lead to 
star formation episodes  in the galactic center are  neglected.
Another basic assumption made in the $CliN$ model is that  \sbh s are already in place 
before \NSC s grow around them; moreover  we  assume  that all galaxies contain initially a  \sbh .

The code is essentially the same as that  used by \citet{2013ApJ...763...62A},
and we direct the reader to that paper for a more detailed description.
Briefly, we represent the galaxy
spheroid by a simple power-law density model: $\rho(r)=\rho_0\left(r/r_0 \right)^{-\gamma}$,
where $\rho_0=(3-\gamma)M_{\rm sph}/4\pi r_0^3$. This latter expression assumes that
the density of the galactic bulge follows a \citet{1993MNRAS.265..250D} profile, $r\sim r^{-4}$ , at  $r\gg r_0$. 
Given a mass for the central  \sbh, $M_{{\rm  \sbh}}$, we compute
the galaxy density profile slope, scale radius, spheroidal mass and galaxy velocity dispersion
by using the corresponding scaling relations that link these properties
to $M_{{\rm  \sbh}}$~\citep{2009ApJ...698..198G, Graham:2012b, 2007ApJ...655...77G}.
The galaxy effective radius, $R_{\rm eff}$,  was derived  from the size-mass 
relation given by equation~(32) of \citet{2003MNRAS.343..978S}.
The ratio $R_{\rm eff}/r_0$ follows from 
equations~(4) and (17) of \citet{1993MNRAS.265..250D}, which
 give $R_{\rm eff}/r_0=(1.8,~1.5,~1)$ for $\gamma =(1,~1.5,~2)$. 

 We assign the stellar cluster masses 
using the cluster initial mass function (CIMF), $d n / d m_{\rm gc} \propto m_{\rm gc}^{-2}$
\citep{bik,degrijs}, 
and limiting mass values of $m_{\rm min} =10^2M_\odot$, $m_{\rm max} =10^7M_\odot$.
 The stellar clusters are assumed to  form  continuously  over the age of the galaxy.
 
As commonly done, we assume that the clusters have initially the same distribution as the stars 
in the galaxy, and that
initially a fixed fraction $f_{\rm gc}=0.05$ of stars form in strongly bound stellar clusters. This 
latter is approximately consistent with  the typical cluster formation
efficiency for the Milky Way  found by \citet{2012MNRAS.426.3008K} and
it is similar to the value adopted in previous studies~\citep{2014ApJ...785...71G}. 
Gnedin et al. showed that 
by adopting a  power-law mass function  $d n / d m_{\rm gc} \propto m_{\rm gc}^{-2}$ and 
 a fraction $f_{\rm gc}=0.04$ of clustered  star formation, one can reproduce  the observed density profile of clusters in M87.

We assume that  the central properties of a stellar cluster remain 
unchanged during inspiral and that $r_t>r_K$, where
$r_K$ is the cluster core radius and  $r_t$ is the cluster tidal~(limiting) radius given by~\citep{K62}:
\begin{equation}\label{rtrking}
r_t= \alpha \frac{\sigma_K}{\sqrt{2}}\left(\frac{3}{r}\frac{ {d} \phi}{{d} r} -4\pi G \rho  \right)^{-1/2}~,
\end {equation}
with $\phi$  the sum of galactic,  \sbh\ and \NSC\ potentials, and $\alpha$ a ``form factor'' that depends on the
 density distribution  within the cluster.
The mass of a King model is then related to its tidal radius and velocity dispersion via the expression
\begin{equation}\label{mt0}
Gm_{\rm gc}\approx \frac{\sigma_K^2r_t}{2}~.
\end{equation} 
 We note that although at first order 
the central properties of the clusters are not expected to change during inspiral,
 the stars in a \NSC\ formed from cluster infalls should arguably end up 
 with the same root-mean-square velocity as the galaxy host, which appears 
 to be consistent with observations
 \citep[see][]{2015MNRAS.451.5378L}.

Given a cluster of central velocity dispersion $\sigma_K,$
the time evolution of its orbital radius is
\begin{eqnarray}\label{sfc2}
r(t)=\left[ r_{in}^{3-\gamma}-\frac{(3-\gamma)^3}{(4-\gamma)\sqrt{2\gamma}}\frac{ 
r_0^{-\gamma}}{4 \pi G \rho_0 }F(\gamma)\ln \Lambda\sigma_k^3 \times t \right]^{\frac{1}{3-\gamma}}
\end{eqnarray}
where $r_{in}$ is the initial separation from the center, for the Coulomb logarithm we take $\ln \Lambda=6$, and $F(\gamma)$ is given in
equation~(1) of \citet{MP04}. We approximately account for the dynamical dissolution of the clusters due to their
collisional evolution and interaction with the external tidal field of the galaxy
by only evolving clusters with dynamical friction time shorter than their dissolution
 time~[equation~(\ref{diss-t}) below].

The cluster orbits are evolved using equation~(\ref{sfc2})
up to a maximum integration time of $10{\rm Gyr}$, and the \NSC\ mass is
obtained by summing up all the mass transported by the infalling clusters
within a radius $r_{\rm \NSC}$. The \NSC\ radius, $r_{\rm \NSC}$, is computed 
from the recursive relation given by equations (14)-(16) of \citet{2013ApJ...763...62A}.
As we evolve the cluster orbits, we compute their tidal disruption radius due to the galaxy,  \sbh\ and
a pre-existing \NSC\ gravitational fields,  adopting
a cluster core radius $r_K=1{\rm pc}$, roughly 
equal to the median value of  the core radii listed in the 
Harris' compilation of Galactic star clusters~\citep{Harris}.
When the clusters reach their tidal disruption radius, the remaining core mass
is dispersed around the radius of disruption. 
 When evaluating the final \NSC\ density distribution  
we make the  assumption that the stars from the disrupted clusters are  isotropically distributed around the center. 
We note that in reality the clusters will distribute initially in a disk-like structure,
but  this will morph into a more spherical  distribution over a fraction of the
nuclear relaxation timescale\ \citep{2014ApJ...794..106A}. 

A  \sbh\ binary formed during a galaxy merger
leaves an imprint on the galactic nucleus in the form of a mass deficit, $M_{\rm ej; bin}$, a
decrease in the mass of the nucleus due to ejection of stars on intersecting orbits~\citep{BG:10}.
We account for this by subtracting 
a mass 
$M_{\rm ej; bin}\approx 0.5 \times N_{\rm m} M_{{\rm  \sbh} }
$ \citep{2006ApJ...648..976M},
where $N_{\rm m}$ is the predicted number of major mergers after the last major accretion event,
from the final accumulated mass inside $ r_{\rm \NSC} $.
The number of mergers $N_{\rm m}$ is obtained from the galaxy mass using the distributions shown in Figure~(2) of \citet{2002MNRAS.336L..61H},
where in order to convert 
$V-$luminosity to mass we used 
a constant mass-to-light ratio $M/L_V=3M_\odot/L_{\odot}$, as  sometimes  adopted in the literature for an old stellar population\ \citep[e.g.,][]{2014ApJ...785...71G}.

 Finally we
compare the resulting \NSC\ density profile, obtained after $10~$Gyr of evolution, 
to the density profile, $\rho(r)$, of the galaxy, and define a galaxy as \emph{nucleated}
if at some radius the \NSC\ stellar density was found  to be higher than that of the background galaxy.
This latter condition is  based on the fact that if the \NSC\ densities are below the galaxy density at all radii,
then it would be difficult to observationally distinguish the two components.

 \subsection{Galaxy-formation model~(GxeV)}\label{GxeV}

Our galaxy formation model (hereafter $GxeV$) is based on that of~\citet{2012MNRAS.423.2533B}, which was further improved in~\citet{2014ApJ...794..104S}. 
More precisely,~\citet{2012MNRAS.423.2533B} presented a semi-analytical galaxy formation model
tracking the evolution of baryonic structures along Dark Matter merger trees. These trees
are produced with an extended Press-Schechter formalism, modified in order to reproduce 
the results of N-body simulations of Dark-Matter halos~\citep{2008MNRAS.383..557P}, while the baryonic structures
include the hot, largely unprocessed intergalactic medium,
galactic disks and spheroids (in both their stellar and cold, chemically-enriched interstellar medium components),
a low angular momentum reservoir of cold nuclear gas available for accretion onto the central MBH~\citep{2004ApJ...600..580G,2014ApJ...782...69L}, and the 
MBH itself. These baryonic components are interconnected by a plethora of interactions, schematically summarized in Figure~\ref{Fig_model}.
Also included are environmental and tidal interactions between galaxies during mergers, following simple semi-analytical recipes~\citep{2003MNRAS.341..434T,2008MNRAS.383...93B}.
Moreover,~\citet{2014ApJ...794..104S} later improved this model by refining the star formation law
-- adding in particular an explicit dependence on the metallicity~\citep{2009ApJ...699..850K}, as well as a distinction between
pseudo-bulges forming from bar instabilities and classical bulges created by major mergers -- and by devising more realistic prescriptions
for the evolution of the spins of MBHs with cosmic time. 

For this work, we have further ameliorated the model by including the formation
and evolution of NSCs (cf. Figure~\ref{Fig_model}). 
More precisely, we consider two possible formation channels for
these objects, namely one where NSCs form from the dynamical friction-driven migration to the nuclear regions
of star clusters created further out in the galaxy, and one in which NSCs form \textit{in situ} as a result
of star formation in the nuclear regions.
 Note that we assume no high redshift seeds for the NSCs, i.e.
the NSC mass grows from zero at high redshifts to its present-day value through the two aforementioned channels (this growth
being modulated/hindered by the effect of galaxy and black-hole mergers, cf. section \ref{sec:mergers}).

In the $GxeV$ models described below we define as nucleated any galaxy containing a central 
cluster more massive than $10^4~M_{\odot}$, as this value corresponds approximately to
the lower  limit of the observed \NSC \ mass distribution~\citep[e.g.,][]{F}.  We also
define an early- (late-)type galaxy as one with bulge to total mass ratio $B/T>0.7$ ($<0.7$).
For each of the $GxeV$ models that we present, we simulate about 1300 galaxies with Dark-Matter masses ranging from $10^{10} M_\odot$ to $10^{15} M_\odot$.

\subsubsection{Migration channel}

To implement the dynamical friction-driven migration of star clusters to the central nucleus, we first assume that star clusters are created during 
star-formation events with efficiency $f_{\rm gc}$, both in galactic bulges and disks. In bulges, \citet{2012MNRAS.423.2533B} and \citet{2014ApJ...794..104S} assumed a volumetric star-formation rate
$\dot{\rho}_\star$, while in disks they considered a vertically-averaged
rate of star formation $\dot{\Sigma}_\star$. Therefore, we assume that star clusters form
in bulges and disks, and that the total mass of the star-cluster population (in both the bulge and disk), $M_{\rm gc}$, changes
with rate
\begin{equation}\label{mdotGC}
\dot{M}^{\rm gc}_{\rm formation}= f_{\rm gc} \times \left(4 \pi \int \dot{\rho}_\star  r^2 \mbox{d}r+ 2 \pi \int \dot{\Sigma}_\star r \mbox{d}r\right)\,,
\end{equation}
where we choose $f_{\rm gc}=0.07$, which is 
the typical value in the Milky Way~\citep{2012MNRAS.426.3008K}. We stress that
plausible values $f_{\rm gc}=0.05$ -- $0.2$ would only mildly impact the normalization of our 
results for the $M_{\rm NSC}-\sigma$ relation, and would not affect its slope significantly.
In addition  we have tried a model with a variable
 $f_{\rm gc}$ set to 0.07, 0.04 and 0.5 in disk, quiescent and starburst galaxies respectively;
 these correspond approximately to the values of Figure~4 in ~\citet{2012MNRAS.426.3008K}. 
 The results of this model were found to not significantly differ from 
 those of models with a fixed  $f_{\rm gc}$. Thus, for the sake of simplicity,  we  present in what follows
only the results of a model with fixed  $f_{\rm gc}=0.07$.

We then assume that the star clusters formed in the bulge are spatially distributed 
in the same way as the bulge stellar population, i.e. if that population has density profile $\rho^\star_{\rm bulge}(r)$, 
the volumetric number density  $n^{\rm bulge}_{\rm gc}$ of star clusters is assumed to satisfy\begin{equation}
p^{\rm bulge}_{\rm gc}(r)\equiv\frac{n^{\rm bulge}_{\rm gc}(r)}{N^{\rm bulge}_{\rm gc}}=\frac{\rho^\star_{\rm bulge}(r)}{M_{\rm bulge}^{\star}}
\,,\label{bulge}
\end{equation}
where $N_{\rm gc}^{\rm bulge}$ is the total number of star clusters in the bulge,
and $M_{\rm bulge}^{\star}$ is the total bulge mass in stars.
 Note that following \citet{2012MNRAS.423.2533B}, we assume a Hernquist distribution for $\rho^\star_{\rm bulge}(r)$; we refer to \citet{2012MNRAS.423.2533B}
for more details on the choice of the scaling radius for this density profile.

Similarly, we assume that the disk's star clusters are distributed according to
the same surface density profile as the overall stellar content of the disk,
i.e. the superficial number density $\nu^{\rm disk}_{\rm gc}(r)$ will satisfy
\begin{equation}\label{disk}
p^{\rm disk}_{\rm gc}(r)\equiv\frac{\nu^{\rm disk}_{\rm gc}(r)}{N^{\rm disk}_{\rm gc}}=\frac{\Sigma^\star_{\rm disk}(r)}{M_{\rm disk}^{\star}}\,,
\end{equation}
where again $N_{\rm gc}^{\rm disk}$ is the total number of star clusters in the disk, 
and $M_{\rm disk}^{\star}$ is the mass of the stellar disk.
 Again, for $\Sigma^\star_{\rm disk}(r)$
we follow \citet{2012MNRAS.423.2533B} and adopt an exponential profile (see \citet{2012MNRAS.423.2533B} for more details on
the calculation of the scaling radius for this distribution).

As for the mass distribution of the newly formed star clusters,
we assume a power-law mass function $\mbox{d} p^{\rm X}_{\rm gc}/{\mbox{d} m_{\rm gc}}
\propto {m_{\rm gc}}^{-2}$ (with $X=$ bulge, disk). Combined with equations~\eqref{bulge} and \eqref{disk}, this gives
the distribution functions for the bulge and disk star clusters 
\begin{gather}
\pi^{\rm bulge}_{\rm gc}\equiv \frac{1}{N^{\rm bulge}_{\rm gc}} \frac{\mbox{d}^2 N^{\rm bulge}_{\rm gc}}{\mbox{d} m_{\rm gc} \mbox{d} V} =\frac{\mbox{d} p^{\rm bulge}_{\rm gc}}{\mbox{d} m_{\rm gc}}=A\,\frac{ p^{\rm bulge}_{\rm gc}(r)}{m_{\rm gc}^2}\,\\
\pi^{\rm disk}_{\rm gc}\equiv \frac{1}{N^{\rm disk}_{\rm gc}} \frac{\mbox{d}^2 N^{\rm disk}_{\rm gc}}{\mbox{d} m_{\rm gc} \mbox{d} S} =\frac{\mbox{d} p^{\rm disk}_{\rm gc}}{\mbox{d} m_{\rm gc}}=A\,\frac{ p^{\rm disk}_{\rm gc}(r)}{m_{\rm gc}^2}\,,
\end{gather}
where $\mbox{d} V$ and $\mbox{d} S$ are units of volume and surface.
We assume that these distribution functions are valid for masses $m_{\rm gc}$ for individual star clusters between
$m_{\min}=10^2 M_{\odot}$ and $m_{\rm max}=10^6$ -- $10^7 M_{\odot}$, and the  normalization factor 
\begin{equation}
A= \frac{ m_{\rm max}\, m_{\rm min}}{{m_{\rm max}}-{m_{\rm min}}}
\end{equation}
ensures that the integral over all masses and over the whole volume (or surface) is 1.
Note that in our runs we assume $m_{\rm max}=10^{7} M_{\odot}$; 
using the dissipationless model described in Section~\ref{Mod1},  we 
find that reducing $m_{\rm max}$ to $10^{6} M_{\odot}$
has the effect of changing the normalization of the scaling relations  by a factor $\sim 10$,
but did not change their overall slopes.

These distribution functions can then be used to calculate approximately how much mass is lost by the population of star clusters
due to their collisional evolution and interaction with the external tidal field of the galaxy (``dynamical dissolution''):
\begin{multline}\label{Mdot_diss}
\dot{M}^{\rm gc}_{\rm dissolution}=\\\frac{M^\star_{\rm bulge}}{M^\star_{\rm bulge}+M^\star_{\rm disk}} M_{\rm gc}\int \frac{\pi^{\rm bulge}(r,m_{\rm cl})}{t_{\rm tid}(r,m_{\rm cl})} 4\pi r^2\, \mbox{d}r \,\mbox{d} m_{\rm cl}\\
+\frac{M^\star_{\rm disk}}{M^\star_{\rm bulge}+M^\star_{\rm disk}} M_{\rm gc}\int \frac{\pi^{\rm disk}(r,m_{\rm cl})}{t_{\rm tid}(r,m_{\rm cl})} 2\pi r\, \mbox{d}r\, \mbox{d} m_{\rm cl}\,,
\end{multline}
with
\begin{equation} \label{diss-t}
t_{\rm tid}\approx 10 {\rm Gyr} \left( \frac{m_{\rm cl}}{2\times10^5 M_{\odot}} \right)^{\alpha} P(r)
\end{equation}
and 
\begin{equation}
P(r)=41.4\left( \frac{r}{\rm kpc}\right)  \left( \frac{V_{\rm vir}}{\rm km/s} \right)^{-1}\,,
\end{equation}
$V_{\rm vir}$ being the halo's virial velocity~\citep[e.g.,][]{2014ApJ...785...71G}.
Recent $N$-body simulations show that $\alpha \approx 2/3$ \citep{2008MNRAS.389L..28G}, which we adopt here.
Also, note that for simplicity  
we do not track separately the mass in star clusters in the bulge and that in the disk, but simply follow the evolution of the total mass in star clusters $M_{\rm gc}$ [cf. equation \eqref{mdotGC}]. In equation~\eqref{Mdot_diss} [and in equations~\eqref{evaporation} and~\eqref{infallrate} below] we therefore 
simply assume that the mass of star clusters in the disk 
is $\approx [M^\star_{\rm disk}/({M^\star_{\rm bulge}+M^\star_{\rm disk}})] M_{\rm gc}$, and that the mass of star clusters in the bulge is
$\approx [M^\star_{\rm bulge}/({M^\star_{\rm bulge}+M^\star_{\rm disk}})] M_{\rm gc}$.

In a similar fashion, we account for the evaporation of star clusters in isolation through
\begin{multline}\label{evaporation}
\dot{M}^{\rm gc}_{\rm evaporation}=\\\frac{M^\star_{\rm bulge}}{M^\star_{\rm bulge}+M^\star_{\rm disk}} M_{\rm gc}\int \frac{\pi^{\rm bulge}(r,m_{\rm cl})}{t_{\rm ev}(m_{\rm cl})} 4\pi r^2\, \mbox{d}r \,\mbox{d} m_{\rm cl}\\
+\frac{M^\star_{\rm disk}}{M^\star_{\rm bulge}+M^\star_{\rm disk}} M_{\rm gc}\int \frac{\pi^{\rm disk}(r,m_{\rm cl})}{t_{\rm ev}(m_{\rm cl})} 2\pi r\, \mbox{d}r\, \mbox{d} m_{\rm cl}
\end{multline}
where 
\begin{equation}
t_{\rm ev}=   \frac{17 m_{\rm cl}}{2\times 10^5 M_\odot}\, \mbox{Gyr}\,.
\end{equation}
Note that because the evaporation time-scale does not depend on the star cluster's location (since it describes evaporation in isolation), equation \eqref{evaporation} can be simplified
to
\begin{equation}\label{evaporation2}
\dot{M}^{\rm gc}_{\rm evaporation}= A M_{\rm gc}\int \frac{m_{\rm gc}^{-2}}{t_{\rm ev}(m_{\rm cl})} \,\mbox{d} m_{\rm cl}\,.
\end{equation}
 Note that our models do not include the possibility of close encounters with giant molecular clouds, which in some cases could significantly reduce the clusters' lifetime.
Also they rely on necessarily  simplified models for the galaxy. 
A more realistic model for the galactic potential might somewhat change 
the timescales introduced above. However, we note that analytical models of \NSC \ formation
indicate  that the final \NSC\ mass is not much affected by the
assumed slope of the galactic background density 
profile\ \citep[see Section 4 of][]{2013ApJ...763...62A}.

The total mass in star clusters
also decreases because the individual star clusters fall toward the central NSC, under the effect of dynamical friction, i.e.
\begin{multline}\label{infallrate}
\dot{M}^{\rm gc}_{\rm infall}=\\\frac{M^\star_{\rm bulge}}{M^\star_{\rm bulge}+M^\star_{\rm disk}} M_{\rm gc}\int \frac{\pi^{\rm bulge}(r,m_{\rm cl})}{t_{\rm df,bulge}(r,m_{\rm cl})} 4\pi r^2\, \mbox{d}r \,\mbox{d} m_{\rm cl}\\+\frac{M^\star_{\rm disk}}{M^\star_{\rm bulge}+M^\star_{\rm disk}} M_{\rm gc}\int \frac{\pi^{\rm disk}(r,m_{\rm cl})}{t_{\rm df,disk}(r,m_{\rm cl})} 2\pi r\, \mbox{d}r\, \mbox{d} m_{\rm cl}\,.
\end{multline}
The dynamical-friction timescale in the bulge 
is dominated by the interaction with the stellar background, and we therefore follow
Binney and Tremaine (2008) [equation (8.12)]:
\begin{equation}
t_{\rm df,bulge}(r,m_{\rm cl})\approx{15 {\rm Gyr}}\frac{6}{\log \Lambda} \left( \frac{r}{5{\rm kpc}} \right)^2 \frac{\sigma}{100{\rm km/s}}
\frac{10^7  M_{\odot}}{m_{\rm cl}}\,,
\end{equation}
where $\sigma$ is the bulge velocity dispersion (related to the halo's virial velocity by $\sigma\approx 0.65 V_{\rm vir}$~\citep{2002ApJ...578...90F}) 
and we choose $\log \Lambda=6$. As for the dynamical friction in the disk, we account for both the interaction with the stellar and gas component, i.e.
\begin{gather}\label{disk_DF}
{t_{\rm df,disk}(r,m_{\rm cl})}^{-1}={t_{\rm df,disk}^\star(r,m_{\rm cl})}^{-1}+{t_{\rm df,disk}^{\rm gas}(r,m_{\rm cl})}^{-1}\\
t_{\rm df,disk}^\star=\frac{L}{\dot{L}_{\star}}\,,\quad 
t_{\rm df,disk}^{\rm gas}=\frac{L}{\dot{L}_{\rm gas}}\,,\quad
L\approx m_{\rm cl} V_{\rm vir} r\,,\\
\dot{L}_{\star}=\left(\frac{G m_{\rm cl}}{\sigma_{\star}}\right)^2 \Sigma_{\star}\,,\quad
\dot{L}_{\rm gas}=\left(\frac{G m_{\rm cl}}{\sigma_{\rm gas}}\right)^2 \Sigma_{\rm gas}
\end{gather}
where $L$ is the orbital angular moment of a star cluster of mass $m_{\rm cl}$ belonging to the disk (and thus 
moving with velocity $\sim V_{\rm vir}$ at a separation $r$ from the galactic center),
while the angular momentum loss due to dynamical friction is expressed~\citep{2011ApJ...729...35A} in terms of
the surface densities $\Sigma_{\star}$ and $\Sigma_{\rm gas}$ of the disk's stellar and gaseous components,
the velocity dispersion of the stars in the disk  $\sigma_{\star}$, and the velocity dispersion/sound speed of the gas in the disk
$\sigma_{\rm gas}\approx 0.1 \sigma_{\star}$~\citep{2009MNRAS.396..141D}. 
Note that for $\sigma_{\star}$
we follow \citet{2005MNRAS.358..503K} and assume $\sigma_{\star}\approx 0.29 V_{\rm max}$, where $V_{\rm max}$ is the maximum rotational velocity
inside the disk. Also observe that in equation
\eqref{disk_DF} we are summing the inverses of the timescales, so that
the infall rates due to the dynamical friction from the gas and that due to the stars  get summed in equation \eqref{infallrate}.

Not all of the mass of the star clusters falling toward the NSC eventually accretes onto it, because star clusters get truncated
due to the galactic tidal field, and tidally disrupted by the central MBH. To account for these effects we assume that the rate of change of
the NSC mass due to infall of star clusters is
\begin{equation}
\dot{M}^{\rm NSC}_{\rm infall}=\dot{M}^{\rm gc}_{\rm infall} \mathcal{F}\,,
\end{equation}
where clearly the major difficulty lies in computing the factor $0\leq\mathcal{F}\leq1$. Since it would be computationally prohibitive 
to evolve the infall of the individual star clusters self-consistently within our semi-analytical galaxy formation model, we calculate
$\mathcal{F}$ with the ``monolithic'' $CliN$  model of Section~\ref{Mod1}, in order to derive an easy-to-use analytical expression. 

More specifically, approximating a star cluster as a King model,
its tidal radius at a distance $r_{\rm \NSC}$ from the 
center of a galaxy containing a MBH 
and a \NSC\ at its center is:
\begin{eqnarray}\label{rtr2}
r_t&\approx&
\frac{\sigma_K}{\sqrt{2}}\Big[4\pi G \rho_0 \left(\frac{r_{\rm \NSC }}{r_0} \right)^{-\gamma} \frac{\gamma}{3-\gamma}  \\ 
&&+\frac{3G(M_{\rm CMO})}{r_{\rm \NSC}^3}\Big]^{-1/2}~, \nonumber
\end {eqnarray}
where
$M_{\rm CMO}=M_{{\rm  \sbh}}+M_{\rm \NSC}$ is the total mass of the central massive objects  (i.e.,  \sbh\ plus \NSC\ mass),
 and  for the galaxy density profile we adopted the power-law model 
$\rho(r)=\rho_0\left(r/r_0\right)^{-\gamma}$.  
Note that to be consistent with the Hernquist profile used in $GxeV$, we set $\gamma=1$ in equation~\eqref{rtr2}.
Also, the distance $r_{\rm \NSC}$ is to be set to the outer radius of the NSC, i.e. the distance from the galaxy center 
below which the star cluster is assumed to have become part of the NSC. We assume $r_{\rm \NSC}=5 r_{\rm h}$, where
$r_{\rm h}$ is the NSC half-mass radius. Our detailed prescription for $r_{\rm h}$ will be presented in equation \eqref{eq:rh} below.
For a King model, the truncated mass $m_{\rm gc}(\sigma_K)$ of the star cluster is then related to its tidal radius via equation~(\ref{mt0}).

Using  the $CliN$ model described in Section~\ref{Mod1}, we find that more than $90\%$ of the \NSC\ mass 
comes from star clusters with initial mass $\gtrsim 0.1 m_{\rm max}$, and that the (mass-weighted)
mean central velocity dispersion of the star clusters contributing to the nucleus growth is 
$\langle \sigma_K \rangle \approx 20$ km/s. When 
setting 
$m_{\rm max}=10^7~M_{\odot}$, the average initial mass of these star 
clusters is $\langle m_{\rm gc,\,in}\rangle\approx 2.5 \times 10^6M_{\odot}$.
Assuming that all star clusters that decay to the center
 have similar central properties ($\sigma_K$), and that these properties 
do not change during the infall, we have
\begin{equation}\label{eq:tidal_disruption}
\mathcal{F} \approx \frac{m_{\rm gc} (\langle \sigma_K \rangle)}{ \langle m_{\rm gc,\,in}\rangle }\,,
\end{equation}
where $m_{\rm gc}(\langle \sigma_K \rangle)$ is given by equations \eqref{mt0} and \eqref{rtr2}.
Finally, to account for the possibility that a star cluster may undergo complete tidal disruption
\textit{before} it decays  to  a radius $r_{\rm \NSC}$, we compare the tidal radius given by equation \eqref{rtr2} with the core radius $r_{K}\approx 1$ pc of the
star cluster. If $r_t<r_{K}$, we set $\mathcal{F}=0$.

\subsubsection{Star-formation channel}
As mentioned above, NSCs may also form by star formation in the cold gas accumulating in the galactic center. In the galaxy formation
model of \citet{2012MNRAS.423.2533B} (and in the improved version of this model used by~\citet{2014ApJ...794..104S}), 
transfer of cold gas to a low-angular momentum nuclear reservoir -- available for accretion onto the central  \sbh\ --
is assumed to be correlated with star-formation events in the galactic bulge through a relation~\citep{2004ApJ...600..580G,2014ApJ...782...69L,2004ApJ...606..763H}
\begin{equation}\label{res_feed}
\dot{M}^{\rm res}_{\rm infall}= A_{\rm res} \psi_{\rm b}\,,
\end{equation}
where  $\psi_{\rm b}$ is the star formation rate in the gaseous bulge, and
$A_{\rm res}\sim 10^{-2}$ -- $10^{-3}$ is a free parameter, which we set to $A_{\rm res}\approx 6\times 10^{-3}$ in this paper (as in \citet{2014ApJ...794..104S}). 
In the model of \citet{2012MNRAS.423.2533B} and \citet{2014ApJ...794..104S}, star formation in the bulge is associated with major galactic mergers, and (less importantly) with bar instabilities in the galactic disk.

Note that the physical mechanism responsible for the loss of angular momentum of the cold gas and its transfer to this nuclear reservoir
may be the radiation drag caused by the newly formed stars~\citep{2001ApJ...560L..29U,2002MNRAS.329..572K,2003ApJ...583...85K}, or the reshuffling/shocks of the gas as a consequence of 
disk instabilities or mergers. More generally, a correlation between
bulge star formation and \sbh\ accretion/growth is expected based on the $M-\sigma$ relation for \sbh s and on the parallelism between the quasar luminosity and star formation history~\citep{2004ApJ...600..580G,2014ApJ...782...69L,2004ApJ...606..763H}.

The reservoir's gas is then made available to accrete onto the \sbh\ on the viscous timescale. Whenever the resulting viscous accretion rate exceeds $A_{\rm Edd} \dot{M}_{\rm Edd}$
($\dot{M}_{\rm Edd}$ being the Eddington mass accretion rate and $A_{\rm Edd}$ a free parameter), we truncate the accretion rate to that value. To allow for moderately 
super-Eddington accretion in the case in which MBHs form from light popIII star seeds (of mass $\sim 200 M_{\odot}$ at $z\sim 15-20$)~\citep{2001ApJ...551L..27M}, 
we set $A_{\rm Edd}=2.2$ for that seed model. This is
because some amount of super-Eddington accretion is known to be needed to reconcile light \sbh\ seeds with the quasar luminosity function at
high redshift~\citep{2014ApJ...784L..38M}. We also consider two ``heavy'' seed models (namely that of \citet{2008MNRAS.383.1079V} and that of \citet{2004MNRAS.354..292K}) whereby \sbh s grow from seeds of $\sim 10^5 M_\odot$
at $z\sim 15-20$, in which case we set $A_{\rm Edd}=1$. 

On top of these prescriptions, in this paper we assume that the gas in the nuclear reservoir also forms stars, which are assumed to contribute to the NSC.
To compute the star-formation rate, we need to choose a size for the reservoir. It is natural to assume that this size is comparable to
the observed size of NSCs. More specifically, we assume that the reservoir is disk-like, and has a (vertically-averaged) exponential density profile, whose scale radius we choose
such that the reservoir has the same half-mass radius $r_h$ as the NSC. The latter is assumed to be
\begin{equation}\label{eq:rh}
r_h=3 \mbox{ pc}\,\max\Bigg(\sqrt{\frac{M_{\rm dyn}}{10^6 M_{\odot}}},1\Bigg),
\end{equation}
with $M_{\rm dyn}=M_{\rm res}+M_{\rm NSC}$ the dynamical mass of the nucleus (including the gas in the nuclear region). Note that this scaling is inspired by observations of the size of NSCs, which is found to scale with
the square root of the luminosity~\citep{2012ApJS..203....5T}.

Once  a density profile for the reservoir is specified, we apply 
a star formation law similar to that
used by \citet{2014ApJ...794..104S} for galactic disks. Note that
\citet{2014ApJ...794..104S} improved on \citet{2012MNRAS.423.2533B} by considering different modes of star formation in
classical bulges forming from major mergers -- where star formation is assumed to take 
place in ``bursts'' of duration comparable to the bulge dynamical time -- as opposed to disks and pseudobulges arising from bar-instabilities of disks
-- where star formation is assumed to happen in a ``quiescent'' mode,
described by the prescription of \citet{2009ApJ...699..850K}.
Since star formation in the central region of our Galaxy is known to be weaker by at least a factor ten relative to what would be expected 
based on the observed gas densities~\citep{2014MNRAS.440.3370K},
it seems more appropriate to use the ``quiescent'' star-formation mode for our nuclear reservoir.\footnote{Observe
 that although star formation takes place quiescently in our model's nuclear reservoir, we also assume [cf. equation \eqref{res_feed}] that 
the reservoir's feeding is triggered by star-formation events in the bulge, i.e. mainly by turbulence-driven bursts of star formation following major mergers.}
Following therefore \citet{2014ApJ...794..104S} and \citet{2009ApJ...699..850K}, we assume that the star formation in the reservoir
takes place on a timescale $t_{\rm SF}$ and involves only a fraction $f_{\rm c}$ of the reservoir's gas, thus allowing us to write
an approximate star formation rate in the reservoir as
\begin{equation}\label{eq:sf_res}
\dot{M}^{\rm sf}_{\rm res}= \frac{f_{\rm c} M_{\rm res}}{t_{\rm SF}}\,.
\end{equation} 
Note that this simple prescription does not account for the spatial distribution
of the star formation activity, e.g. the observational fact that
younger stellar components seem to be more centrally concentrated than the rest of the NSC 
[see for example \citet{2014MNRAS.441.3570G}].

The fraction $f_{\rm c}$ of cold gas available for star formation
is set by the fraction of molecular gas for metallicities  $Z'>0.01$ (in solar units), when star-formation
happens in molecular clouds. At lower metallicities 
$Z'<0.01$, star formation takes place in the atomic phase~\citep{2012ApJ...759....9K}. In general, $f_{\rm c}$ is expected to
decrease as the metallicity decreases, but recent observations of nearby  spirals and dwarfs \citep{bigiel10}, as well as the Small Magellanic Cloud \citep{bolatto11}
show that it levels off at 2\%~\citep{2015ApJ...800...20G}.

We can thus summarize the dependence of $f_{\rm c}$ on the metallicity
by the explicit expression~\citep{2014ApJ...794..104S,2009ApJ...699..850K}
\begin{equation}
f_{\rm c} =  \left\{ \begin{array}{l l}
 1-\left[1+\left(\frac{3}{4} \frac{s}{1+\delta}\right)^{-5}\right]^{-1/5},  & ~{\rm ~if}~f_{\rm c} > 2\%,\\
 2\%, &  ~{\rm otherwise}
 \end{array} \right.
\label{eq:fh2}
\end{equation}
with
$$ s = \ln{(1+0.6 \chi)}/(0.04 \Sigma_{\rm 1} Z^{'}),$$
$$ \chi = 0.77 (1+ 3.1 Z^{'0.365}),$$
$$\delta = 0.0712 \left(0.1 s^{-1} + 0.675\right)^{-2.8},$$
$$\Sigma_{\rm 1} = \Sigma_{\rm res}/ (M_{\sun}~ \rm{pc}^{-2})\,.$$
The timescale $t_{\rm SF}$ is instead given by~\citep{2014ApJ...794..104S,2009ApJ...699..850K}
\begin{equation}
 t_{\rm SF}^{-1} =  (\rm 2.6 ~Gyr)^{-1} \times  \left\{ \begin{array}{l l}
 \left(\frac{\Sigma_{\rm res}}{\Sigma_{\rm th}}\right)^{-0.33}, &  \Sigma_{\rm res} <  \Sigma_{\rm th},\\
  \left(\frac{\Sigma_{\rm res}}{\Sigma_{\rm th}}\right)^{0.34}, &  \Sigma_{\rm res} > \Sigma_{\rm th}
\end{array} \right.
\label{eq:tsf_disc}
\end{equation}
with $\Sigma_{\rm th}=85 M_\odot/{\rm pc}^2$.
This expression is obtained by assuming that star formation happens in clouds, and the two branches appear 
according to whether the cloud density is determined by internal processes, or by the external pressure (in galaxies
with sufficiently high surface densities). Also, note that the typical cloud mass does appear in this expression because
it is related to the gas surface density by identifying it with the local Jeans mass, and by assuming marginal gravitational stability of the reservoir.
For the surface density of the reservoir, in both equation \eqref{eq:fh2} and \eqref{eq:tsf_disc}, we choose
 $\Sigma_{\rm res}\approx M_{\rm res}/(2\pi r_{\rm res}^2)$, $r_{\rm res}$ being the
scale radius of the reservoir's exponential surface density profile. This is the central value of the density for an exponential surface density profile. While 
this choice is quite arbitrary, we note that at high surface densities the combination $f_{\rm c}/t_{\rm SF}$ has a weak dependence on $\Sigma_{\rm res}$, namely $f_{\rm c}/t_{\rm SF}\propto\Sigma_{\rm res}^{0.34}$. Therefore, a different choice of $\Sigma_{\rm res}$ by a factor 2 (5) only changes $\dot{M}^{\rm sf}_{\rm res}$ as given by equation \eqref{eq:sf_res} by 0.1 (0.24) dex.

\begin{figure}
 \centering
 \begin{overpic}[scale=0.65]{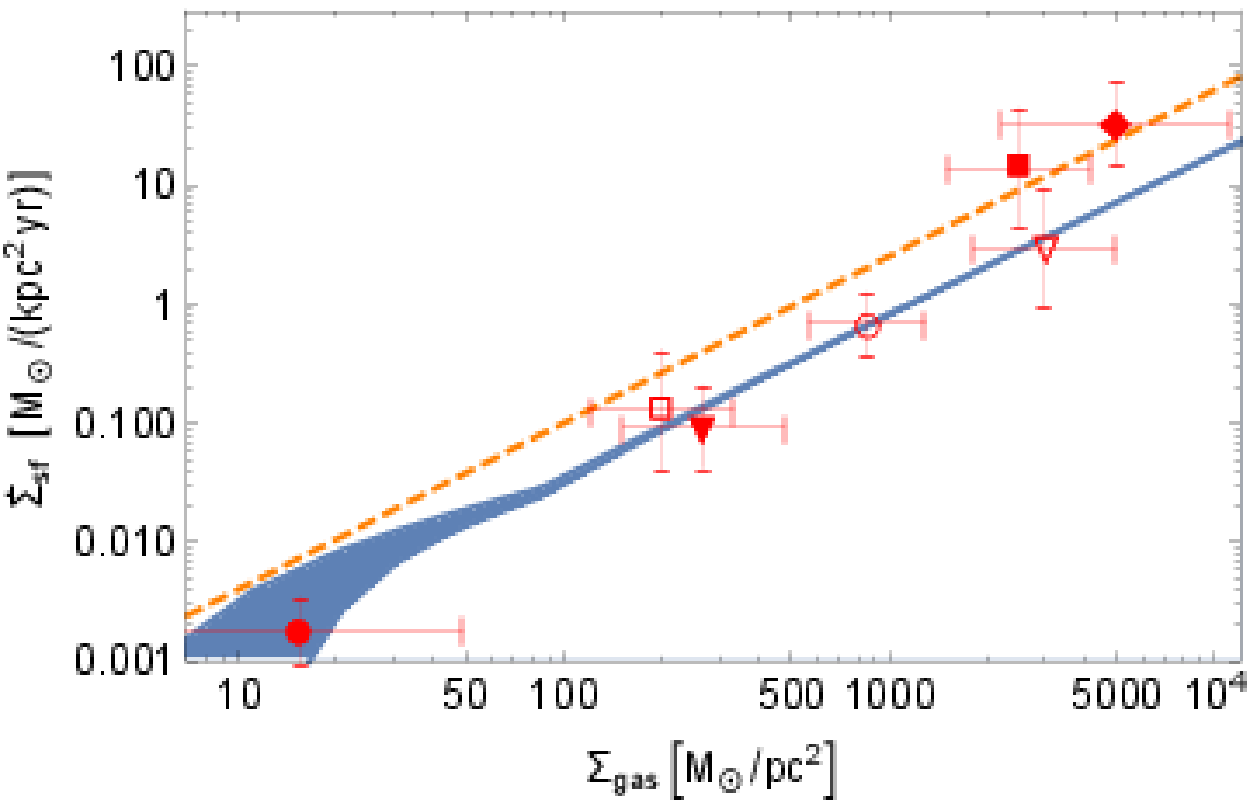}
     \put(16,43.5){\includegraphics[scale=0.43]{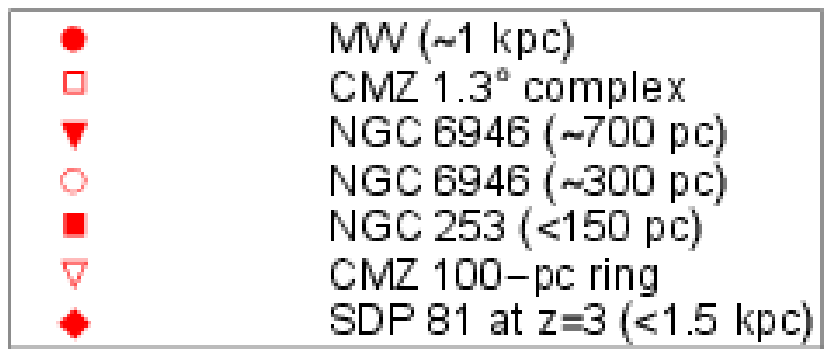}}
  \end{overpic}
 \caption{\label{SF}
 A comparison between our star-formation prescription in the nuclear region -- denoted by a blue line, the width of the line
 representing the scatter of the predicted star-formation rate with metallicity, which we assume in this figure to vary between 1 and 4 solar units --
 and the measured star formation in the Milky Way's central molecular zone (CMZ) as reported by \citet{2014MNRAS.440.3370K}; the average 
star formation rate in the central 150 pc of NGC 253,
as reported in Table 6 of \citet{leroy}; the Milky Way's star formation rate at 
galactocentric distance of $\sim 1$ kpc and that of NGC 6946 at galactocentric distances of $\sim 300$ and $\sim 700$ pc, as reported by \citet{2012ARAA..50..531K} in their Figure 7;
and the measured nuclear star formation in the star-forming galaxy SDP 81 at $z\approx 3$~\citep{2015arXiv150308720D,2015arXiv150505148S}.
For comparison, we also show by a dashed orange line the prediction of the Kennicutt-Schmidt law~\citep{Kenn}.}
 \end{figure}

We stress that equation \eqref{eq:sf_res}  is admittedly a rough prescription for the star formation in the nuclear regions. However,
it seems to work reasonably well 
when compared 
 to observations of the star formation rates in the central 500 pc of the Milky Way (the ``central molecular zone''). Indeed,
 in Figure~\ref{SF} we compare the predictions of our star formation prescription [where in equations \eqref{eq:fh2} and \eqref{eq:tsf_disc} we identify $\Sigma_{\rm res}$
with the surface gas density represented in the horizontal axis]
 with measurements of the star formation 
 in the central molecular zone at different angular scales (and thus different average gas surface densities).

In the light of this reasonable agreement, we chose not to include any feedback effects (AGNs or supernovae)
when considering star formation in the nuclear regions. We stress however that both kinds of feedback are
included when considering the bulge and disk star formation [cf. Figure \ref{Fig_model}, and \citet{2012MNRAS.423.2533B}]. They therefore indirectly affect
the star formation in the nuclear region, since the feeding of the reservoir depends on the larger-scale galactic evolution via equation
\eqref{res_feed}.

\subsubsection{Effect of galaxy mergers and black-hole binaries}
\label{sec:mergers}

\paragraph{\bf Delays between galaxy and black-hole mergers.} 

When two dark matter halos merge, the two galaxies, being more compact than the halos, initially retain their identity,
and are slowly brought together by dynamical friction. During this evolution, environmental effects such as tidal
stripping and tidal evaporation become important and contribute to remove mass from the smaller galaxy.\footnote{We refer
to \citet{2012MNRAS.423.2533B} for more details about how we compute the dynamical-friction timescale and these environmental effects.}
When the galaxies finally merge, within a few dynamical-friction time-scales after the halo merger,
 the MBHs of the two galaxies
are expected to form a binary system. 

This binary is slowly brought together by a variety of effects, i.e. dynamical friction at large separations,
and at smaller $\lesssim$ pc separations by gas interactions (if sufficient gas is present in the nuclear regions) or by
interaction with stars if the nuclear star cluster has a sufficiently dense core, or by triaxiality of the central potential. 
 Gas interactions (and in particular planet-like migration) might bring a MBH binary to merger on
time-scales $\lesssim 10^7-10^8$ yr~\citep{bence,2014SSRv..183..189C}, while three-body interactions with stars 
might require timescales 
up to $\sim$ Gyr~\citep{yu} or more. 
In fact, it is not completely obvious that a MBH binary would merge at all (within a Hubble time)
in the presence of three-body interactions with stars alone. This is known as the ``final-parsec'' problem~\citep{Begelman80}, but recent numerical simulations
suggest that in triaxial merger remnants
 (such as those that would be expected from a recent galaxy merger) 
 MBH binaries might coalesce on timescales of a few Gyr~\citep{lastPc1,2014arXiv1411.1762V,lastPc2,2015arXiv150505480V}.
Rotation in flattened galaxy models has also been suggested to help drive MBH 
binaries to coalescence, cf. \citet{2015arXiv150506203H}.
Also, if a MBH binary stalls at $\sim$ pc separation, a later galaxy merger may add another MBH to the system, forming a MBH triple.
At least for large $\sim 10^8 M_\odot$ MBH masses, triple MBH interactions
may trigger binary mergers (and possibly ejection of the smallest MBH) on timescales $\lesssim 10^8$ yr~\citep{hoffman}.
Because of the uncertainty about the mechanisms described above, we adopt here a set of minimal simplified prescriptions to
estimate the ``delay'' between galaxy and MBH mergers. These prescriptions are presented in the following.

In gas-rich environments, defined by the criterion $M_{\rm res}> M_{\rm  \sbh,1}+M_{\rm  \sbh,2}$,
we assume that the delay is simply given by the viscous time of the nuclear gas. At a distance $r$ from the central MBH,
the viscous time is simply $t_\nu\sim r^2/\nu$, $\nu$ being the kinematic viscosity. The latter can be approximately expressed in terms of
the gas velocity $v$ and the Reynolds number ${\cal R}$ as $\nu\sim r v/{\cal R}$~\citep{APA,betavisc}.
Since the kinematic viscosity is huge for accretion disks, it is natural to assume that ${\cal R}$ is the critical Reynolds number ${\cal R}_c\sim 10^2-10^3$
marking the onset of turbulence. Indeed, laboratory experiments find that once this critical Reynolds number is reached, the viscosity
increases significantly~\citep{APA,betavisc}.
Assuming now that $r\sim G (M_{\rm  \sbh,1}+M_{\rm  \sbh,2})/\sigma^2$
is the binary's influence radius and that $v\sim \sigma$, we can write
the ``delay'' in gas-rich nuclear environments as~\citep{2004ApJ...600..580G}
\begin{equation}
t_{\rm delay,\,gas}\sim t_\nu \sim {\cal R}_ct_{\rm dyn}
\end{equation}
where 
$t_{\rm dyn}=G (M_{\rm  \sbh,1}+M_{\rm  \sbh,2})/\sigma^3$ is the dynamical time at the influence radius.
In this paper we set ${\cal R}_c=10^3$. 
This prescription does indeed yield delays $\lesssim10^7-10^8$ yr in gas-rich environments, as expected.

In gas-poor enviroments (i.e. $M_{\rm res}< M_{\rm  \sbh,1}+M_{\rm  \sbh,2}$), three-body interactions with stars dominate and bring the MBHs
together on a timescale~\citep{Begelman80}
\begin{multline}\label{tgw}
 t_{\rm delay,\,stars}\sim 5 \,{\rm Gyr}\left(\frac{a_{\rm gr}}{4.5\times 10^{-2} \mbox{ pc}}\right)^4 \\\times\left[\frac{q}{(1+q)^2} \left(\frac{M_{\rm  \sbh,1}+M_{\rm  \sbh,2}}{10^8}\right)^3\right]^{-1}\,,
 \end{multline}
where $q=M_{\rm  \sbh,2}/M_{\rm  \sbh,1}\leq1$, and $a_{\rm gr}$ is the radius at which gravitational-wave emission becomes dominant over three-body interactions with stars
 at driving the binary's evolution, i.e.
\begin{multline}\label{eq:agr}
a_{\rm gr}\approx\; 1.65 \times 10^{-2} {\rm pc} \,\times \left[\frac{q}{(1+q)^2} \left( M_{\rm  \sbh,1}+M_{\rm  \sbh,2}\over {10^8 M_{\odot}}\right)^3\right.\\\times\left.\left( \sigma \over 200\, {\rm km/s} \right) 
\left( \rho_{\rm \star} \over 10^3 M_\odot {\rm pc^{-3}}\right)^{-1}\right]^{1/5}\,.
\end{multline}
Here, the density $\rho_{\rm \star}$ of the stellar background in which the binary moves is given by
the maximum of the stellar-bulge density and the density of the NSC. For the latter, we 
use the average NSC density within its half-mass radius, i.e. $\rho_{\rm NSC} \approx 
[M_{\rm NSC}-2(M_{\rm  \sbh,1}+M_{\rm  \sbh,2})]/[(8/3) \pi r_h^3]$, where we decrease the mass of the NSC 
by twice the mass of the binary to approximately account for the mass deficit it causes on the NSC during its inspiral
[cf. equation \eqref{ejmass} below]. 
As for the stellar-bulge density, our model assumes
a Hernquist profile, as mentioned earlier. Comparisons to N-body simulations~\citep{sesanaprep} 
show that the appropriate radius 
where this density profile needs to be evaluated in order to estimate $a_{\rm gr}$ correctly is the binary's mass influence radius, i.e. 
the radius at which the enclosed bulge mass (in stars) equals twice the binary mass. 

In order to model triple MBH systems when they form, we utilize the results of \citet{hoffman}. By using numerical simulations, \citet{hoffman} found that MBH triples typically
trigger the merger of the two more massive components on timescales of $10^8$ yrs (on average) for binary masses of $\sim 6\times 10^8 M_\odot$ (and roughly comparable
masses for the two components), while the lightest MBH may be ejected from the galaxy or be left wandering far from the galaxy's center, 
or less likely come back and coalesce with the remnant of the inner binary's merger. Moreover, the time-scales
for the merger of the inner binary present a log-normal scatter of about $1.4$ dex around the median value of $10^8$ yrs, as shown in Figure 8 (upper panel) in \citet{hoffman}. To extend these results to arbitrary masses, we rescale these time-scales
with the system's dynamical time at the binary's hardening radius, and we thus obtain an average ``delay'' due to MBH triple interactions given by
\begin{equation}
t_{\rm delay,\,triple}\sim 10^8 \,{\rm yr} \left(\frac{M_{\rm  \sbh,1}+M_{\rm  \sbh,2}}{6\times 10^8 M_\odot}\right)^{1/4}\times\frac{8 q^{3/2}}{(1+q)^3}\,.
\end{equation}
Whenever a triple forms, we then extract the delay with which the two more massive objects merge from a log-normal distribution centered on this timescale, and with r.m.s. of 1.4 dex. For simplicity, we also assume that the lightest MBH is ejected from the galaxy and therefore lost to the subsequent galaxy evolution.

Finally, \citet{hoffman} noted that triple MBH interactions are less effective at driving MBHs to merger in low-mass systems, where the third incoming MBH is ejected from the galaxy before it can
shrink the inner binary to separations at which gravitational-wave emission is important. Indeed, the escape velocity from a galaxy scales as the velocity dispersion, i.e. $v_{\rm esc}\sim \sigma\sim M^{1/4}$,
where we defined the mass of the inner binary, $M\equiv M_{\rm  \sbh,1}+M_{\rm  \sbh,2}$, and we used the
Faber-Jackson relation $M\sim \sigma^4$.
The separation $a_{\rm gw}$ 
at which a MBH binary is driven to coalescence by gravitational-wave emission within a Hubble time $t_{\rm H}$
can be obtained from Eq.~\eqref{tgw} (with the replacements $a_{\rm gr}\to a_{\rm gw}$ and $t_{\rm delay,\,stars}\to t_{\rm H}$), and scales as $a_{\rm gw}\sim M^{3/4}q_\nu^{1/4}$
(with $q_\nu=M_{\rm  \sbh,1}M_{\rm  \sbh,2}/(M_{\rm  \sbh,1}+M_{\rm  \sbh,2})^2$ the symmetric mass ratio). In order for the third MBH to
be able to shrink the inner binary to the separation $a_{\rm gw}$, 
the velocity $v_3$ of the third MBH 
should be lower than $v_{\rm esc}$
when the binary's separation is  $a_{\rm gw}$.
The energy of the third MBH can be estimated simply by energy equipartition as 
$E_3\sim M_{\rm  \sbh,3} v_{\rm 3}^2 \sim G M_{\rm  \sbh,1} M_{\rm  \sbh,2}/a_{\rm gw}$, from which one gets $v_{\rm 3}\sim M^{1/8}q_\nu^{3/8}q_3^{-1/2}$, with $q_3=M_{\rm  \sbh,3}/M$. This in turn
gives $v_{\rm 3}/v_{\rm esc}\sim M^{-1/8}q_\nu^{3/8}q_3^{-1/2}$. The normalization of this ratio can be estimated using the results of \citet{hoffman}, who find $v_{\rm esc}\sim 1400$ km/s and $v_{\rm 3}\sim 750$ km/s
for $M\sim 6\times 10^8 M_{\odot}$, $q_\nu \sim 1/4$ and $q_3\sim 0.25-0.3$ (cf. their Figs. 5 and 11), thus giving 
\begin{equation}\label{vratio}
\frac{v_{\rm 3}}{v_{\rm esc}}\sim 0.5 \times \frac{q_\nu^{3/8}}{q_3^{1/2} [M/(6\times 10^8)]^{1/8}}\,.
\end{equation}
Therefore, whenever this equation gives $v_{\rm 3}> v_{\rm esc}$, we assume that the third MBH is ejected without driving the merger of the inner binary.
This implies that triple MBH interactions become ineffective at driving the merger of binary
systems with total mass $M\lesssim 2\times 10^6\times (q_\nu^3/q_3^4)\, M_\odot$.

\paragraph{\bf Mass deficit caused by black-hole binaries}
The presence of a MBH binary has profound effects on the  NSC, causing a ``mass-deficit'' in the central regions of the galaxy. 
Indeed, during the binary's inspiral, three-body interactions with the stars in the nuclear region
result in an exchange of energy between the binary (which becomes more and more bound) and the stars, to which large velocities are imparted that are 
capable of ejecting them from the galactic nucleus. Also, when the MBH binary finally merges, the resulting \sbh\ remnant acquires  
a kick velocity up to a few thousands km/s due to the anisotropic emission of gravitational waves~\citep{2007PhRvL..98w1102C}, and this 
also contributes to remove stars from the galactic core. A similar mass deficit may be caused by the
ejection of the lightest MBH in a triple system.

We estimate the 
mass-ejection rate from the NSC due to the MBH binary's inspiral as
\begin{equation}\label{ejmass}
\dot{M}^{\rm insp}_{\rm ej}\approx (M_{\rm  \sbh,1}+M_{\rm  \sbh,2})  \frac{0.7q^{0.2} +0.5 \ln \left({a_{\rm h}/ a_{\rm gr}}\right)}{t_{\rm delay,\,stars}}
\end{equation}
where $q=M_{\rm  \sbh,2}/M_{\rm  \sbh,1}\leq 1$ is the binary's mass ratio and
 $a_{\rm h}$ is the the semi-major axis at which the binary becomes ``hard'' (i.e. tightly bound).

The first term in the numerator of equation~\eqref{ejmass} represents the mass scoured by the  \sbh\ binary before it becomes hard,
where we have 
identified the ejected mass with the mass deficit as defined in~\citet{2006ApJ...648..976M}; the second term
represents instead the mass ejected from $a_{\rm h}$ to $a_{\rm gr}$~\citep{2013degn.book.....M}; 
the explicit expressions for  $a_{\rm h}$ is given for instance in~\citet{2013degn.book.....M}, i.e.
\begin{equation}
a_{\rm h}\approx\; 0.27 (1+q)^{-1} \left(M_{\rm  \sbh,2}\over {10^7 M_{\odot}} \right) \left( \sigma \over 200\, {\rm km/ s} \right)^{-2} {\rm pc}~,
\end{equation}
while for $a_{\rm gr}$ we utilize equation \eqref{eq:agr}.
Also,  in the denominator of equation \eqref{ejmass}, we note the presence of the timescale $t_{\rm delay,stars}$, computed via equations \eqref{tgw} and \eqref{eq:agr}.
That timescale accounts for the fact that both terms in the numerator where computed for MBH binaries in gas-poor environments,
hence the mass deficit at the numerator should be ``spread'' over the timescale characterizing stellar interactions.
Of course, if the MBH binary lives in a gas-poor enviroment, in our model the inspiral lasts exactly $t_{\rm delay,\,stars}$, and the final mass deficit caused
by the binary is $\sim (M_{\rm  \sbh,1}+M_{\rm  \sbh,2})  [0.7q^{0.2} +0.5 \ln \left({a_{\rm h}/ a_{\rm gr}}\right)]$.
If the binary evolution is instead driven by gas interactions or by the formation of a MBH triple, that mass deficit is suppressed
by a factor respectively $t_{\rm delay,\,gas}/t_{\rm delay,\,stars}$ or $t_{\rm delay,\,triple}/t_{\rm delay,\,stars}$.\footnote{Note 
that we have also tried setting the delays between  galaxy and MBH mergers
to very small values $t_{\rm delay}\sim 10^6$ yr irrespective of the mechanism driving the binary's evolution prior to the gravitational-wave dominated phase.
In this case, the mass deficit is always $\sim (M_{\rm  \sbh,1}+M_{\rm  \sbh,2})  [0.7q^{0.2} +0.5 \ln \left({a_{\rm h}/ a_{\rm gr}}\right)]$. This
test confirms that our results for  the NSC evolution are reasonably robust, and that the  overall conclusions of this paper do not depend on our particular model for the delays.}
 
In addition to the mass deficit caused during the inspiral, when the MBH binary finally merges the resulting MBH remnant 
acquires a kick,
which, as mentioned above, can further remove mass from the NSC. We estimate this mass 
deficit as~\citep{2008ApJ...678..780G}
\begin{equation}\label{ejkick}
M_{\rm ej}^{\rm kick}\approx 5 (M_{\rm  \sbh,1}+M_{\rm  \sbh,2}) \left(V_{\rm kick}/V_{\rm esc} \right)^{1.75}\,,
\end{equation}
where  $V_{\rm kick}$ is
 the recoil velocity of the \sbh\ remnant and $V_{\rm esc}$ is the escape velocity from the galactic center. 
  The latter can be easily calculated within our model, from the mass and density profiles of the bulge and NSC.
 As for  $V_{\rm kick}$, we follow \citet{2012MNRAS.423.2533B} and use the analytical formula of \citet{2010ApJ...719.1427V}, which fits the results of numerical-relativity simulations. 

Similarly, in the case of triple MBH systems, if the lightest MBH gets ejected from the system before it can cause the inner binary to merge [i.e., in our model, if 
the ratio $v_3/v_{\rm esc}$ given by equation \eqref{vratio} is larger than one], we assume that the ejected MBH causes a mass deficit $M_{\rm ej}^{\rm kick,\,triple}\sim 5 M_{\rm MBH,3}$, with $M_{\rm MBH,3}$ the ejected MBH's mass. We stress that for simplicity we neglect the mass deficit caused by the recoil of the lightest MBH in the cases in which the triple interactions trigger the merger of the inner binary (i.e., in our model, when $v_3/v_{\rm esc}<1$). 
Summarizing, the total mass deficit due to MBH coalescences is 
$M_{\rm ej}=M_{\rm ej}^{\rm insp}+M_{\rm ej}^{\rm kick}+M_{\rm ej}^{\rm kick,\,triple}$.

In addition to the effect of MBH binaries, we also account for the possible 
tidal disruption of NSCs by MBHs during galaxy mergers. Indeed, if a merger takes place between a galaxy (``1'') that contains
a NSC (but not a MBH) and another galaxy (``2'') hosting a MBH (and possibly a NSC), the NSC of ``galaxy 1'' will be dragged by dynamical friction toward
the MBH (and NSC if present) of galaxy 2, and will therefore be tidally truncated/disrupted, as in the case of star clusters falling toward the nucleus
of an isolated galaxy. To model this effect, we assume that the NSC resulting from such a galaxy merger has mass 
\begin{equation}\label{tidesInMergers}
M_{\rm NSC}=M_{\rm NSC,\,1}\times\mathcal{F}+M_{\rm NSC,\,2}\,.
\end{equation}
The fraction $0\leq\mathcal{F}\leq 1$ accounts for the tidal truncation/disruption effects, and is calculated via equations \eqref{mt0}, \eqref{rtr2}, and~\eqref{eq:tidal_disruption} with
$\sigma_K $ taken to be the velocity dispersion of ``NSC 1'' (which one can compute in terms of its mass and size). 
Note that calculating $\mathcal{F}$
in the same way as for star clusters is justified, at least to first approximation. Indeed,
as mentioned earlier, the star clusters that contribute the most to equation \eqref{eq:tidal_disruption}
have average initial mass $\langle m_{\rm gc,\,in}\rangle\approx 2.5 \times 10^6M_{\odot}$, which is of the same
order of magnitude as a typical NSC mass.
As in the case of star clusters, if
$r_t<r_{K}=r^{\rm NSC\,1}_h/c$ (with $c=30$ a typical NSC concentration parameter), we assume that ``NSC 1'' is fully disrupted, and set $\mathcal{F}=0$. 
On the other hand, if both ``galaxy 1'' and ``galaxy 2'' have MBHs, we assume that the gravitational field of each massive 
BH protects its NSC from tidal truncation and disruption, and we set $\mathcal{F}=1$. Similarly, if neither galaxy contains a MBH, we set $\mathcal{F}=1$.

\begin{figure*}
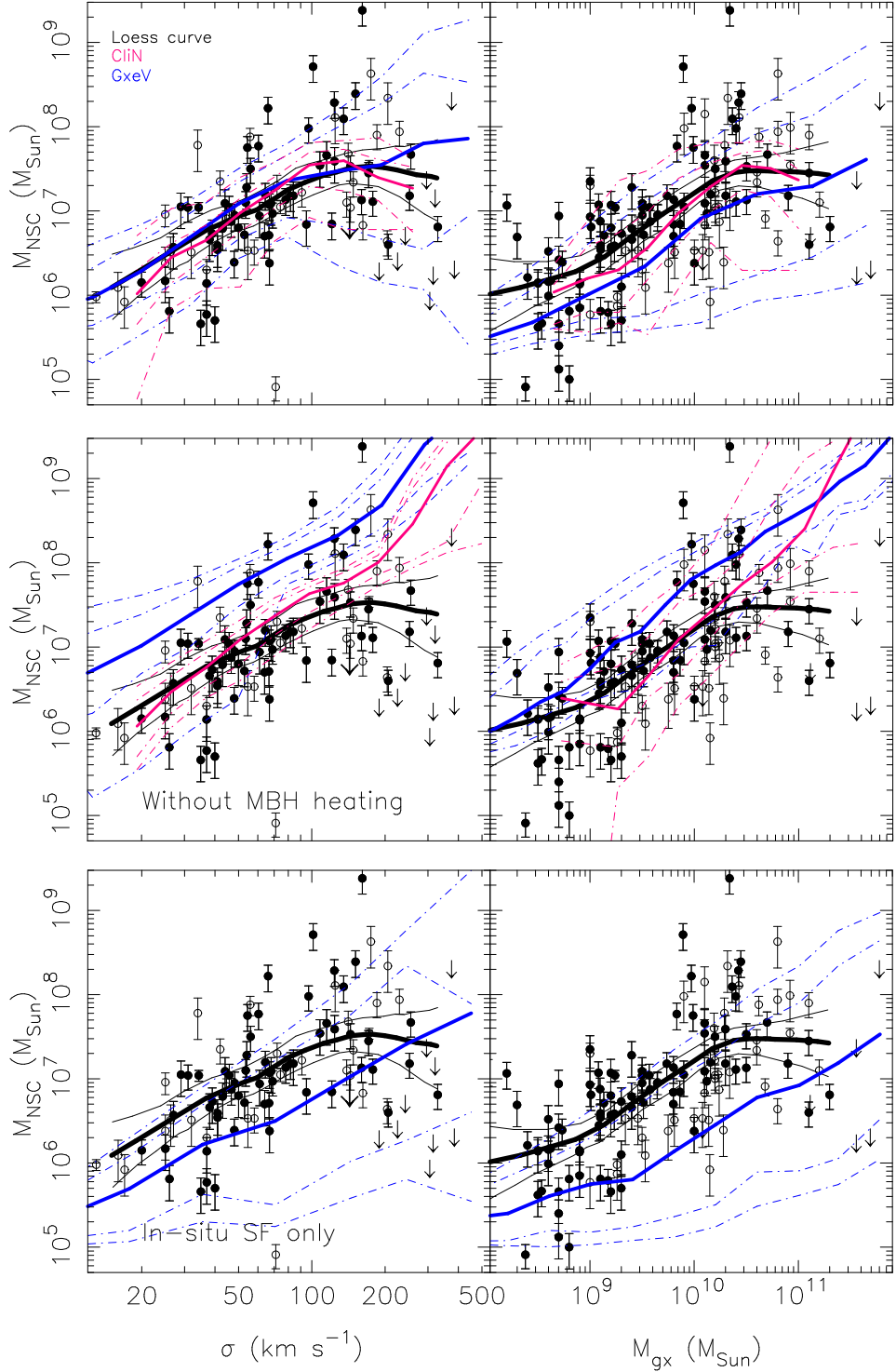

\centering
\subfigure{\includegraphics[width=2.28in,angle=270.]{Fig3a.eps}}  \\
\subfigure{\includegraphics[width=2.3in,angle=270.]{Fig3b.eps}} \\
\subfigure{\includegraphics[width=2.83in,angle=270.]{Fig3c.eps}} \\
\caption{$M_{\rm \NSC}$ against galaxy velocity dispersion (left panel) and galaxy total mass (right panel).
Filled  circles represent early-type galaxies (S0 or earlier) 
while open  circles are late-type systems. 
Short  (long) arrow symbols
are \NSC s in early (late) type galaxies with 
only an upper limit on their mass~\citep{Neumayer:2012}. 
Since it is likely that these systems do not contain a \NSC\,
we decided to not  include these upper limits in our regression analysis.
Black lines are the computed $Loess$ curves and thin black 
lines give the corresponding  $95\%$ variance bands.
Red lines give the median of the $CliN$ model where only cluster
inspirals are considered and corresponding 
$70\%$ and $90\%$ confidence-level regions. Blue lines are the results of our semi-analytical model $GxeV$
including cluster inspirals, in-situ star formation and processes related to hierarchical galaxy and MBH evolution.
Dashed lines indicate the $70\%$ and $90\%$ confidence-level regions.
Middle panel: same as top panel, but with  \sbh \ dynamical effects not included in the 
semi-analytical computations (see Section~\ref{bh-eff} for details) . 
Bottom panel: same as upper panels, but \NSC s
are allowed to grow only via in-situ star formation~(see Section~\ref{isvsm} for details).
\label{main}}
\end{figure*}

\begin{figure}
\centering
\includegraphics[width=2.8in,angle=0.]{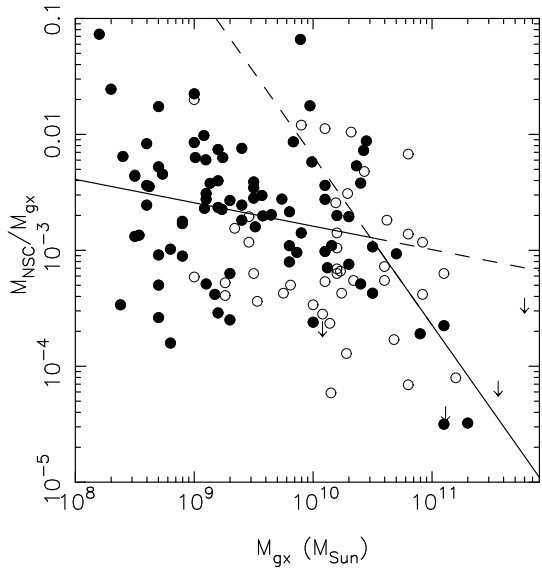}
\caption{ \NSC \ mass fraction as a function of the total stellar mass of the host galaxy.
 Filled circles are early type galaxies, open circles are late-type galaxies.
 Black lines give simple linear fits to the data with
 $M_{\rm gx}\lesssim 10^{10.5}M_{\odot}$ and with
 $M_{\rm gx}\gtrsim 10^{10.5}M_{\odot}$.
In the low mass regime the \NSC \ to galaxy mass ratio shows a weak correlation with $M_{\rm gx}$, 
while in the more massive systems ($M_{\rm gx}\gtrsim 10^{10.5}M_{\odot}$) there is
 a clear trend with $M_{\rm gx}$, with the \NSC\ mass fraction 
 decreasing  rapidly with the galaxy mass.
 This is consistent with the flattening of the scaling relations found
 at high galaxy masses in Figure~\ref{main}.
  \label{mass-fr}}
\end{figure}

\section{Data Sample}\label{data-an}
We compare our models to observational data in order to gain insights on the reliability of such models. We then
  use our models to put constraints on
  the origin and evolution of \NSC s and to understand how
  their properties are linked  to the evolution of  \sbh s and their host galaxies.

Our sample of \NSC\ objects was obtained by combining  data from 
\citet{2013ApJ...763...76S}, \citet{Erwin:2012}, \citet{Neumayer:2012} and by estimating  \NSC\ 
object masses of the galaxies in the Fornax cluster catalog of   
 \citet{2012ApJS..203....5T}.
     
\citet{Erwin:2012}  give mass estimates for 
a total of 18  galaxies that cover Hubble types from S0 to Sm with dynamically determined \NSC\ masses, and
15 Sa and later type galaxies  from \citet{2006AJ....132.1074R} with \NSC\ mass estimates based on high-resolution spectroscopy. Velocity dispersions for 9 of these galaxies were obtained from Table 2 of \citet{Neumayer:2012},
the rest from the  HyperLeda database when available.
\citet{2013ApJ...763...76S} constructed their  sample of \NSC s using photometric data from \citet{Ferrarese:2006},
\citet{Balcells:2007}, \citet{Graham:2009} and \citet{Graham:2012b}.  Table~1 of \citet{2013ApJ...763...76S} reports $76$
galaxies containing a \NSC\ with a well determined mass, and total mass and velocity dispersion estimates
for the majority of these.
The uncertainties on the nuclear object masses 
are given by \citet{Ferrarese:2006},  \citet{Balcells:2007} and \citet{Graham:2009}  as $45\%$, $33\%$, and a factor of two, respectively.

We  additionally obtained estimates for the stellar masses of the full galaxy,  $M_{\rm gx}$,
and   the NSC, using the photometric data for the 43 early-type galaxies in
the Fornax cluster catalog of  \citet{2012ApJS..203....5T}. 
To determine, $M_{\rm gx}$, we multiplied the total galaxy luminosity in the $K$ band given in the HyperLeda database 
by a mass to light ratio of $M/L_K=0.8$~\citep{Bell:2001}, based on the typical colors of the
bulge population. Galaxy total B-magnitudes, $M_{\rm B}$, and velocity dispersions, $\sigma$, were also taken from the  HyperLeda database. The \NSC\ masses were obtained by multiplying  the 
published luminosities by the appropriate mass-to-light ratio, obtained by using the empirical correlations between color and mass-to-light-ratio  given in  \citet{Bell:2003}. Typical errors on $M_{\rm \NSC}$
are $\sim 35\%$.

 After eliminating  duplicate
galaxies contained in more than one of the samples -- for each galaxy we included 
 in the collective sample the mass estimate
with the smallest uncertainty -- we collected a total of 89 galaxies with reliable estimates for
  both $M_{\rm \NSC}$ and $\sigma$, 127 objects 
     with determined $M_{\rm \NSC}$ and $M_{\rm gx}$, and
     208 galaxies with  measured \NSC\ mass and host galaxy total B-magnitude.
 
   Although  we were not able to obtain all galaxy properties for every object from the literature,
      our data collectively represent  the largest sample to date of \NSC\ and host galaxy physical properties.  In Figure~\ref{main} we plot the \NSC\ masses as a function of the velocity dispersion
      of the galaxy spheroid as well as the galaxy mass, for early~(filled circles) and late~(open circles) type systems.
       
\subsection{ Data analysis}
In order to investigate the correlation between  \NSC s and their host galaxy properties
we use a  locally-weighted linear regression model~(\emph{Loess})
to fit non-parametric curves to the data~\citep{cleveland79,cleveland+devlin88}.
\emph{Loess} does not require one to specify a global function of any form to fit a model to the data, 
as it combines multiple low-order polynomial  regression models  
in a $k$-nearest-neighbor-based metamodel.  
Unlike a more ``standard'' simple linear regression analysis, 
no assumption is made that the data can be fitted by a straight line.
\emph{Loess} scatter-plot smoothing can therefore be used to reveal complex relationships that could be overlooked
with traditional parametric estimation strategies.
The obvious trade-off  is that  in general it is not possible to express a \emph{Loess} model with a simple mathematical formula.

The smoothness of the \emph{Loess} regression function is determined by the amount of data points used
for the local fits, a quantity controlled by the span parameter, $\alpha$.
Here we select an optimal value of $\alpha$ by using the generalized cross 
validation criterion~\citep[GCV;][]{craven79}.
The basic idea is that the optimal representation of the data is obtained by adopting the
smoothing parameter  that minimizes the mean-squared error of the fit~\citep{golub79,li85}.

The solid thick black curves in Figure~\ref{main} are $Loess$ interpolations 
  obtained with the optimal smoothing parameters $\alpha=0.63$ and $0.71$ for
  $M_{\rm \NSC}~{\rm vs}~\sigma$ and $M_{\rm gx}$ 
 respectively. The $95\%$  variance  bands of the \emph{Loess} curves 
are shown as thin black curves.

The first noteworthy point is that
the $Loess$ curves shown in Figure~\ref{main}  are characterized by a  significant bending 
 at $\sigma\gtrsim 100~{\rm km/s}$ and $M_{\rm gx}\gtrsim 10^{10.5} M_{\odot}$, indicating
 NSC - host galaxy scaling  relations that are flat or even declining  for the 
 most luminous galaxies.
 Moreover, the reconstructed variance bands follow the general trend of the $Loess$ curves, indicating that 
it is unlikely that the non-linear features present in the curves are due to random fluctuations in the data alone.

We tested whether or not a linear parametrization of the relations provides 
an adequate description of the data by using a $F$-test to compare  the $Loess$ fits to
 simpler weighted linear fits~\citep{fox99}. The $F$ tests showed that the 
 null hypothesis -- i.e., that the bivariate $Loess$ model yields no improvement in the fit over the linear regression --
can be rejected at a high level of significance. 
For all three relations investigated here,
the $Loess$ curves provide a better description to the data than linear models at $>99\%$ confidence level.
This confirms  that the non-linear features  present in the functional dependence of  $M_{\rm \NSC}$ on the galaxy 
properties are likely not due to noise variability in the data.

 A complementary view is provided in Figure~\ref{mass-fr}, where we plot the \NSC \ mass fraction as a function of the total stellar mass of the host galaxy, i.e., $M_{\rm NSC}/M_{\rm gx}~vs~M_{\rm gx}$.
 Evidently,  in the low mass regime ($M_{\rm gx}\lesssim 10^{10.5}\ M_\odot$) the 
 \NSC \ to galaxy mass ratio  shows a weaker correlation  with total galaxy mass
(with Kendall's rank  correlation coefficient $\tau_B=-0.28$ and associated probability value $p=0.06$), 
while in the more massive systems ($M_{\rm gx}\gtrsim 10^{10.5}M_{\odot}$) there is
 a clear and steeper trend with $M_{\rm gx}$, in the sense that the \NSC\ mass fraction 
 decreases rapidly with the galaxy mass. This is in contrast with the results of 
\citet{2013ApJ...763...76S}, who claimed quite a steep decrease of the \NSC \ mass
ratio based on fits to the entire sample distribution. 

We finally note that although our analysis demonstrates that current data are
consistent with a significant bending of the \NSC -host galaxy scaling relations,
further observations and analysis, combined with theory, will be needed in order to refine our results.
On the other hand, the broadening of the \NSC \ mass distribution for
the highest luminosity galaxies, as well as the presence of 
low mass nuclei ($\sim 10^6~M_{\odot}$) in high mass galaxies ($M_{\rm gx}\gtrsim 10^{11}M_{\odot}$)
certainly poses a
serious concern for previous  claims of the existence of fundamental correlations between 
\NSC\ and host galaxy properties. Moreover, as we show in the following, the particular form 
of these relations is in agreement
with what is expected on the basis of our semi-analytical models of \NSC\ formation.

\begin{figure}
\centering
\includegraphics[width=3.in,angle=270.]{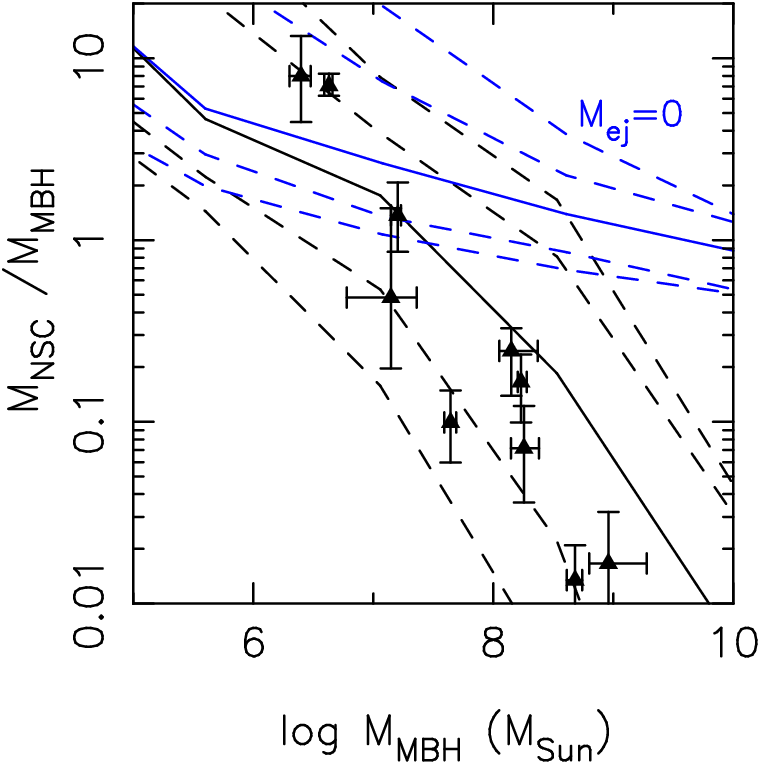}
\caption{Fraction of \NSC\ to  \sbh\ mass as a function of  \sbh \ mass. This
plot measures the relative importance of the two types of central objects as
one proceeds from \NSC -dominated to  \sbh\-dominated nuclei. 
The points are observational data and represent galaxies with measured  \sbh\ and \NSC\ mass~\citep{Graham:2009,Erwin:2012}.
The black curve and corresponding $70\%$ and $90\%$ confidence bands are obtained 
through our fiducial galaxy formation model where all the relevant effects described 
in Section~\ref{GxeV} were included. The blue lines correspond to a model in which the
scouring effect of  \sbh\ binaries forming during galaxy mergers was ignored, i.e., we
set $M_{\rm ej}=0$ in these integrations. 
Note how the  predictions of the two models start to diverge at
$M_{\rm  \sbh}\approx 10^8M_{M_\odot}$ where  \sbh s and \NSC s
have comparable masses in the model with $M_{\rm ej}=0$; 
at $M_{\rm  \sbh}\gtrsim 10^8M_{M_\odot}$  \sbh \ mergers become efficient at
eroding the surrounding \NSC.
\label{M-NSC}}
\end{figure}

\section{results}

The red and blue lines in Figure~\ref{main}
show the resulting  \NSC\ mass as a function of $\sigma$ obtained respectively with $CliN$
after $10~$Gyr of evolution and with  $GxeV$ at $z=0$.
These curves give the median output at a given $\sigma$ or $M_{\rm gx}$
as well as its $70\%$ and $90\%$
confidence-level regions, i.e. the regions containing respectively
$70\%$ and $90\%$ of the \NSC s produced by our models at a given
stellar mass.
The $CliN$ model median can be fit  by 
$M_{\NSC} \approx 10^{7} M_{\odot}(\sigma / 50~{\rm km/s)^{2}}$ and
$M_{\NSC} \approx 10^{7} M_{\odot} \left(M_{\rm gx} / 10^{10} M_\odot\right)^{0.9}$ at 
$\sigma \lesssim 100 {\rm km/s}$ and $M_{\rm gx} \lesssim 10^{10.5} M_{\odot}$
respectively. 
This appears  to be in good agreement
 with analytical expectations -- \citet{2013ApJ...763...62A}
 finds 
$M_{\rm NSC} \approx 10^7 M_{\odot} \left({\sigma}/{50~{\rm km/s}}\right)^{3/2}$,
adopting an idealized  isothermal sphere galaxy model. 
Similar scalings and a similar scatter around the median values
are found with the galaxy evolution model $GxeV$, which takes into account both
cluster inspirals and in-situ star formation. The similarity between the results of the two models
with and without in-situ star formation 
suggests that dynamical friction migration of star clusters   is a fundamental  process for
the growth of the nuclei. However, as  we show below in Section~\ref{isvsm},  dissipative processes 
also play an  important role, contributing a significant fraction of the
total \NSC\ masses in our models.
 
Figure~\ref{main} 
shows that at  $\sigma \gtrsim 100 {\rm km/s}$ and $M_{\rm gx}\gtrsim 10^{10.5} M_{\odot}$,
the scaling relations produced with $CliN$ (i.e., a purely dissipationless formation model) 
appear to flatten  toward the most luminous galaxies in agreement with 
the results of the data analysis of Section~\ref{data-an}.
However, a  more careful analysis of the plot also shows that such a model
fails to explain the existence  of  \NSC s of mass $\gtrsim 10^8M_{\odot}$
which are clearly present in the observational data.
The reason for this discrepancy is that in $CliN$
we have assumed that all galaxies contain a  \sbh , and that these  \sbh s are 
in place at the center of all  initial galaxy models before the \NSC s  grow around them. 
These assumptions are not fully supported by our galaxy formation model~(see Section~\ref{Z-ev-sec})
and  artificially lead to an enhanced mass removal from 
 the stellar clusters accreting onto the central NSC, because of the
tidal field of the \sbh .

The galaxy evolution model $GxeV$  produces  scaling relations at $z=0$
that are in remarkably good agreement with the observed correlations.
The scatter around the median value 
clearly increases at the high mass end of the \NSC\ distribution.
A large population of \NSC s are produced 
 at $\sigma \gtrsim 100 {\rm km/s}$ that are significantly underweight with
 respect to what the same model would predict
 by simply extrapolating the  scaling correlations and scatter 
 from low ($M_{\rm gx}\lesssim 10^{10.5} M_{\odot}$) 
 to high galaxy masses. In the following section we argue that 
 the formation of such ``underweight"  \NSC s and,
consequently, the apparent bending of the scaling correlations 
can be explained in terms of the interaction of the \NSC s with 
their host galaxy  \sbh s.

\subsection{The Role of Massive Black Holes}\label{bh-eff}
In our semi-analytical models  \sbh s affect the formation and evolution of 
\NSC s in two important ways:
(i) the strong tidal field of the central  \sbh\ accelerates the mass loss from stellar clusters
as they enter  its sphere of influence and eventually disrupts them as they come closer than
 a distance $r_\mathrm{disr}$, where
\begin{equation}\label{eq:rtidal}
\frac{M_{\rm  \sbh}}{\frac43\pi r_\mathrm{disr}^3} \approx \rho(0)
\approx \frac{9}{4\pi G}\frac{\sigma_K^2}{r_K^2} .
\end{equation}
After a cluster is disrupted, most of its stars  are dispersed around $r_\mathrm{disr}$,
which will limit the nuclear density within this radius.  Thus, a preexisting  \sbh \
will  limit the amount of mass that can be transported to 
the center by decaying stellar clusters~\citep{2013ApJ...763...62A}.
As mentioned earlier [cf. equation \eqref{tidesInMergers}], 
a similar effect applies to a NSC falling toward a central MBH after a galaxy merger.
(ii) during galaxy mergers,   \sbh \ binaries form and  harden by
ejecting surrounding stars \citep{2001ApJ...563...34M}. 
 By dynamically heating the \NSC , a \sbh \ binary 
will  lower the NSC central stellar density, 
or fully destroy the NSC when the binary's mass is 
significantly larger than the \NSC\ mass~\citep{BG:10}. 
The kick imparted to the merger remnant
by asymmetric emission of gravitational wave radiation will remove additional mass from the
galaxy nucleus \citep{2008ApJ...678..780G}.

To demonstrate  the role of  \sbh s  in the evolution of the nuclei,  we
performed additional 
simulations where the two dynamical effects due to  \sbh s mentioned above were not included.
(Hereafter we refer to these models as models ``without \sbh\ heating''.)
The  blue and red curves in the middle panels of Figure~\ref{main} show the 
scaling correlations resulting from such models.
When compared to  our fiducial models (upper panels), these
new integrations produce steeper and significantly tighter 
$M_{\rm NSC}-\sigma$
and $M_{\rm NSC}-M_{\rm gx}$ relations.
This indicates that the break in the \NSC \ scaling relations,
for which we found evidence in the observational data,
can be attributed  to  the interaction of \NSC s with their companion
 \sbh s.
 
Simple arguments can indeed be used to understand why the 
break of the \NSC \ scaling relations occurs at
$\sigma \approx 100 {\rm km/s}$.
 From equation (\ref{eq:rtidal}), by requiring
  $r_\mathrm{disr}\gtrsim r_{\rm \NSC}\approx 10~$pc, and adopting
$r_K ={\rm 1 pc}$ and $\sigma_K = {\rm20~km/s}$, 
 we find that   \sbh s more massive than roughly $\approx 10^8 M_{\odot}$ 
 will suppress the further growth  
 of a \NSC\ via accretion of stellar clusters. 
Thus, as the  \sbh\ grows, the contribution
of star cluster inspirals  to \NSC\ growth decreases and eventually 
stops for $M_{\rm  \sbh} \gtrsim 10^8 M_{\odot}$. Using ~\citep{2005SSRv..116..523F,Ferrarese:2006}: 
\begin{eqnarray}\label{sigmabh}
M_{{\rm  \sbh}} \approx 8.6 \times 10^6 \left( \sigma \over {100 \rm km/s}  \right)^{4.41}~M_\odot,
\end{eqnarray}
a  \sbh \ mass larger than $\sim 10^8 M_{\odot}$ corresponds to stellar spheroids 
with velocity dispersion $\sigma \gtrsim 150~{\rm km/s}$. This is consistent with the  value of  $\sigma$
at which the \NSC -host galaxy scaling correlations appear to flatten in the data sample.

\begin{figure}
\centering
\subfigure{\includegraphics[width=1.55in,angle=270.]{Fig6a.eps}} \\
~\subfigure{\includegraphics[width=1.92in,angle=270.]{Fig6b.eps} }\\
\caption{ 
The local fraction of nucleated galaxies of our models compared 
to observational constraints on the nucleated fraction of 
early type galaxies in Virgo~\citep{Cote:2006}, 
Fornax~\citep{2012ApJS..203....5T}, and Coma~\citep{2014MNRAS.445.2385D}.
The lower panels are for late type galaxies. 
The left panels
show the results of our fiducial models with 
all the relevant effects included; the 
right panels correspond to  models with no dynamical 
heating due to  \sbh s (see text for details). 
A comparison between these latter models and our fiducial models
indicate that the lack of \NSC s in galaxies more massive than
$\approx10^{11}~M_{\odot}$ is due to 
  \sbh s that 
fully destroy the surrounding clusters during galaxy mergers,
and also quench their
growth by disrupting migrating clusters due to their strong central tidal field.
\label{fract}}
\end{figure}

\begin{figure}
\centering
\includegraphics[width=1.8in,angle=270.]{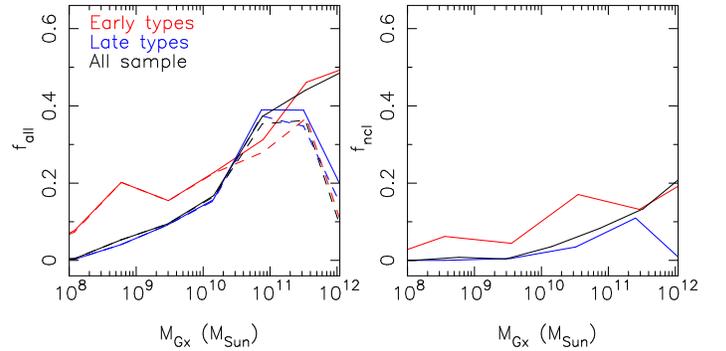}
\caption{ 
The left panel displays  the predicted  fraction of all galaxies, $f_{\rm all}$, containing 
a  \sbh \ ($M_{\rm  \sbh}\gtrsim 10^5M_{\odot}$; solid lines), and  containing a  \sbh \ plus a \NSC \  (dashed lines).
The right panel shows the fraction of nucleated galaxies, $f_{\rm ncl}$, that contain 
a central AGN, identified here as a  \sbh \ with bolometric luminosity $L>10^{10}~L_{\odot}$. 
}
\label{fract2}
\end{figure}

Another    argument leads us  to identify $M_{{\rm  \sbh}} \approx 10^8 M_{\odot}$
as the critical value of  \sbh \ mass above which \NSC s are significantly affected.
The impact of a   \sbh \ binary
on a central cluster depends on the $M_{\rm \NSC}/M_{{\rm  \sbh}}$ mass ratio.
Given that during a merger a  \sbh \ binary will eject a mass comparable to its total mass, 
 a requirement for a  \sbh\ binary to affect significantly the nuclear structure of a galaxy is 
  that its mass is of the order of or larger than the mass of the surrounding cluster, i.e., 
 $M_{\rm  \sbh}/M_{\rm \NSC} \gtrsim 1$. 
 
 Using a standard weighted linear regression
on our data gives the relation $\sigma \approx 50~{\rm km/s}(M_{\rm \NSC}/10^7{M_{\odot}})^{1.6}$;
using equation~(\ref{sigmabh})
 we find the new relation:
 \begin{equation}\label{nsc-bh}
 \frac{M_{{\rm  \sbh}}}{6\times 10^7 M_\odot} \approx \left( \frac{M_{\rm \NSC}}{6\times 10^7 M_\odot}\right)^{2.7}.
 \end{equation}
 Thus, the mass of the  \sbh \ grows faster then the mass of the \NSC , with the
transition from \NSC \ dominated to  \sbh-dominated galaxies occurring at
$M_{{\rm  \sbh}} \approx 10^8 M_{\odot}$ or $\sigma \approx 150~{\rm km/s}$.
This  is roughly the value of $\sigma$ at which \NSC s appear to be significantly affected by 
their host galaxy  \sbh s in our models.

 The total mass ejected from the center depends also
 on the number of stages in the merger hierarchy that
have occurred since the  \sbh s first formed --
i.e., after $N_m$ mergers, the mass deficit is ~$\sim 0.7 N_m  M_{\rm  \sbh}$,
 with $M_{ \sbh}$ the final (current) black hole mass~\citep{2006ApJ...648..976M}. 
Some low-mass galaxies today might have never experienced 
a major merger. By contrast, more massive galaxies form 
via  mergers of primordial lower mass halos, and
underwent an above average number of mergers.
In addition, since the occupation fraction
of   \sbh s increases with galaxy mass, 
 the scouring effect of  \sbh \ binaries is
 enhanced in the higher mass spheroids.

From Figure~\ref{main}, we also see that
the $GxeV$ model generates two distinct populations of \NSC s at 
$\sigma \gtrsim 100~{\rm km/s}$, which can be easily identified by looking at
the model confidence bands.
Nuclei  with  $M_{\rm \NSC}\lesssim 10^{7} M_{\odot}$ have masses that do not
strongly correlate with their host galaxy mass and velocity dispersion --
the mass of these \NSC s lies significantly below the  mass
  that we would  obtain
  by extrapolating the  scaling correlations   from low  to high galaxy masses.
   This  population of underweight nuclei 
 is a result of the disruptive 
 \sbh \ binary mergers  that have
 partially  eroded the surrounding clusters.
 \NSC s  above the model median seem instead to follow the general trend outlined in the low mass galaxy region of the plot, where $M_{\rm NSC} \sim \sigma ^2$.
 The host galaxies of such nuclei  did not  experience
a major merger event since the epoch of the last major gas accretion event, 
so that their  \NSC s remained  essentially unaffected by  \sbh \ binaries up to the present epoch.
This picture is in agreement with expectations based on  hierarchical models of galaxy formation
-- for example, \citet{2010MNRAS.406.2267F} find that
 for a dark halo mass of $10^{12}~M_{\odot}$ 
 only  31, 53 and 69 per cent of these halos have experienced
a major merger since $z =$1, 2 and 3, respectively. 

We can now ask which of the two processes, tidal disruption of star clusters or
mass ejection from  \sbh\ binaries, is responsible for the broadening of
the scaling correlations found in the data. 

We find that both $CliN$ models with $M_{\rm ej}=0$ and ones without the tidal truncation/disruption of clusters by the central  \sbh\ 
produce scaling correlations that appear to flatten at $\sigma\gtrsim 100~{\rm km/s}$.
The \NSC \ scaling correlations produced by 
$GxeV$  models in which we set $M_{\rm ej}=0$  (but in which we included the tidal disruption of migrating 
clusters) are instead at odds with observations, as they
show no broadening of the \NSC \ host galaxy property correlations for high velocity dispersion galaxies;
 only $GxeV$  models that do  take into account  the scouring effect of \sbh\ binaries were found to be in  
good agreement with the observed relations.
We conclude that in $GxeV$ the crucial ingredient 
to reproduce   the $M_{\rm \NSC}$-host galaxy property scaling relations 
is  the scouring effect of  \sbh\ binaries,
while the tidal stripping of migrating clusters by central \sbh s is only a secondary effect in these models.
 We remark  that the $GxeV$ model has a clear advantage over the $CliN$ model,
because the former follows the \textit{hierarchical} evolution of NSCs and MBHs along merger-trees, while the latter assumes
a \textit{monolithic} evolution. Although the results of
the two models are generally in good agreement with one another, the differences
outlined above 
are mainly a result of the implicit assumption made  
in the $CliN$ model  that the \NSC s \textit{always} grow around \textit{pre-existing} \sbh s.
This leads to an artificially enhanced mass removal from 
 the stellar clusters accreting onto the central \NSC, because 
of the  \sbh \ tidal field.

To illustrate more clearly  the  role of  \sbh\ binaries in shaping their companion \NSC \ properties,
we compare in Figure\ \ref{M-NSC}  the observationally constrained \NSC \ to  \sbh\ mass ratio 
in galaxies containing both types of central objects to  the same ratio obtained (i)~in our fiducial 
 $GxeV$ model containing all the relevant  effects described in Section~\ref{GxeV}, and
(ii)~a model in which we set $M_{\rm ej}=0$, i.e., in which the scouring effect of  \sbh\ binaries was not included in 
 the calculation. Clearly, the scouring effect due to  \sbh\ binaries 
 described by equations~(\ref{ejmass}) and (\ref{ejkick}) is the
 crucial ingredient to 
 reproduce the observed correlation. Also, note that the 
correlations produced by the two models start to diverge from each other at 
$M_{\rm  \sbh}\approx 10^8M_{\odot}$, in agreement with our predictions\ \citep[see also][]{2015ApJ...806L...8A}.

\subsubsection{\NSC\ and  MBH occupation fraction}
Observationally, the frequency of
nucleation in early type galaxies
is found to increase sharply from zero for spheroids brighter than $M_B = -19.5$
to $\gtrsim 90\%$
for galaxies fainter than this magnitude~\citep{Cote:2006,2012ApJS..203....5T}. 
In a sample of 332 late type galaxies, 
\citet{2014MNRAS.441.3570G}
found that  $\gtrsim 80\%$ of these galaxies harbor a well defined \NSC . 
Hence \NSC s are found in most galaxies of all Hubble type, but tend to disappear in
the brightest spheroids.

 Figure~\ref{fract} compares the frequency of nucleation obtained in our models
 to that of  late and early-type galaxies, as derived from observations.
 We define here early-type galaxies as systems with bulge-to-total mass ratio $M_{\rm bulge}/ M_{\rm gx} \gtrsim 0.7$, and late-type galaxies 
as systems with $M_{\rm bulge}/ M_{\rm gx} \lesssim 0.7$.
In agreement with observations,
our models predict that almost all early-type galaxies 
that are less massive than $\approx 10^{11}~M_{\odot}$ contain a \NSC ,
and that the frequency of nucleation is nearly zero for galaxies with mass larger than this
value. 
We note that the exact occupation number
obtained through $GxeV$ should be considered somewhat approximate --
the identification of a galaxy as nucleated in this model is  uncertain given
 that we cannot directly reconstruct the density profile of the \NSC \ and
 compare it to the density profile of the galaxy background. Nevertheless, the 
 occupation fraction is found to decrease significantly  for early type galaxies with mass  
 larger than $10^{11}~M_{\odot}$,
in good agreement with observational constraints.
The bottom panel of   Figure~\ref{fract} shows the frequency of nucleation  in late type systems.
Approximately $\sim 90\%$ of these galaxies were found to contain a \NSC, also in 
fairly good agreement with
observations.

The observed drop in the nucleation frequency at
high galaxy masses is often attributed  to the disruptive effects of   \sbh \  binaries
forming during galaxy mergers~\citep{BG:10} and to the tidal disruption of migrating  clusters
at large galactocentric distances by  central  \sbh s~\citep{2013ApJ...763...62A}.
In the right panels of Figure~{\ref{fract}} we show the nucleated fraction 
in  models where the dynamical heating due to  \sbh s on migrating clusters
and during mergers was not included. In these models the percentage of nucleation
remains close to $100\%$ regardless of galaxy luminosity,  demonstrating
that  \sbh s are responsible for the absence of \NSC s in the most luminous 
galaxies, as also argued in previous work.

\NSC s and  \sbh s are known to coexist in some galaxies
across a wide range of masses and Hubble types~\citep{2008ApJ...687..997S}.
However, not all \NSC s contain a  \sbh ; an example is 
the M33 \NSC \ that has no central  \sbh\ down to highly
constraining limits~\citep{2001Sci...293.1116M,2001AJ....122.2469G}.
Studies indicate that at least some \NSC s can host a central  \sbh ,
but the overall occupation fraction of  \sbh s in \NSC s remains unknown.
The left panel of Figure~\ref{fract2} displays the fraction of all galaxies in the $GxeV$ model that
contain a  \sbh\ and the fraction of galaxies containing 
both a  \sbh\ and a \NSC. Our models predict that a high fraction of galaxies of
intermediate and low luminosity contain a central \NSC , but that the NSCs tend to disappear 
in massive galaxies. Thus, the total fraction of mixed systems with both a \NSC\ and a  \sbh\
is roughly  equal to the  \sbh\ occupation fraction for galaxies with mass $M_{\rm gx}\lesssim 10^{11}~M_{\odot}$,
and is roughly equal to the \NSC \ occupation fraction for galaxies more massive than  $10^{11}~M_{\odot}$.
In the right panel of Figure~\ref{fract2} we show the fraction of nucleated galaxies that contain 
a  \sbh\ with bolometric luminosity $L>10^{10}L_{\odot}$.
We find that the fraction of  galaxies with a \NSC \ that also have an active  \sbh\ increases from
$\sim 5 \%$ to $30\%$  going  from $M_{\rm gx}\sim 10^9~M_{\odot}$ 
to  $\sim 10^{12}~M_{\odot}$. These results are in fairly good agreement with the observational
results from \citet{2008ApJ...687..997S}, who found that about 
$10\%$ of \NSC s in their spectroscopic sample also host an AGN.

\begin{figure}
\centering
\mbox{\subfigure{\includegraphics[width=2.5in,angle=270.]{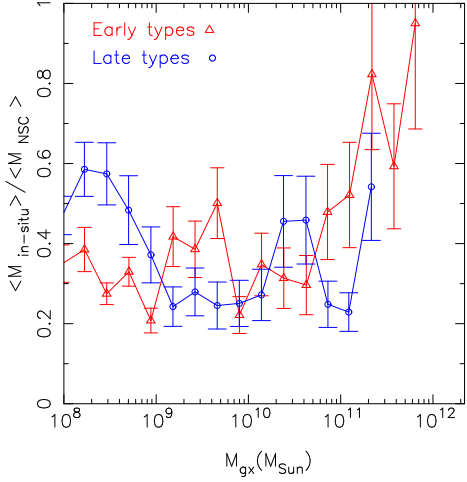}}\quad
}
\caption{Ratio of the median mass obtained in the only in-situ star formation 
model to the median mass of our fiducial model which  also includes cluster 
inspiral processes~($\langle M_{\rm in-situ} \rangle / \langle M_{\rm \NSC} \rangle$).
Error bars represent $1\sigma$ uncertainties. A large fraction of the mass
of our model \NSC s is built up locally from episodes of star formation triggered by infalling gas. 
\label{insitu-fr} }
\end{figure}

\begin{figure}
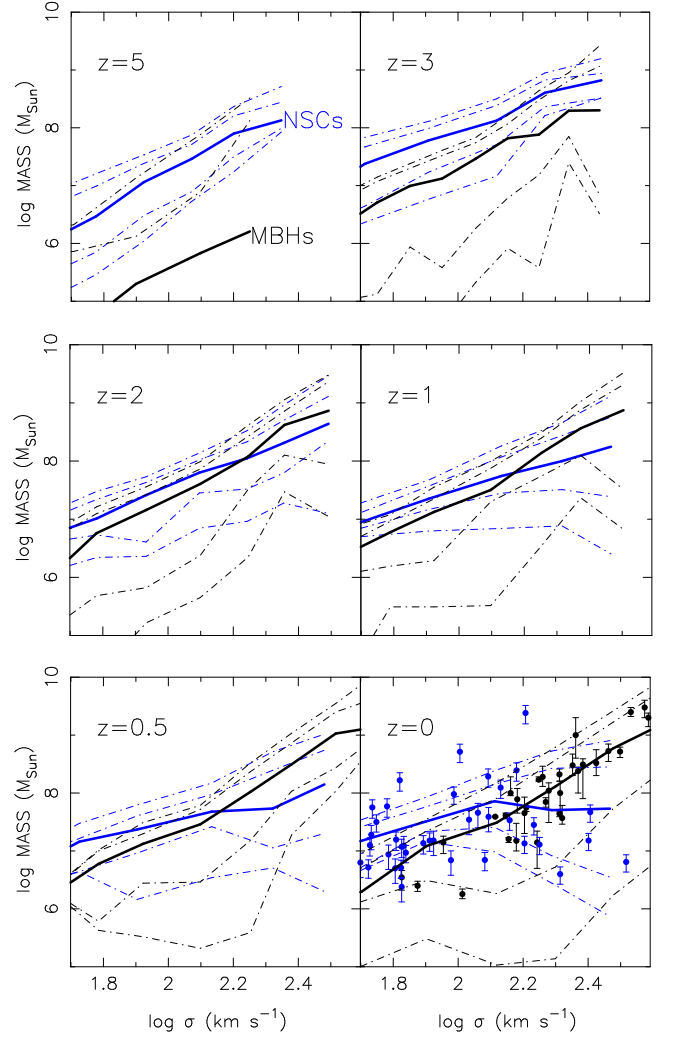

\centering
\subfigure{\includegraphics[width=1.59in,angle=270.]{Fig9a.eps}}  \\
\subfigure{\includegraphics[width=1.59in,angle=270.]{Fig9b.eps}} \\
\subfigure{\includegraphics[width=1.92in,angle=270.]{Fig9c.eps}}\\
\caption{ Scaling correlations of \NSC s and  \sbh s in our galaxy formation model at different redshifts.
At high redshift the \NSC s are the dominant central component of galaxies; between
$z=3$ and $z=1$ the  \sbh s grow faster 
 and by $z=2$ they are the dominant central component in galaxies 
with $\sigma \gtrsim 100~{\rm km/s}$.
After this point,  \sbh \ mergers become efficient at carving out the pre-existing nuclei.
This effect induces the bending/broadening 
of the $M_{\rm \NSC}$-$\sigma$ relation toward high galaxy masses at low redshifts, which can be clearly 
seen in the bottom panels of the figure. In the bottom right 
blue points represent \NSC s and black points represent  \sbh s \citep{2002ApJ...574..740T}
\label{Z-ev}}
\end{figure}

\begin{figure}
\includegraphics[width=3.in,angle=270.]{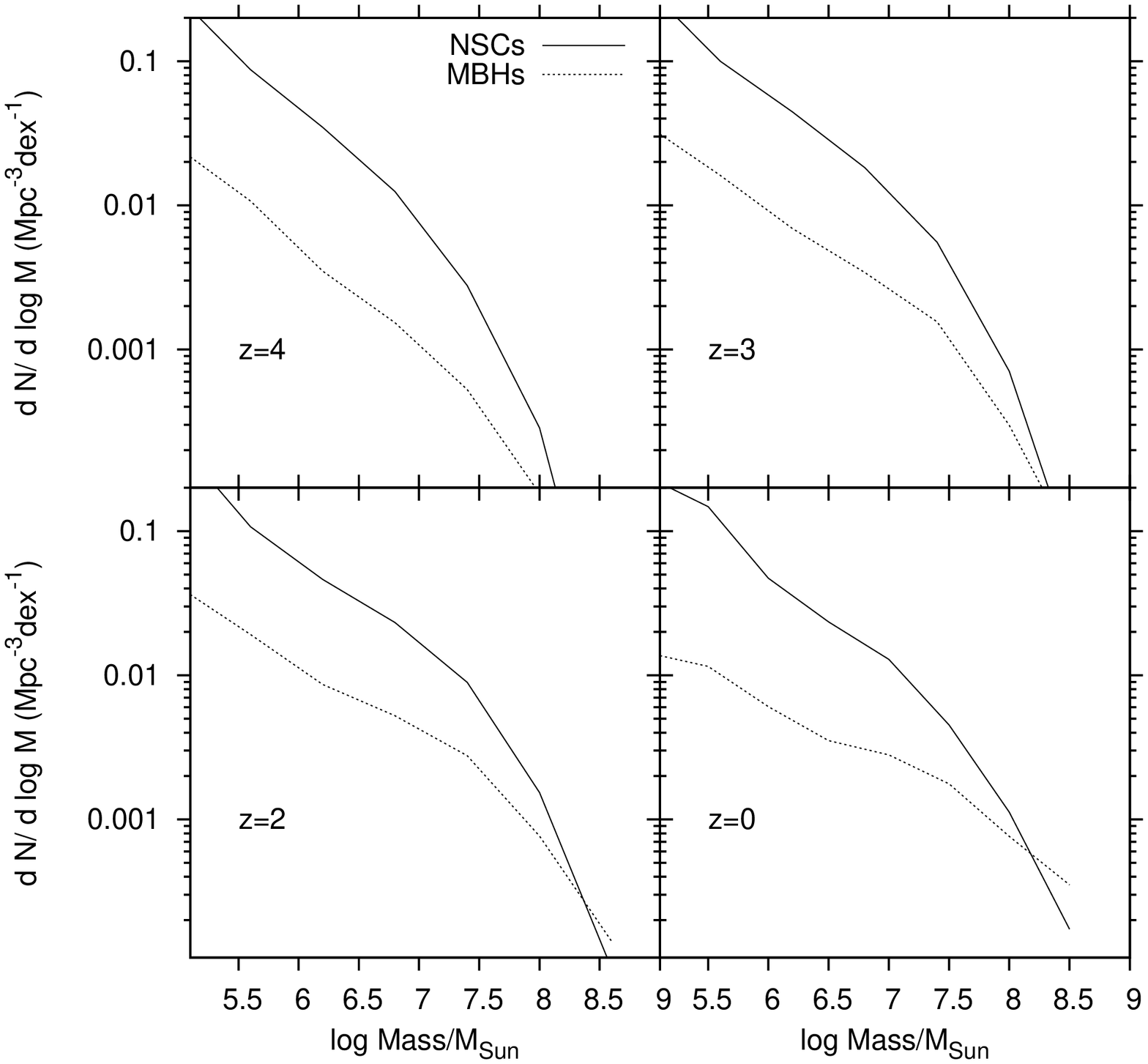}
\caption{Mass distribution of  \sbh s and \NSC s at various redshifts. 
At high redshift, $z\lesssim 4$, the \NSC s are typically more massive than  \sbh s.
At $z=0$ \NSC \ masses are typically of the order $10^{6-7}M_{\odot}$, with only a few rare clusters
having masses above $10^8M_{\odot}$. Note the transition from \NSC\ dominated to \sbh\ dominated galaxies
 for  $M_{\rm MBH} \gtrsim 10^8M_{\odot}$ occurring at  $z\lesssim 2$.
\label{Z-ev-M}}
\end{figure}

\subsection{In situ {\it vs} migration: the relative contribution to \NSC\ growth}\label{isvsm}
The lower panels of Figure~\ref{main} show the scaling relations generated by
a version of the  $GxeV$ model with no contribution from cluster inspirals. 
In these models, the only mechanism responsible  for the formation and growth
of the nuclei is local fragmentation of gas that leads to star formation and
 in-situ build up of a central compact cluster.

The lower panels of 
Figure~\ref{main} demonstrate that  even a purely dissipative  model
provides a quite good description of the observed \NSC-host galaxy scaling relations.
The  median of the mass distribution as a function of
$\sigma$
 can be  fit by $M_{\rm \NSC}\approx 6.5\times 10^6  (\sigma/100)^{1.4}M_{\odot}$.
Hence,  in this model we obtain scaling relations that are  consistent with the observation that 
\NSC s follow scaling relations that are  shallower than the corresponding ones
for  \sbh s.

We can  derive analytical 
scaling  relations for the in-situ formation
model by considering the  \sbh \ and reservoir characteristic  timescales of evolution.
Let us consider a situation in which the 
MBH is hosted by a gas-rich nucleus [resulting from a recent starburst, cf. equation \eqref{res_feed}], i.e. $M_{\rm MBH}\ll M_{\rm res}$. The MBH will
then grow approximately at the Eddington rate, i.e. $\dot{M}_{\rm MBH} \approx M_{\rm MBH}/t_{\rm salp}$, with $t_{\rm salp}$ the Salpeter timescale
\begin{equation} \label{tsp}
t_{\rm salp} =
\frac{k \epsilon c}{4\pi G}=4.5 \times 10^7 \left({\epsilon \over 0.1}\right) ~{\rm yr}\,,
\end{equation}
where $\epsilon$ is the radiative efficiency and $k\equiv 0.398{\rm cm^2~g^{-1}}$ the electron scattering opacity.
Gas accumulates at the center of the galaxy mainly during starbursts [cf. equation \eqref{res_feed}]. Bursts of star formation
will take place on the dynamical timescale of the galactic bulge, i.e. 
\begin{equation} \label{tdy}
t_{\rm dyn} = \frac{R}{\sigma} =10^7 \left( \frac{\sigma}{100 {\rm km/s}}\right)^{2.06}~{\rm yr}~,
\end{equation}
with $R$ the galactic bulge scale radius. In the last expression we have 
used the fact that both $R$ and $\sigma$ are related to the total galaxy luminosity
through the empirical correlations~\citep{2003AJ....125.1849B,2006MNRAS.370.1445D}:
\begin{eqnarray}
R\approx2.6\left(L \over 1.6\times 10^{10} L_\odot \right)^{0.7} {\rm kpc}
\end{eqnarray}
and 
\begin{eqnarray}\label{Lscale}
\sigma\approx150 \left(L \over 1.6\times 10^{10} L_\odot \right)^{0.23} {\rm km/s}~.
\end{eqnarray}
We can thus write, from equation \eqref{res_feed}, $\dot{M}_{\rm res}\approx A_{\rm res} M^{\rm gas}_{\rm bulge}/t_{\rm dyn}$ during the starburst event, and therefore
\begin{equation}\label{int_eq}
\frac{\Delta M_{\rm res}}{\Delta M_{\rm MBH}} \approx \frac{\dot{M}_{\rm res}}{\dot{M}_{\rm MBH}} \approx 
\frac{t_{\rm salp}}{t_{\rm dyn}} \frac{A_{\rm res} \Delta M^\star_{\rm bulge}}{M_{\rm MBH}}\,,
\end{equation}
where $\Delta M_{\rm res}$ is the total mass of cold gas that falls to the nucleus during the star formation event,
and $\Delta M_{\rm MBH}$ and $\Delta M^\star_{\rm bulge}$ denote the changes in the MBH and stellar bulge masses during the starburst.
(Note that eventually all the gas of the bulge is transformed into stars, if feedback is neglected, i.e. $M^{\rm gas}_{\rm bulge}\approx \Delta M^\star_{\rm bulge}$.)

Now, if we assume that the MBH approximately satisfies the Magorrian relation~\citep{magorrian1,magorrian2} 
$M_{\rm MBH}\approx A_{\rm M} M^\star_{\rm bulge}$ [with $A_{\rm M}\approx 1.4 \times 10^{-3}$~\citep{magorrian2}], we can write
\begin{equation}\label{int_eq2}
\frac{\Delta M_{\rm res}}{\Delta M_{\rm MBH}} \approx \frac{t_{\rm salp}}{t_{\rm dyn}} \frac{A_{\rm res} \Delta M^\star_{\rm bulge}}{A_{\rm M} M^\star_{\rm bulge}}
\approx \frac{t_{\rm salp}}{t_{\rm dyn}} \frac{A_{\rm res}}{A_{\rm M}} f_{\rm crit}\propto\sigma^{-2.06}\,,
\end{equation}
where have assumed $\Delta M^\star_{\rm bulge}/M^\star_{\rm bulge} \approx f_{\rm crit}$, because starbursts are typically triggered by major
galaxy mergers, i.e. ones with mass ratio between the baryonic components of the two galaxies larger than $f_{\rm crit} \approx 0.25$.~\footnote{Note that the same approximate scaling $\Delta M_{\rm res}/{\Delta M_{\rm MBH}}\propto \sigma^{-2.06}$ can be obtained from equation \eqref{int_eq}  
by assuming that the MBH satisfies the $M-\sigma$ relation $M_{\rm MBH}\sim \sigma^4$, if one observes that $M^\star_{\rm bulge}\sim (M^\star_{\rm bulge}/L) L \sim \sigma^4$ (where we have used equation \eqref{Lscale} and assumed constant mass-to-light ratio). Indeed, this latter relation
yields $\Delta M^\star_{\rm bulge}\sim \sigma^3 \Delta \sigma$, which replaced in equation \eqref{int_eq} gives
$\Delta M_{\rm res}/{\Delta M_{\rm MBH}}\propto \sigma^{-2.06}$, if one assumes $\Delta\sigma \sim \sigma$ in major galactic mergers.}
Eventually all the gas of the reservoir will either form stars or accrete onto the MBH, i.e. $\Delta M_{\rm res}=\Delta M_{\rm MBH}+\Delta M_{\rm NSC}$. However,
since we assumed $M_{\rm MBH}\ll M_{\rm res}$, most of the nuclear gas will end up in the NSC rather than in the MBH, hence $\Delta M_{\rm res}\approx \Delta M_{\rm NSC}$, and we can write, for $M_{\rm MBH}\ll M_{\rm NSC}$,
\begin{equation}
\frac{\mbox{d} M_{\rm NSC}}{\mbox{d} \sigma} \approx\frac{t_{\rm salp}}{t_{\rm dyn}} \frac{A_{\rm res}}{A_{\rm M}} f_{\rm crit} \frac{\mbox{d} M_{\rm MBH}}{\mbox{d} \sigma}\,,
\end{equation}
where we used $\Delta M_{\rm NSC}\approx (\mbox{d} M_{\rm NSC}/\mbox{d} \sigma) \Delta \sigma$ and $\Delta M_{\rm MBH}\approx (\mbox{d} M_{\rm MBH}/\mbox{d} \sigma) \Delta \sigma$.\footnote{{ Note
that $\mbox{d} M_{\rm MBH}/\mbox{d} \sigma$ and $\mbox{d} M_{\rm NSC}/\mbox{d} \sigma$ are essentially the rates of
change of the MBH and NSC masses during a  starburst (time being parametrized by the host galaxy's velocity dispersion $\sigma$).}}

 We can then integrate this equation to give 
\begin{multline}
 M_{\rm NSC} \approx 1.9 \times \frac{t_{\rm salp}}{t_{\rm dyn}} \frac{A_{\rm res}}{A_{\rm M}} f_{\rm crit} M_{\rm MBH}\\\approx 7.8\times 10^7 M_\odot \left(\frac{\sigma}{100 \,\mbox{km/s}}\right)^{2.35}\,,
\end{multline}
where we have used equation \eqref{sigmabh} and assumed a radiative accretion efficiency $\epsilon\approx 0.1$. 
 Although very simplified,  this model approximately reproduces the scaling
 relations generated by our dissipative 
$GxeV$ model (for $M_{\rm MBH}\lesssim M_{\rm NSC}$) in slope, and also (within a factor of a few) in normalization.

The mass growth of \NSC s is regulated by two processes: inspiral of 
star clusters and in-situ star formation. 
 In order to quantify 
 the amount of \NSC \ mass that is contributed by local star formation in our models,
 we computed the ratio of
the median \NSC \ mass obtained  by including both cluster inspirals
and in-situ star formation to that obtained with only the contribution of in-situ star formation. We
plot this quantity as a function of total galaxy mass in Figure~\ref{insitu-fr}.

In our models, about half of the mass of  \NSC s is contributed by local star formation 
 for galaxies less massive than $\sim 3\times10^{11} M_{\odot}$, while for galaxies more massive than this
 value, the contribution from dissipative processes becomes progressively more important. 
Some provisional evidence for a preferential dissipative mode in  galaxies of progressively
larger masses was presented by \citet{Cote:2006} and \citet{2012ApJS..203....5T}.
These authors showed  that the brightest  nuclei which reside in the 
most luminous hosts have colors that are significantly redder
than expected on the basis of  a star cluster merger scenario,
suggesting that an additional process --
e.g., dissipational infall of metal-rich gas during mergers -- likely begins
to dominate the formation of nuclei in galaxies of higher masses.

We finally note that the relative contribution of the two formation channels depends
on the adopted value of star cluster formation efficiency $f_{\rm gc}$ which remains
a quite uncertain parameter of our models.  However, given that 
plausible values of  $f_{\rm gc}$ were found to only  impact the normalization of our results 
for the empirical relations and to not affect their slope, assuming $f_{\rm gc}\approx$ const we can simply express 
the relative contribution of the two formation channels 
 using the   general
formulation:
\begin{equation} 
\frac{\langle M_{\rm cl} \rangle }  {\langle M_{\rm in-situ} \rangle} \approx 1.5 \times \left(f_{\rm cl} \over 0.07\right)~,
\end{equation}
with $\langle M_{\rm in-situ} \rangle $ the typical  \NSC \ mass that originates  in-situ from episodes
   of star formation and $\langle M_{\rm cl} \rangle$ the mass brought in by migrating clusters.

   \section{Cosmological evolution of scaling correlations}~\label{Z-ev-sec}
   The existence of a fundamental connection between  \sbh s and \NSC s was 
first suggested by \citet{Ferrarese:2006}. These authors showed that
\NSC s and  \sbh s follow similar scaling 
relations with their host galaxy properties, and thus argued that they
 are two different manifestations of
the same astrophysical type of system, which they called
``central massive object''. 
Later, \citet{G12} used a large sample of nucleated galaxies to show  
that the scaling relations of  \NSC s and  \sbh s
are quite different from each other, with the former following much shallower 
 correlations with their host galaxy properties. Graham's findings  
 might suggest 
 that  \NSC s and  \sbh s formed through different 
 physical processes. This could be the case if for example  \NSC s originated elsewhere in the galaxy and
 then migrated to the center through dynamical friction processes~\citep[e.g.,][]{2013ApJ...763...62A}.

\citet{Neumayer:2012}  presented a first 
$M_{\rm  \sbh}$ versus $M_{\rm \NSC}$ diagram and
found, in agreement with our study, the existence of three different regimes: (a) \NSC \ dominated nuclei
at $\sigma \lesssim 100~{\rm km/s}$, (b) a transition region, and (c)  \sbh -dominated nuclei at
$\sigma \gtrsim 150~{\rm km/s}$. \citet{Neumayer:2012} argued
that this is consistent with a picture in which black holes form inside \NSC s with a low-mass fraction. They subsequently grow much faster than the \NSC , destroying it completely when the ratio $M_{{\rm  \sbh}}/M_{\NSC}$ grows above $\sim 100$. 

 \citet{Neumayer:2012} also argued against  \sbh \ mergers as responsible for the 
disruption of the nuclei in the highest mass galaxies. 
These authors pointed out that if \NSC s are disrupted during mergers, 
elliptical galaxies -- thought to be the product of galaxy mergers -- should rarely host a \NSC .
The fact that most early-type galaxies have a \NSC \ 
would therefore suggest that mergers 
do not play a pivotal role in leading to \NSC \ disruption.
We do not agree with this interpretation.

As  discussed in Section~\ref{bh-eff}, a  \sbh \ binary will eject from the galaxy center a mass comparable to its own mass, so that only mergers with  \sbh \ to 
NSC mass ratio larger than unity 
can significantly heat the \NSC \ and make it susceptible to destruction during the merger
event. 
\NSC s can therefore survive and indeed grow during the morphological transformation from disk-dominated to bulge-dominated  galaxies if the progenitor galaxy \NSC s are significantly more massive 
than their central  \sbh s. 
Indeed, from the \NSC\  and  \sbh\  scaling correlations, 
we find that \NSC s dominate the nuclei of galaxies with velocity dispersion 
 $\sigma \lesssim 150{\rm km/s}$, so that mergers of galaxies  at the low end of the $\sigma$  distribution
 will not lead to the complete disruption of the host galaxy \NSC s, while mergers of massive galaxies, 
 characterized by a larger  \sbh \ to \NSC \ mass ratio, will lead to their full disruption.
 This picture is consistent with both the
abundance of \NSC s in early type galaxies of intermediate luminosity, and with the fact that 
\NSC s tend to disappear in the brightest spheroids  hosting the  most massive  \sbh s.
     
\begin{figure}
\centering
\subfigure{\includegraphics[width=1.59in,angle=270.]{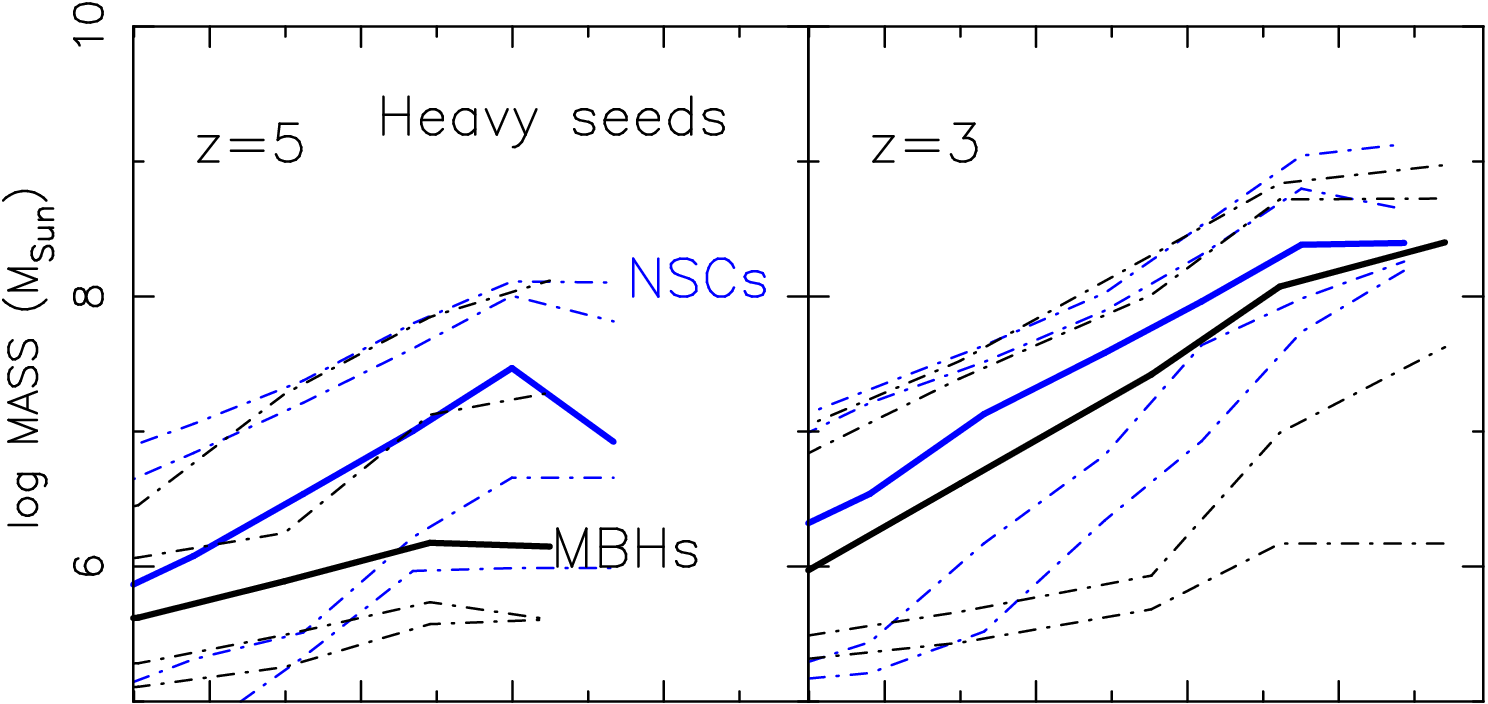}}  \\
\subfigure{\includegraphics[width=1.59in,angle=270.]{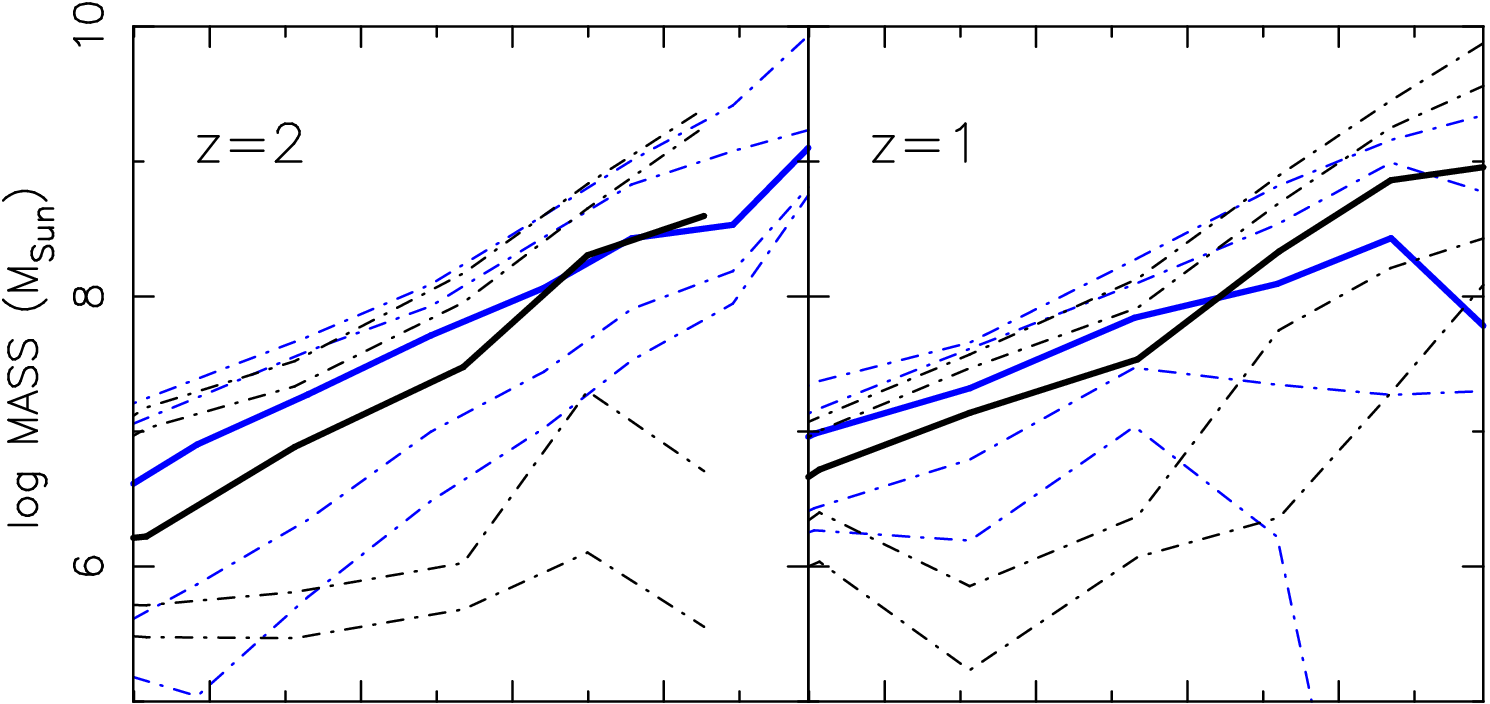}}\\
\subfigure{\includegraphics[width=1.925in,angle=270.]{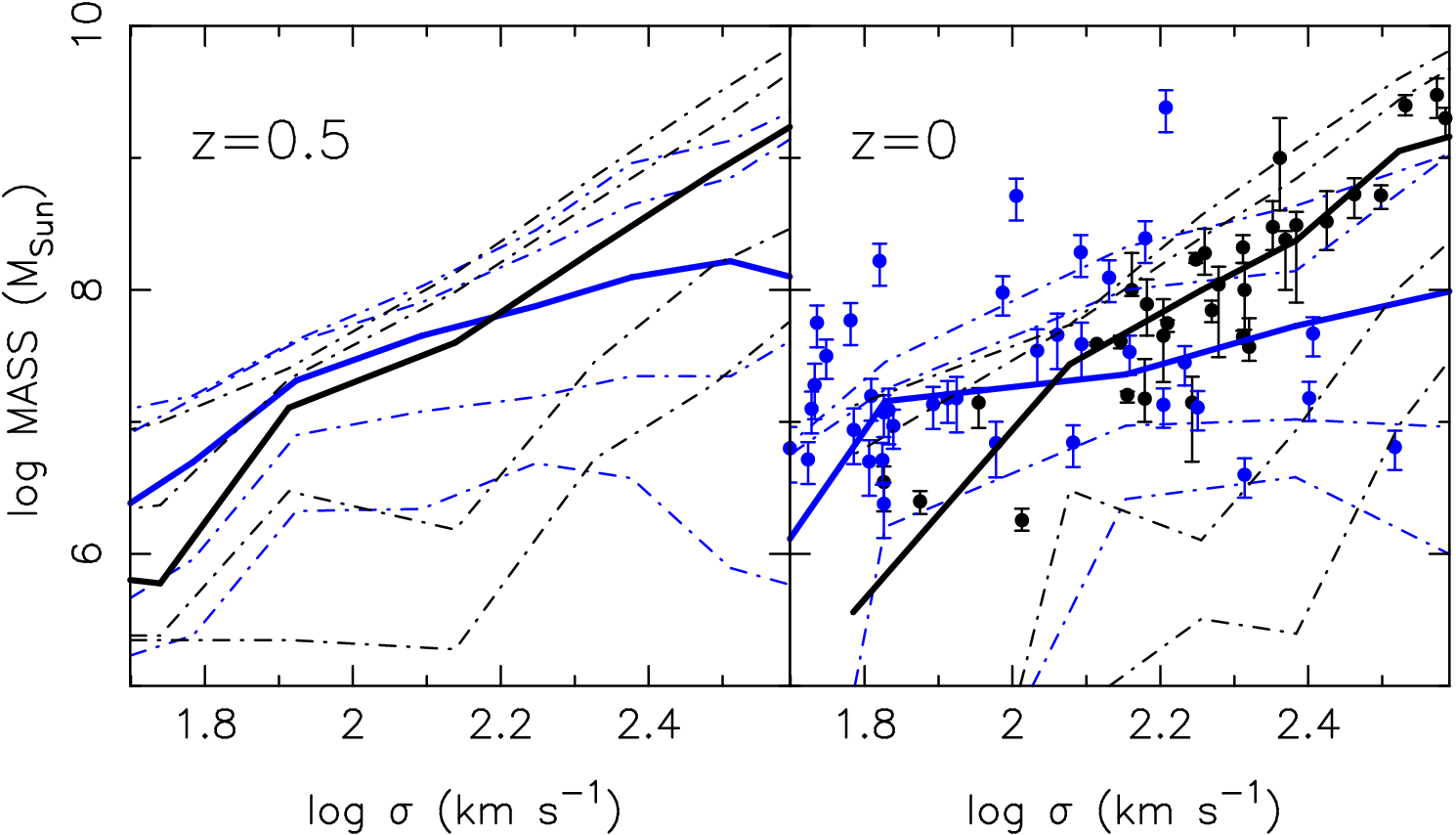}}\\
\caption{Same as Figure~\ref{Z-ev} but for the heavy  \sbh \ seed scenario of~\citet{2004MNRAS.354..292K}.
\label{Z-ev-HEAVY}}
\end{figure}

     To  illustrate the simultaneous evolution of  \sbh  s and \NSC s in our semi-analytical
     galaxy formation models,  we plot in Figure~\ref{Z-ev}   the  \sbh \ and \NSC \ scaling correlations at different 
     redshifts and the corresponding  mass distributions in Figure~\ref{Z-ev-M}. 
    Specifically, in order to calculate the median and confidence regions in our model
     we have      have only considered the  \sbh s residing in bulge-dominated galaxies (which we identify  
     with ones having bulge to total mass ratio larger than $0.7$).\footnote{We make this choice to correct
for the observational bias that selects galaxies with a significant bulge component, so that measurements
of $\sigma$ are possible in the first place.}
     
     We find that at high redshifts, $z\gtrsim 3$, the nuclei of galaxies are dominated by \NSC s.
     Between redshift $4$ and $2$ the  \sbh s  grow faster than the NSCs,
      becoming  by $z\approx 3$ the dominant nuclear component in galaxies with $\sigma\gtrsim 100{\rm km/s}$. 
      After this point,  the \NSC \ scaling correlations start
to flatten at high values of $\sigma$, as  \sbh \ binaries forming during mergers are now efficient at eroding the surrounding clusters.
In addition,  the \NSC s can no longer grow efficiently in the 
most massive galaxies,  since
inspiraling stellar clusters are tidally disrupted at larger galactocentric distances  
in  galaxies with progressively more massive   \sbh s. 

Figure~\ref{Z-ev-M} displays the redshift evolution of the mass distribution of  \sbh s and \NSC s.
In these plots we include all galaxies in our models regardless of their specific value of  bulge to total mass ratio.
At high redshift the \NSC s are typically more massive than  \sbh s; after $z\approx 2$
the mass distribution of \NSC s intersects at about $10^8~M_{\odot}$ the distribution of \sbh s.
This sets the transition between \NSC \ dominated and \sbh \ dominated galaxies seen also in the observational data.
We also compared our synthetic \NSC \ mass distributions  for early type galaxies
with the observed mass distribution of \NSC s for 
the catalogs of early type galaxies of \citet{Cote:2006} and \citet{2012ApJS..203....5T}.
Anderson-Darling and  Kolmogorov-Smirnov statistical  tests used to 
compare the observed to the \sbh \ model cumulative distributions
 give   $p$-values of $0.35$ and $0.42$ respectively, indicating that the 
  hypothesis that the simulated and observed distributions are significantly different from each other can be rejected
at a high level of confidence.
For a comparison with the observed MBH mass function at $z=0$, see instead \citet{2012MNRAS.423.2533B} and \citet{2014ApJ...794..104S}.

\begin{figure}
\includegraphics[width=3.in,angle=270.]{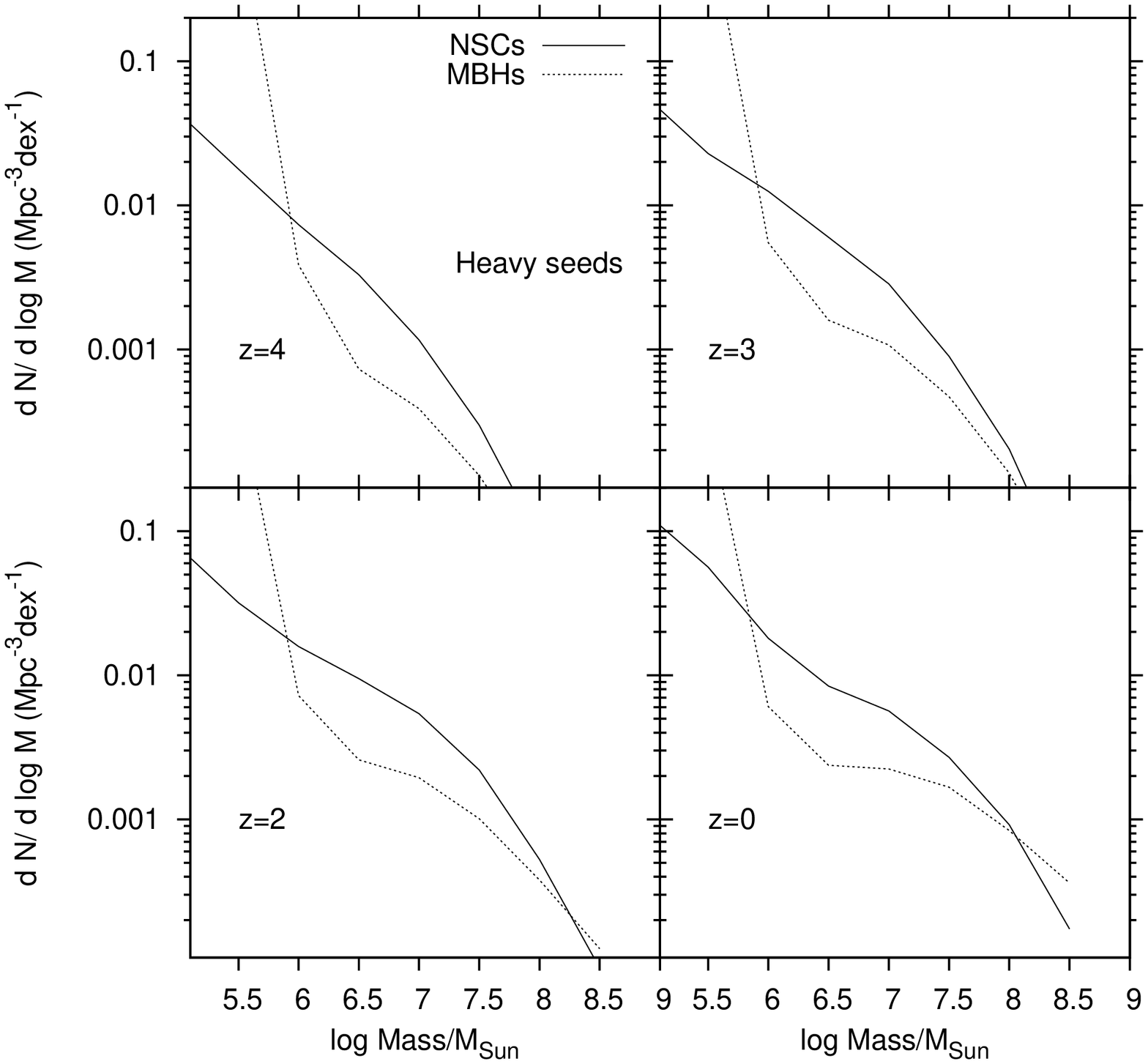} \\
\caption{Same as Figure~\ref{Z-ev-M} but for the heavy  \sbh \ seed scenario of \citet{2004MNRAS.354..292K}.\label{Z-ev-M-HEAVY} 
}
\end{figure}

  \subsection{Dependence on the  MBH seed model}
In the previous sections, we have discussed the results of
a galaxy formation model in which  \sbh s are evolved starting at $z\sim 15-20$
from initial ``light'' masses $M_{\rm seed}\sim 200~M_{\odot}$.
This could be the case if  \sbh s formed as remnants of Pop~III stars~\citep{2001ApJ...551L..27M}.

In order to test the dependence of our results on the assumed scenario for  the formation of  \sbh s,  we
evolved  additional  models in which halos are initially populated by \sbh \ seeds of mass
 $M_{\rm seed}\sim 10^5~M_{\odot}$, which could for example  form as a result of  the collapse of
massive protogalactic disks 
at redshifts $z\gtrsim 10-15$\ \citep{2004MNRAS.354..292K,2006MNRAS.370..289B, 2006MNRAS.371.1813L,2008MNRAS.383.1079V}. In particular, we consider a model in which
these ``heavy'' seeds form with high halo occupation number~\citep{2004MNRAS.354..292K}, and one
in which the halo occupation number is moderate (namely the model of~\citet{2008MNRAS.383.1079V},
where we set the critical Toomre parameter below which the formation of a MBH seed becomes possible to $Q_c=2$).

 Figure~\ref{Z-ev-HEAVY} shows the redshift evolution of the $M_{\rm \NSC}-\sigma$ and
$M_{{\rm  \sbh}}-\sigma$ relations
in the case of the  \sbh \ heavy seed scenario of~\citet{2004MNRAS.354..292K}. Figure~\ref{Z-ev-M-HEAVY} 
displays  the corresponding   \sbh \ and \NSC \ mass distributions 
from $z=4$ to $z=0$. A comparison of these plots with those in Figures~\ref{Z-ev} and \ref{Z-ev-M}
demonstrates that the local mass distribution of \NSC s and their evolution from high redshift 
is not affected by the \sbh \ seed model in any important way.
(Of course, the MBH mass function in the heavy-seed models
differs from that in the light-seed model at the low-mass end, which is dominated by
MBHs that have not evolved significantly from their seeds. However, at intermediate and high masses,
the MBH mass functions are very weakly dependent on the seed model, since memory of the initial
conditions has been lost due to accretion and mergers.)
Note that also in
the heavy-seed scenarios, we find that the \NSC \ cumulative mass distributions at $z=0$ 
for  early type galaxies are consistent with 
the observed mass distribution in \citet{Cote:2006} and \citet{2012ApJS..203....5T}   at a high
level of significance (similar conclusions hold for the seed model of \citet{2008MNRAS.383.1079V}).

The fact that our results about \NSC \ evolution are not sensitive to the particular model chosen 
for the formation of  \sbh \ seeds is primarily a consequence of the high \NSC \ to  \sbh \ mass ratio 
at high redshift. At $z \gtrsim 3$, in 
both the heavy and light seed scenarios,  the \NSC s are the dominant central component of galaxies (with the exception of small
NSC masses $\lesssim 10^6 M_\odot$, at which MBHs may dominate if they are seeded with high halo occupation number at high redshift).
It is only after  the peak of the quasar activity at $z\approx2$ that  \sbh s become massive enough 
to significantly affect the \NSC s at intermediate and high masses. However,
by this time the  \sbh\ mass distributions in  different seed scenarios
are very similar to each other (again, with the exception of the low-mass end),
 and therefore the subsequent evolution of the \NSC s is also
very similar. The mass growth of the  \sbh \ population is
in fact  dominated by the mass accreted during 
the quasar epoch at $z\approx 2$, thus washing out the imprint of the initial conditions, cf. also \citet{2012MNRAS.423.2533B}.

Since to date the  
formation process of  \sbh s  remains largely unconstrained, 
the fact that our results 
are robust against  different initial  seed formation scenarios is 
important  if we want to make robust predictions about the overall evolution of the \NSC \ population.

\begin{figure*}
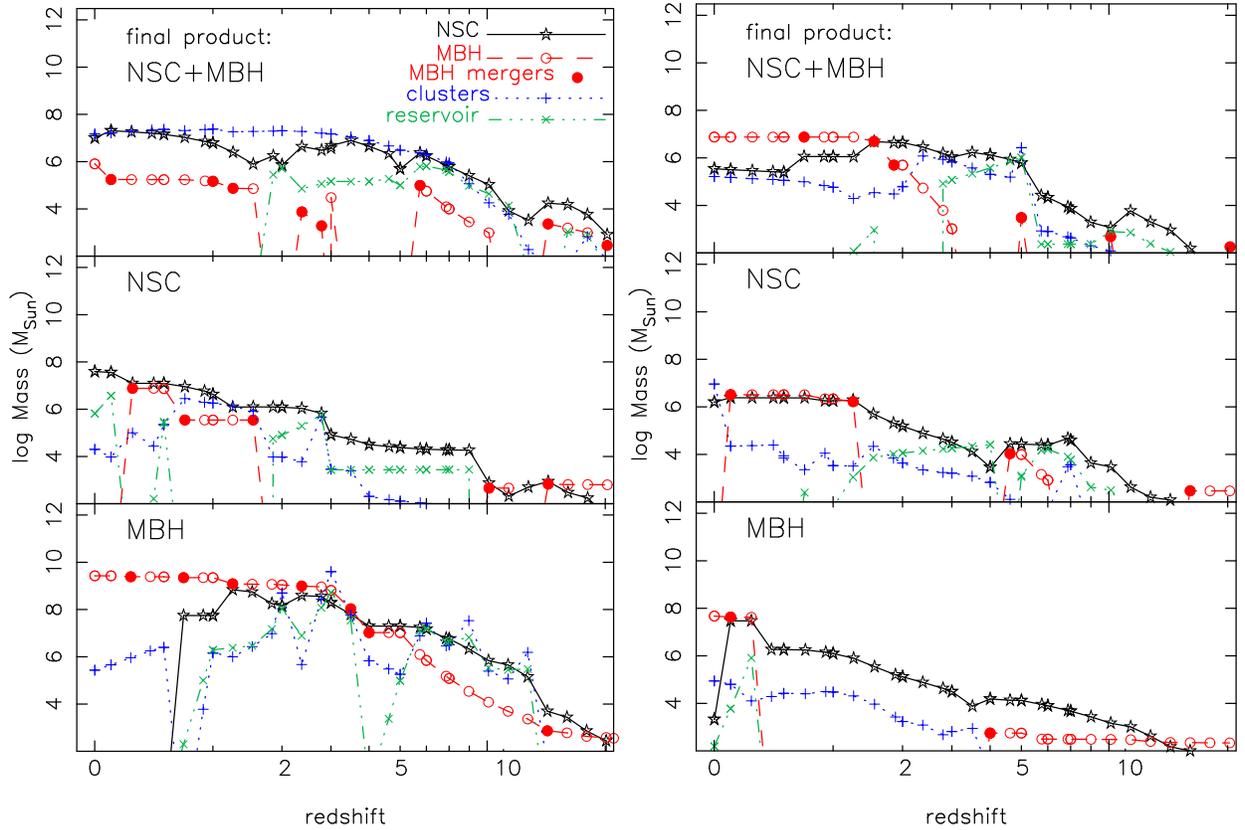

\centering
\includegraphics[width=3.2in,angle=0.]{Fig13a.eps}
\includegraphics[width=3.2in,angle=0.]{Fig13b.eps}
\caption{\label{nsc-ev-TR} Examples of possible main-progenitor evolutions of NSCs and MBHs, as predicted by our model and for different
final products at $z=0$.
At upper left we indicate the final product of the evolution, a composite \NSC\ +\sbh\ nucleus (upper panels), a \NSC\
without \sbh\ (middle panels), or a \sbh\ dominated nucleus lacking a \NSC\ (bottom panels).
 We consider the evolution of the NSC hosted in the central galaxy at $z=0$, and proceed back in time, following
the main NSC progenitor at each merger. However, when no NSC is present in the central galaxy at $z=0$ (namely in the bottom-left panel), we follow
the MBH main-progenitor history.}
\end{figure*}

\section{Discussion }

\subsection{Formation of the central regions of galaxies}
   In this paper, we have presented a study aimed at understanding 
  how the central regions of galaxies formed, and how the
  evolution of   \sbh s and \NSC s is connected to 
that of their host galaxy.
  
 In Figure \ref{nsc-ev-TR}, we show examples of NSC and MBH evolution predicted by the $GxeV$ model. 
These cases should be seen as qualitative because of the great variety of possible NSC and MBH histories
that are possible within our model, but we have chosen examples that are roughly representative of the various 
possibilities mentioned below. In more detail, we consider the evolution of NSCs back in time following their main 
progenitor, i.e. we start from the NSC hosted in the central galaxy 
at $z=0$, and at each galaxy merger we follow its more massive NSC progenitor. When
no NSC is present at $z=0$ in the central galaxy (namely in the bottom-left panel), we follow the main-progenitor history of the MBH.
As can be seen, NSCs grow by the combined action of {(i)} stellar-cluster infall, 
which happens continuously throughout their history, but is particularly enhanced
in starburst galaxies (cf. the blue line in Figure \ref{nsc-ev-TR}, which denotes the mass in stellar clusters, and which 
shows a smooth evolution with superimposed spikes due to starbursts); and {(ii)} 
in-situ star formation, which takes places for the most part in starburst galaxies formed by major mergers (cf. the
spikes in the mass of the low-angular momentum reservoir available for nuclear star formation, i.e. the green line in
Figure \ref{nsc-ev-TR}). Note also that MBH mergers have a prominent effect on the MBH and NSC history,  not only because MBH binaries erode or even completely
destroy the NSC if the MBHs are sufficiently massive, but also because the merger can eject the remnant MBH from the galactic nucleus, when the kick velocity 
imparted by the anisotropic gravitational-wave emission is larger than the escape velocity from the central parts of the galaxy. 
Note instead that the apparent ``re-growth'' of the MBH after an ejection, which can be seen in Figure \ref{nsc-ev-TR}, is simply a consequence 
 of our choice of following the NSC main-progenitor history: at certain galaxy mergers, the galaxy containing
the main NSC (but no MBH) may merge with one carrying a MBH.

Below we discuss in more detail how the evolution of the galaxy, and its merger history 
can give rise to the variety of galactic central structures observed today.

{\it Nucleated galaxies containing a MBH (cf., upper panels of Figure~\ref{nsc-ev-TR}). \\ }
If  the  \sbh\ mass  does not grow above   $\approx 10^7~M_{\odot}$ throughout the evolution of the
galaxy,  the central \NSC\ cannot be significantly 
eroded during galaxy mergers by inspiraling  \sbh \ binaries,
 because of the large \NSC \ to \sbh \ mass ratio.
Moreover, stellar clusters will migrate to the center essentially 
undisturbed by the tidal field  of the central  \sbh .

If during the last merger event the  \sbh \ remnant is retained, the end
product will be a galaxy  containing both a central \NSC \ and a  \sbh ;
An example of such a galaxy is the Milky Way which hosts a  \sbh \ of mass $\approx 4\times 10^6M_{\odot}$~\citep{1998ApJ...509..678G,2009ApJ...692.1075G}
and a \NSC \   of mass $\approx 3 \times 10^7~M_{\odot}$~
\citep{2014AA...566A..47S,2014AA...570A...2F}.
A handful of external  galaxies are also know to contain both 
a \NSC \ and a  \sbh , which are often  found to have comparable masses~\citep{2008ApJ...678..116S}. These galaxies 
lie  near the transition region between  \sbh \ dominated and \NSC \ dominated systems.
 As the MBHs of these galaxies grow, MBH mergers will partially destroy their host \NSC s in the process (e.g., upper right panel of Figure~\ref{nsc-ev-TR}).

 {\it Nucleated galaxies without MBH (cf., middle panels of Figure~\ref{nsc-ev-TR}). \\ }
 After a merger a  \sbh \ can receive a ``kick'' due to gravitational recoil 
 with velocities as large as a few 
 thousands $\rm km/s$, which may eject the  \sbh \ from its host galaxy.
The ejected   \sbh \ will
carry a cluster of bound stars, a hypercompact stellar system that can appear
 similar in size and luminosity to star clusters or ultracompact dwarf 
 galaxies~\citep{2009ApJ...699.1690M}. 
 If the ejected \sbh \ has a mass ~$\lesssim 10^7M_{\odot}$, the \NSC \  will remain virtually 
 unaffected. If from that point on the galaxy evolves passively without experiencing 
 mergers which might bring a new  \sbh \ to the center, the 
 evolution's end product will be a nucleus 
 containing a  \NSC \ but no  \sbh .
 Examples of such type of evolution are illustrated in the middle panels of Figure~\ref{nsc-ev-TR}.
A similar end product can  be attained if a  \sbh \ seed never formed
 and the galaxy evolved
 without experiencing major mergers during its entire evolution.

We note in passing  that whether a   \sbh \ seed will form might depends
on the properties of the central cluster at high redshift~~\citep{2012ApJ...755...81M}.
If the accumulated cluster 
has  a high velocity dispersion $\sigma \gtrsim 40~{\rm km/s}$,
this might seed the growth  of a central 
 \sbh \ because
kinematic heating from binary stars is insufficient to prevent
complete core collapse, which might then lead to runaway mergers and 
the formation of a central  \sbh\ seed of $\sim 10^3~M_{\odot}$. 
In galaxies hosting clusters with initial velocity dispersion less than 
$40~{\rm km/s}$,  binaries will prevent core collapse
and consequently the formation of a  \sbh \ seed.
Galaxies with  clusters  with initial velocity dispersions below
this limit might therefore never form a   \sbh . 
Such evolutionary path might lead to the formation of galactic nuclei 
such as those of M33 or NGC205, which lack a  \sbh \ and host a 
low velocity dispersion \NSC~\citep{2001Sci...293.1116M,2001AJ....122.2469G}.

{\it MBH \ dominated galaxies without \NSC\ (cf., lower panels of Figure~\ref{nsc-ev-TR}). \\ }
After the  \sbh \ mass grows above $\gtrsim 10^8~M_{\odot}$, any
major merger will be highly disruptive for a central \NSC ; 
 \sbh \ binaries forming during  major mergers
in this high  \sbh \ mass regime will carve out the galactic center of stars destroying
a pre-existing \NSC , eventually producing a central mass-deficit~\citep{2001ApJ...563...34M,BG:10}.

 The accretion of a nucleated  dwarf galaxy by a low-density giant galaxy,
 could  bring a \NSC \ to the center of the latter galaxy.
However, \citet{2001ApJ...551L..41M} showed that
  the secondary galaxy is disrupted during the merger by the giant galaxy  
  \sbh \ tidal field, producing a remnant with a central density that is only slightly higher than that of the giant galaxy initially.
Moreover,  as shown in \citet{2013ApJ...763...62A}, after a \NSC \ is disrupted in a dry merger
it  will have a hard time to regrow,  
because the  \sbh \ tidal field is effective at disrupting migrating clusters.
Thus, once the galactic  \sbh \ mass grows above $\gtrsim 10^8~M_{\odot}$,
  the central \NSC \ is likely to be disrupted and the central core is likely to be preserved
  during the subsequent evolution of the galaxy.

{ \it Low mass spheroids without  \NSC .\\ }
\NSC s tend to disappear in Galaxies fainter than $M_V\sim -12$~\citep{VdB}.
While the purely dissipationless formation model $CliN$ predicts 
that the \NSC \ occupation fraction decreases substantially in low mass
systems in agreement with observations, 
the $GxeV$ model results in a high occupation fraction of \NSC s in low mass
galaxies, and it is therefore in tension  with observational findings.
We believe that the reason for this discrepancy is that discreteness effects are not
accounted for in $GxeV$ as argued in the following. 
In this latter model,  we compute the 
flux of mass accreting onto the nucleus by \textit{averaging} over the initial mass function of the stellar cluster population.
 In $CliN$, we simulate  instead  the inspiral of each of the formed clusters, 
 drawing their masses from the same initial mass function.
 Because of the initial small number of stellar clusters  in low mass galaxies,
and because of the steep  power law that we adopt for their initial  mass function,
some of these galaxies will host no cluster that is massive enough to decay to the center before
being disrupted in the process. This suggests that the lack of
\NSC s in spheroids less luminous than $M_V\sim -12$ is a consequence of the initial low number of
massive clusters in these galaxies. Clearly, this effect cannot be accounted for in the $GxeV$ model, 
which does not follow the evolution of individual stellar clusters but rather the average mass infall to the nucleus. 

\subsection{Galaxy cores and cusp regeneration}
Our analysis focuses on the mass evolution of \NSC s  and ignores effects
due to dynamical relaxation that can change the
density profile of the clusters over time. 
We have shown that  the merger history of galaxies after $z\approx 2$ plays a key role in determining 
the structure of \NSC s observed today and their disappearance in the brightest galaxies. 
 In our models, we did not account for 
relaxation processes that can rebuild an eroded cusp 
as stars diffuse into the  \sbh \ sphere of influence.
Even if a \NSC \ is fully disrupted during  a merger,
 a stellar cusp 
of mass  of order $10\%$ the mass of the black hole and
extending out to one tenth its sphere of influence will reform over
the relaxation time~\citep{2006ApJ...648..890M}. Although cusp regrowth is not expected to have an important effect on the mass evolution of 
the \NSC s, it is worth discussing it, as
the ability of a  \sbh \ to shape a galactic nucleus and 
the observational consequences  depend on whether 
cores are regenerated or preserved after mergers.

\citet{BG:10} performed $N$-body simulations to investigate the evolution of merging star clusters
 with seed black holes.
Using black holes masses of $1-5\%$ their host \NSC\ ones, they investigated
 how the binary coalescence  dynamically heats and destroys the surrounding clusters.
While \citet{BG:10} argued that this mechanism provides a 
pathway to explain the observed reduction in the
nucleus-to-galaxy stellar mass ratio as one proceeds from dwarf to giant elliptical galaxies,
they did not discuss how their results would be affected by 
two-body relaxation  after the  \sbh \ binary merges, and between mergers.

During galaxy mergers,  cores are carved out
in the stellar distribution, with radii of order the influence radius of the massive binary.
This makes \NSC \ susceptible to disruption.
On the other hand, if the supply of stars is continuously replenished, the damage to the \NSC \ can be much smaller.
Furthermore, even if a large core is formed, a stellar density cusp can regrow 
via energy exchanges between stars moving in the gravitational field of the  \sbh \ remnant. Black holes with masses of $1-5\%$ that of the \NSC , such as 
 those considered in \citet{BG:10},
correspond to low mass galaxies with a short ($\lesssim 10^9\ $yr) relaxation time. 
From equation~(\ref{nsc-bh}) one finds that such a low mass ratio corresponds to 
a  \sbh \ mass of $\approx 5\times10^5~M_{\odot}$. Using equation $\eqref{sigmabh}$,
this corresponds to
 a host stellar spheroid velocity dispersion of $\approx 50~{\rm km/s}$.
In such low mass  galaxies, 
the cusp will regenerate  itself  in about $1~$Gyr from the infall of a second black hole,
as collisional relaxation drives the stellar distribution toward its steady state form~\citep{2013degn.book.....M}.
We conclude that in low mass spheroids the effect of  \sbh \ binaries on the host \NSC \
density distribution is expected to be minimal, and their imprint on 
the nuclear properties is expected  to be erased after a short time.

The probability that a galaxy of the size of the Milky Way 
experienced a major merger (i.e., with mass ratio larger than 0.3) after
$z=2$ is about $50\%$~\citep{2010MNRAS.406.2267F}.
Galaxies of such  luminosities  have relaxation times that
are longer than the age of the galaxy, but which are short enough that  two-body relaxation
might have affected their central clusters.  These galaxies might have nuclei 
that are still evolving toward their steady state after they were dynamically heated 
by a  \sbh \ binary.  Thus,  the results of our models imply that ``under-massive'' \NSC s in galaxies with
bulge mass $M_{\rm bulge}\approx 10^{10}~M_{\odot}$ might still carry
an imprint of the merger history of their host galaxy, in the form of a
density core
that extends out to a small fraction of the central  \sbh \ influence radius. 
Whether these \NSC s will turn out to harbor parsec-scale cores will be  addressed in
the future with high resolution imaging, which  will enable 
to resolve the crowded stellar environment of \NSC s\ \citep{2014AA...568A..89G}.

\section{Conclusions}
NSCs and MBHs represent important central components of galaxies that bear witness to their formation history.   Our aim has been to generate a numerical framework to better understand how the central structures of galaxies formed, and how the
evolution of   \sbh s and \NSC s is connected to that of their host galaxies.

We have  studied the evolution of NSCs in a cosmological context, by taking into account the growth of MBHs by merging  subhalos containing both MBHs and NSCs.  A semi-analytical galaxy formation model is applied to follow the evolution of dark matter halos along merger trees, as well as that of the baryonic components. Specifically, the evolution  of dark matter halos along the merger trees includes their baryonic components  such as hot gas, stellar and gaseous bulges,  and stellar and gaseous galactic disks. We study the simultaneous evolution of  \sbh  s and \NSC s in our galaxy formation model.
The main results of our work are summarized in the following:

(1) The mass growth of \NSC s is seen to be regulated by inspiral of star clusters, in-situ star formation
as well as galaxy and MBH mergers. We have found that  both cluster inspirals 
and  in-situ star formation
contribute a significant fraction  of the total mass of  NSCs. 

(2) We found that in-situ star formation 
(as well as growth through migrating clusters), generates \NSC \ - host galaxy scaling relations that are
much shallower than the same correlations for  \sbh s, in agreement with observations.

(3) In our model, the merger history of galaxies after $z\sim 2$ plays a key role in determining 
the structure of the \NSC s observed today.
Core-depletion due to gravitational slingshot of host galaxy stars by inspiralling MBHs forming during galaxy mergers has a negligible impact on the evolution of NSCs in low mass galaxies, while it leads to their full disruption in galaxies more massive than $\sim10^{11}~M_\odot$. In galaxies of intermediate luminosity~($\sim 10^{10}~M_\odot$), MBH mergers cause the partial dissolution of the clusters. Such partially eroded clusters appear at $z=0$ significantly underweight relative to linear NSC-host spheroid scaling correlations. The presence of an under-massive NSC population results in an increased
scatter  of the NSC-host galaxy empirical correlations at high galaxy luminosities, for which we find evidence in observational data.

 (4) We identify $M_{{\rm  \sbh}} \approx 10^8 M_{\odot}$ as the critical value of  \sbh \ mass above which \NSC s are significantly affected. Once the galactic  \sbh \ mass grows above this value, 
 \sbh\ binary mergers become
highly disruptive for the central \NSC \ leading to the formation of a
central low density core. For galaxies more massive than the Milky Way, 
after such low density core forms,
the central galactic regions are likely to remain depleted of stars 
throughout  the subsequent evolution of the galaxy.

(5) Our models predict that the fraction of nucleated early type galaxies containing  
an accreting  \sbh\ (with bolometric luminosity $L>10^{10}~L_{\odot}$)
increases from  $5\%$ at $M_{\rm gx}\approx 10^9~{M_{\odot}}$ 
to $30 \%$ at $M_{\rm gx}\approx 10^{11}~M_{\odot}$.
These fractions are found to be  much smaller for late type galaxies,
for which only $\sim10~\%$ of nucleated galaxies with $M_{\rm gx}\approx 10^{11}~{\rm M_{\odot}}$
also contain an active  \sbh . Among galaxies of all types, the fraction of mixed systems containing both
a  \sbh\ and a \NSC\ is about $\sim 20\%$ at  $10^9~M_{\odot}  \lesssim M_{\rm gx} \lesssim  10^{11}~M_{\odot}$.

(6) We found that the results shown in this paper are quite robust 
when we change the cluster formation efficiency, 
$f_{\rm gc}\lesssim 0.2$; when we allow $f_{\rm gc}$ to vary with galactic properties~\citep[e.g. we set it to 0.07, 0.04 and 0.5 
in disk, quiescent and starburst galaxies respectively;][]{2012MNRAS.426.3008K}; when we change 
the  \sbh \ seed model~(i.e. light-seed models vs heavy-seed models, with several halo occupation numbers at high redshift)
as well as other details of our model, such as  merger-tree resolution, initial redshift of the simulations,
prescriptions for AGN feedback, etc.

In conclusion, our study supports a scenario  in which 
black holes form inside \NSC s with a low-mass fraction, so that \NSC s are initially the dominant central
component of galaxies. After their formation, \NSC s 
and  \sbh s grow in parallel on their own scaling correlations with host galaxy properties,
with \NSC s dominating in low mass spheroids and  \sbh s dominating in high mass galaxies.
The simple fact that  \NSC s and  \sbh s
grow on different scaling correlations explains the well known  transition from   \sbh-  to \NSC-dominated 
galaxies  as one proceeds from dwarfs to giant ellipticals,
without the need of invoking competitive feedback processes from 
 young \NSC s  and/or AGN activity.

\bigskip
\acknowledgments
During the course of this work, we have benefited from  conversations
with several colleagues, including 
M.\ Colpi, D.\ Merritt, N.\ Murray, and A.\ Sesana.
We acknowledge support from a CIERA postdoctoral fellowship at Northwestern University
 (to F.A.); from the European Union's Seventh Framework Programme (FP7/PEOPLE-2011-CIG)
through the  Marie Curie Career Integration Grant GALFORMBHS PCIG11-GA-2012-321608 (to E.B.);
from   ERC project 267117 (DARK) hosted by Universit\'e Pierre et Marie Curie - Paris 6 and at
JHU by National Science Foundation grant OIA-1124403  (to J.S.).
F.A. acknowledges hospitality from the Institut d'Astrophysique de Paris,
where the early plan for this work was conceived.
E.B and J.S. acknowledge hospitality from the Lorentz Center (Leiden, NL), where part of this work was carried out.
 Computations were performed on the gpc supercomputer at the SciNet HPC Consortium, as well as on the Horizon Cluster at the Institut d'Astrophysique de Paris.


\begin{thebibliography}{67}
\expandafter\ifx\csname natexlab\endcsname\relax\def\natexlab#1{#1}\fi


\bibitem[Agarwal 
\& Milosavljevi{\'c}(2011)]{2011ApJ...729...35A} Agarwal, M., \& Milosavljevi{\'c}, M.\ 2011, \apj, 729, 35 



\bibitem[Aharon 
\& Perets(2015)]{2015ApJ...799..185A} Aharon, D., \& Perets, H.~B.\ 2015, \apj, 799, 185 


\bibitem[Antonini~et~al.~(2012)]{AM12}Antonini, F., Capuzzo-Dolcetta, R., Mastrobuono-Battisti, A. \& Merritt, D. \ 2012,
ApJ, 750, 111

\bibitem[Antonini(2013)]{2013ApJ...763...62A} Antonini, F.\ 2013, \apj, 763, 62 

\bibitem[Antonini(2014)]{2014ApJ...794..106A} Antonini, F.\ 2014, \apj, 
794, 106 

\bibitem[Antonini et al.(2015)]{2015ApJ...806L...8A} Antonini, F., 
Barausse, E., \& Silk, J.\ 2015, \apjl, 806, L8 

\bibitem[Arca-Sedda 
\& Capuzzo-Dolcetta(2014)]{2014MNRAS.444.3738A} Arca-Sedda, M., \& Capuzzo-Dolcetta, R.\ 2014, \mnras, 444, 3738 

\bibitem[Balcells et al.(2003)]{2003ApJ...582L..79B} Balcells, M., Graham, 
A.~W., Dom{\'{\i}}nguez-Palmero, L., 
\& Peletier, R.~F.\ 2003, \apjl, 582, L79 

\bibitem[{{Balcells} {et~al.}(2007){Balcells}, {Graham}, \&
  {Peletier}}]{Balcells:2007}
{Balcells}, M., {Graham}, A.~W., \& {Peletier}, R.~F. 2007, \apj, 665, 1084

\bibitem[Baldassare et al.(2014)]{2014ApJ...791..133B} Baldassare, V.~F., 
Gallo, E., Miller, B.~P., et al.\ 2014, \apj, 791, 133 

\bibitem[Barausse(2012)]{2012MNRAS.423.2533B} Barausse, E.\ 2012, \mnras, 
423, 2533; Erratum \ 2014, \mnras, 440, 1295 


\bibitem[Begelman et al.(1980)]{Begelman80} Begelman, M.~C., 
Blandford, R.~D., \& Rees, M.~J.\ 1980, \nat, 287, 307 


\bibitem[Bekki et al.(2004)]{2004ApJ...610L..13B} Bekki, K., Couch, W.~J., 
Drinkwater, M.~J., \& Shioya, Y.\ 2004, \apjl, 610, L13 



\bibitem[Bekki \& Graham~(2010)]{BG:10}Bekki, K. \& Graham, A. \ 2010, ApJ, 714, L313

\bibitem[Begelman et al.(2006)]{2006MNRAS.370..289B} Begelman, M.~C., 
Volonteri, M., \& Rees, M.~J.\ 2006, \mnras, 370, 289 

\bibitem[{{Bell} \& {de Jong}(2001)}]{Bell:2001}
{Bell}, E.~F., \& {de Jong}, R.~S. 2001, \apj, 550, 212

\bibitem[{{Bell} {et~al.}(2003){Bell}, {McIntosh}, {Katz}, \&
  {Weinberg}}]{Bell:2003}
{Bell}, E.~F., {McIntosh}, D.~H., {Katz}, N., \& {Weinberg}, M.~D. 2003, \apjs,
  149, 289

\bibitem[Bender et al.(2005)]{2005ApJ...631..280B} Bender, R., Kormendy, 
J., Bower, G., et al.\ 2005, \apj, 631, 280 

\bibitem[Bernardi et al.(2003)]{2003AJ....125.1849B} Bernardi, M., Sheth, 
R.~K., Annis, J., et al.\ 2003, \aj, 125, 1849

\bibitem[Bigiel et al.(2010)]{bigiel10} Bigiel, F., Leroy, A., 
Walter, F., et al.\ 2010, \aj, 140, 1194 

\bibitem[Bik~et al.~(2003)]{bik}Bik, A., Lamers, H.~J.~G.~L.~M., Bastian, N., Panagia,~N., \& Romaniello, M. \ 2003, A\&A, 397, 473

\bibitem[B{\"o}ker et al.(2002)]{2002AJ....123.1389B} B{\"o}ker, T., Laine,
S., van der Marel, R.~P., et al.\ 2002, \aj, 123, 1389

\bibitem[B{\"o}ker et al.(2004)]{2004AJ....127..105B} B{\"o}ker, T., Sarzi, 
M., McLaughlin, D.~E., et al.\ 2004, \aj, 127, 105 

\bibitem[Bolatto et al.(2011)]{bolatto11} Bolatto, A.~D., Leroy, 
A.~K., Jameson, K., et al.\ 2011, \apj, 741, 12 

\bibitem[Boylan-Kolchin et al.(2008)]{2008MNRAS.383...93B} Boylan-Kolchin, 
M., Ma, C.-P., \& Quataert, E.\ 2008, \mnras, 383, 93 


\bibitem[den Brok et al.(2014)]{2014MNRAS.445.2385D} den Brok, M., 
Peletier, R.~F., Seth, A., et al.\ 2014, \mnras, 445, 2385 




\bibitem[Campanelli et al.(2007)]{2007PhRvL..98w1102C} Campanelli, M., 
Lousto, C.~O., Zlochower, Y., 
\& Merritt, D.\ 2007, Physical Review Letters, 98, 231102 


\bibitem[Capuzzo-Dolcetta 
\& Miocchi(2008)]{2008MNRAS.388L..69C} Capuzzo-Dolcetta, R., \& Miocchi, P.\ 2008, \mnras, 388, L69 

\bibitem[Carollo et al.(1998)]{1998AJ....116...68C} Carollo, C.~M.,
Stiavelli, M., \& Mack, J.\ 1998, \aj, 116, 68

\bibitem[Carson et al.(2015)]{2015arXiv150105586C} Carson, D.~J., Barth, 
A.~J., Seth, A.~C., et al.\ 2015, arXiv:1501.05586 

\bibitem[Cleveland~(1979)]{cleveland79} Cleveland, W.  \ 1979 J. Amer. Statist. Assoc. 74, 82936

\bibitem[Cleveland~(1988)]{cleveland+devlin88} Cleveland, W. \&
Devlin,   S. J.,  \ 1988  J. Amer. Statist. Assoc 83, 596


\bibitem[Colpi(2014)]{2014SSRv..183..189C} Colpi, M.\ 2014, \ssr, 183, 189 

\bibitem[{{C{\^o}t{\'e}} {et~al.}(2006){C{\^o}t{\'e}}, {Piatek}, {Ferrarese},
  {Jord{\'a}n}, {Merritt}, {Peng}, {Ha{\c s}egan}, {Blakeslee}, {Mei}, {West},
  {Milosavljevi{\'c}}, \& {Tonry}}]{Cote:2006}
{C{\^o}t{\'e}}, P., {Piatek}, S., {Ferrarese}, L., {et~al.} 2006, \apjs, 165,
  57

\bibitem[Craven~(1979)]{craven79}
Craven, P. and Wahba, G. (1979). Smoothing noisy data with spline functions. Numer. Math., 31: 377-403

\bibitem[den Brok et al.(2014)]{F} den Brok, M., 
Peletier, R.~F., Seth, A., et al.\ 2014, \mnras, 445, 2385 

\bibitem[Dehnen(1993)]{1993MNRAS.265..250D} Dehnen, W.\ 1993, \mnras, 265, 
250

\bibitem[de~Grijs~et~al.~(2003)]{degrijs}de Grijs, R., Anders, P., Bastian, N., Lynds, R., Lamers, H. J. G. L. M., \& O'Neil, E.~J. \ 2003, MNRAS, 343, 1285

\bibitem[Dekel 
\& Cox(2006)]{2006MNRAS.370.1445D} Dekel, A., \& Cox, T.~J.\ 2006, \mnras, 370, 1445 


\bibitem[Gonz{\'a}lez Delgado et al.(2008)]{GD08} 
Gonz{\'a}lez Delgado, R.~M., P{\'e}rez, E., Cid Fernandes, R., 
\& Schmitt, H.\ 2008, \aj, 135, 747 


\bibitem[De Lorenzi et al.(2013)]{2013MNRAS.429.2974D} De Lorenzi, F., 
Hartmann, M., Debattista, V.~P., Seth, A.~C., 
\& Gerhard, O.\ 2013, \mnras, 429, 2974 


\bibitem[Duschl et al.(2000)]{betavisc} Duschl, W.~J., Strittmatter, P.~A., \& Biermann, P.~L.\ 2000, \aap, 357, 1123 

\bibitem[Dutton 
\& van den Bosch(2009)]{2009MNRAS.396..141D} Dutton, A.~A., \& van den Bosch, F.~C.\ 2009, \mnras, 396, 141 


\bibitem[Dye et al.(2015)]{2015arXiv150308720D} Dye, S., Furlanetto, C., 
Swinbank, A.~M., et al.\ 2015, arXiv:1503.08720

\bibitem[{{Erwin} \& {Gadotti}(2012)}]{Erwin:2012}
{Erwin}, P., \& {Gadotti}, D.~A. 2012, Advances in Astronomy, 2012



\bibitem[Fakhouri et al.(2010)]{2010MNRAS.406.2267F} Fakhouri, O., Ma, 
C.-P., \& Boylan-Kolchin, M.\ 2010, \mnras, 406, 2267

\bibitem[Feldmeier et 
al.(2014)]{2014AA...570A...2F} Feldmeier, A., Neumayer, N., Seth, A., et al.\ 2014, \aap, 570, AA2 

\bibitem[Ferrarese(2002)]{2002ApJ...578...90F} Ferrarese, L.\ 2002, \apj, 
578, 90 

\bibitem[{{Ferrarese} {et~al.}(2006{\natexlab{a}}){Ferrarese}, {C{\^o}t{\'e}},
  {Dalla Bont{\`a}}, {Peng}, {Merritt}, {Jord{\'a}n}, {Blakeslee}, {Ha{\c
  s}egan}, {Mei}, {Piatek}, {Tonry}, \& {West}}]{Ferrarese:2006}
{Ferrarese}, L., {C{\^o}t{\'e}}, P., {Dalla Bont{\`a}}, E., {et~al.}
  2006{\natexlab{a}}, \apjl, 644, L21

\bibitem[Ferrarese 
\& Ford(2005)]{2005SSRv..116..523F} Ferrarese, L., \& Ford, H.\ 2005,
\ssr, 116, 523




\bibitem[Fox~(1999)]{fox99}Fox, J. \ 1999  Nonparametric regression analysis. Typescript, McMaster University


\bibitem[Frank et al.(2002)]{APA} Frank, J., King, A., 
\& Raine, D.~J.\ 2002, Accretion Power in Astrophysics, by Juhan Frank and Andrew King and Derek Raine, pp.~398.~ISBN 0521620538.~Cambridge, UK: Cambridge University Press, February 2002.,

\bibitem[Gebhardt et al.(2001)]{2001AJ....122.2469G} Gebhardt, K., Lauer, 
T.~R., Kormendy, J., et al.\ 2001, AJ , 122, 2469


\bibitem[Genzel et al.(2015)]{2015ApJ...800...20G} Genzel, R., Tacconi, 
L.~J., Lutz, D., et al.\ 2015, \apj, 800, 20 


\bibitem[Gnedin et al.(2014)]{2014ApJ...785...71G} Gnedin, O.~Y., Ostriker, 
J.~P., \& Tremaine, S.\ 2014, \apj, 785, 71 


\bibitem[Graham 
\& Guzm{\'a}n(2003)]{2003AJ....125.2936G} Graham, A.~W., \& Guzm{\'a}n, R.\ 2003, \aj, 125, 2936 

\bibitem[Graham~(2012)]{G12}Graham A. W. 2012a, MNRAS, 422, 1586

\bibitem[{{Graham}(2012{\natexlab{b}})}]{Graham:2012b}
 Graham, A. \ 2012{\natexlab{b}}, \mnras, 2608


\bibitem[Georgiev \&  B{\"o}ker(2014)]{2014MNRAS.441.3570G} Georgiev, I.~Y., B{\"o}ker, T.\ 2014, \mnras, 441, 3570 

\bibitem[Ghez et al.(1998)]{1998ApJ...509..678G} Ghez, A.~M., Klein, B.~L., 
Morris, M., \& Becklin, E.~E.\ 1998, \apj, 509, 678 

\bibitem[Gieles 
\& Baumgardt(2008)]{2008MNRAS.389L..28G} Gieles, M., \& Baumgardt, H.\ 2008, \mnras, 389, L28 

\bibitem[Gillessen et al.(2009)]{2009ApJ...692.1075G} Gillessen, S., 
Eisenhauer, F., Trippe, S., et al.\ 2009, \apj, 692, 1075 



\bibitem[Golub~(1979)]{golub79}
Golub, G., Heath, M. and Wahba, G. (1979). Generalized cross validation as a method for choosing a good ridge parameter. Technometrics, 21: 215-224

\bibitem[Graham 
\& Driver(2007)]{2007ApJ...655...77G} Graham, A.~W., \& Driver, S.~P.\ 2007, \apj, 655, 77 

\bibitem[{{Graham} \& {Spitler}(2009)}]{Graham:2009}
{Graham}, A.~W., \& {Spitler}, L.~R. 2009, \mnras, 397, 2148


\bibitem[Granato et al.(2004)]{2004ApJ...600..580G} Granato, G.~L., De 
Zotti, G., Silva, L., Bressan, A., \& Danese, L.\ 2004, \apj, 600, 580 

\bibitem[Gualandris 
\& Merritt(2008)]{2008ApJ...678..780G} Gualandris, A., \& Merritt, D.\ 2008, \apj, 678, 780 

\bibitem[Gullieuszik et 
al.(2014)]{2014AA...568A..89G} Gullieuszik, M., Greggio, L., Falomo, R., Schreiber, L., \& Uslenghi, M.\ 2014, \aap, 568, A89 

\bibitem[G{\"u}ltekin et al.(2009)]{2009ApJ...698..198G} G{\"u}ltekin, K., 
Richstone, D.~O., Gebhardt, K., et al.\ 2009, \apj, 698, 198 

\bibitem[Haehnelt 
\& Kauffmann(2002)]{2002MNRAS.336L..61H} Haehnelt, M.~G., \& Kauffmann, G.\ 2002, \mnras, 336, L61 

\bibitem[Haiman et al.(2004)]{2004ApJ...606..763H} Haiman, Z., Ciotti, L., 
\& Ostriker, J.~P.\ 2004, \apj, 606, 763 

\bibitem[Haiman et al.(2009)]{bence} Haiman, Z., Kocsis, B., 
\& Menou, K.\ 2009, \apj, 700, 1952 

\bibitem[H{\"a}ring 
\& Rix(2004)]{magorrian2} H{\"a}ring, N., \& Rix, H.-W.\ 2004, \apjl, 604, L89 


\bibitem[Harris~(1996)]{Harris}Harris, W. \ 1996, AJ, 112, 1487

\bibitem[Hartmann et al.(2011)]{2011MNRAS.418.2697H} Hartmann, M., 
Debattista, V.~P., Seth, A., Cappellari, M., 
\& Quinn, T.~R.\ 2011, \mnras, 418, 2697 

\bibitem[Hoffman 
\& Loeb(2007)]{hoffman} Hoffman, L., \& Loeb, A.\ 2007, \mnras, 377, 957 


\bibitem[Holley-Bockelmann 
\& Khan(2015)]{2015arXiv150506203H} Holley-Bockelmann, K., \& Khan, F.~M.\ 2015, arXiv:1505.06203 


\bibitem[Kawakatu 
\& Umemura(2002)]{2002MNRAS.329..572K} Kawakatu, N., \& Umemura, M.\ 2002, \mnras, 329, 572 

\bibitem[Kawakatu et al.(2003)]{2003ApJ...583...85K} Kawakatu, N., Umemura, 
M., \& Mori, M.\ 2003, \apj, 583, 85


\bibitem[Khan et al.(2011)]{lastPc1} Khan, F.~M., Just, A., 
\& Merritt, D.\ 2011, \apj, 732, 89 

\bibitem[Sesana \& Khan(2015)]{sesanaprep} Sesana, A., \& Khan, F.~M.\ 2015, arXiv:1505.02062 

\bibitem[Kennicutt(1998)]{Kenn} Kennicutt, R.~C., Jr.\ 1998, \apj, 498, 541 

\bibitem[Kennicutt \& Evans(2012)]{2012ARAA..50..531K} Kennicutt, R.~C., \& Evans, N.~J.\ 2012, \araa, 50, 531 

\bibitem[King~(1962)]{K62} King, I. R., 1962, AJ, 67, 471

\bibitem[Kormendy 
\& Ho(2013)]{2013ARAA..51..511K} Kormendy, J., \& Ho, L.~C.\ 2013, \araa, 51, 511 

\bibitem[Koushiappas et al.(2004)]{2004MNRAS.354..292K} Koushiappas, S.~M., 
Bullock, J.~S., \& Dekel, A.\ 2004, \mnras, 354, 292 

\bibitem[Kregel et al.(2005)]{2005MNRAS.358..503K} Kregel, M., van der 
Kruit, P.~C., \& Freeman, K.~C.\ 2005, \mnras, 358, 503 


\bibitem[Kruijssen(2012)]{2012MNRAS.426.3008K} Kruijssen, J.~M.~D.\ 2012, 
\mnras, 426, 3008 


\bibitem[Kruijssen et al.(2014)]{2014MNRAS.440.3370K} Kruijssen, J.~M.~D., 
Longmore, S.~N., Elmegreen, B.~G., et al.\ 2014, \mnras, 440, 3370 

\bibitem[Krumholz et al.(2009)]{2009ApJ...699..850K} Krumholz, M.~R., 
McKee, C.~F., \& Tumlinson, J.\ 2009, \apj, 699, 850 


\bibitem[Krumholz(2012)]{2012ApJ...759....9K} Krumholz, M.~R.\ 2012, \apj, 
759, 9 

\bibitem[Lapi et al.(2014)]{2014ApJ...782...69L} Lapi, A., Raimundo, S., 
Aversa, R., et al.\ 2014, \apj, 782, 69 

\bibitem[Lauer et al.(2012)]{2012ApJ...745..121L} Lauer, T.~R., Bender, R., 
Kormendy, J., Rosenfield, P., \& Green, R.~F.\ 2012, \apj, 745, 121 

\bibitem[{{Leigh} {et~al.}(2012){Leigh}, {B{\"o}ker}, \& {Knigge}}]{Leigh:2012}
{Leigh}, N., {B{\"o}ker}, T., \& {Knigge}, C. 2012, \mnras, 424, 2130

\bibitem[Leigh et al.(2015)]{2015MNRAS.451.5378L} Leigh, N.~W.~C., 
Georgiev, I.~Y., B{\"o}ker, T., Knigge, C., 
\& den Brok, M.\ 2015, \mnras, 451, 5378 

\bibitem[Leroy et al.(2015)]{leroy} Leroy, A.~K., Bolatto, 
A.~D., Ostriker, E.~C., et al.\ 2015, \apj, 801, 25 

\bibitem[Li~(1985)]{li85}Li, K.C. \ 1985  From Stein's unbaised risk estimates to the method of generalized cross-validation. Annals of Statistics, 13: 1352-1377

\bibitem[Lodato 
\& Natarajan(2006)]{2006MNRAS.371.1813L} Lodato, G., \& Natarajan, P.\ 2006, \mnras, 371, 1813 


\bibitem[Madau et al.(2014)]{2014ApJ...784L..38M} Madau, P., Haardt, F., 
\& Dotti, M.\ 2014, \apjl, 784, LL38 

\bibitem[Madau 
\& Rees(2001)]{2001ApJ...551L..27M} Madau, P., \& Rees, M.~J.\ 2001, \apjl, 551, L27 


\bibitem[Magorrian et al.(1998)]{magorrian1} Magorrian, J., 
Tremaine, S., Richstone, D., et al.\ 1998, \aj, 115, 2285 


\bibitem[Matthews et al.(1999)]{1999AJ....118..208M} Matthews, L.~D.,
Gallagher, J.~S., III, Krist, J.~E., et al.\ 1999, \aj, 118, 208

\bibitem[McLaughlin et al.(2006)]{2006ApJ...650L..37M} McLaughlin, D.~E., 
King, A.~R., \& Nayakshin, S.\ 2006, \apjl, 650, L37 


\bibitem[Merritt 
\& Cruz(2001)]{2001ApJ...551L..41M} Merritt, D., \& Cruz, F.\ 2001, \apjl, 551, L41 

\bibitem[Merritt et al.(2001)]{2001Sci...293.1116M} Merritt, D., Ferrarese, 
L., \& Joseph, C.~L.\ 2001, Science, 293, 1116 

\bibitem[Merritt et al.~(2004)]{MP04}Merritt, D., Piatek, S., Portegies Zwart, S., Hemsendorf, M. \ 2004, ApJ, 608, L25

\bibitem[Merritt(2006)]{2006ApJ...648..976M} Merritt, D.\ 2006, \apj, 648, 
976 

\bibitem[Merritt 
\& Szell(2006)]{2006ApJ...648..890M} Merritt, D., \& Szell, A.\ 2006, \apj, 648, 890 

\bibitem[Merritt et al.(2009)]{2009ApJ...699.1690M} Merritt, D., 
Schnittman, J.~D., \& Komossa, S.\ 2009, \apj, 699, 1690 

\bibitem[Merritt(2013)]{2013degn.book.....M} Merritt, D.\ 2013, Dynamics 
and Evolution of Galactic Nuclei, by David Merritt.~ISBN: 
<ISBN>9780691158600</ISBN>.~544 pp.~| 6 x 9 | 5 halftones.~136 line illus.~Princeton: 
Princeton University Press, 2013,  

\bibitem[Milosavljevi{\'c} 
\& Merritt(2001)]{2001ApJ...563...34M} Milosavljevi{\'c}, M., \& Merritt, D.\ 2001, \apj, 563, 34 


\bibitem[Milosavljevi{\'c}(2004)]{2004ApJ...605L..13M} Milosavljevi{\'c}, 
M.\ 2004, \apjl, 605, L13 

\bibitem[Miller 
\& Davies(2012)]{2012ApJ...755...81M} Miller, M.~C., \& Davies, M.~B.\ 2012, \apj, 755, 81 



\bibitem[{{Neumayer} \& {Walcher}(2012)}]{Neumayer:2012}
{Neumayer}, N., \& {Walcher}, C.~J. 2012, Advances in Astronomy, 2012

\bibitem[Parkinson et al.(2008)]{2008MNRAS.383..557P} Parkinson, H., Cole, 
S., \& Helly, J.\ 2008, \mnras, 383, 557 


\bibitem[Perets 
\& Mastrobuono-Battisti(2014)]{2014ApJ...784L..44P} Perets, H.~B., \& Mastrobuono-Battisti, A.\ 2014, \apjl, 784, LL44 

\bibitem[Phillips et al.(1996)]{1996AJ....111.1566P} Phillips, A.~C.,
Illingworth, G.~D., MacKenty, J.~W., \& Franx, M.\ 1996, \aj, 111, 1566

\bibitem[Pfuhl et al.(2011)]{2011ApJ...741..108P} Pfuhl, O., Fritz, T.~K., 
Zilka, M., et al.\ 2011, \apj, 741, 108 



\bibitem[Rossa et al.(2006)]{2006AJ....132.1074R} Rossa, J., van der Marel, 
R.~P., B{\"o}ker, T., et al.\ 2006, \aj, 132, 1074 

\bibitem[Scott \& Graham(2013)]{2013ApJ...763...76S} Scott, N., \& Graham, A.~W.\ 2013, \apj, 763, 76

\bibitem[Sesana et al.(2014)]{2014ApJ...794..104S} Sesana, A., Barausse, 
E., Dotti, M., \& Rossi, E.~M.\ 2014, \apj, 794, 104 

\bibitem[Seth et al.(2006)]{2006AJ....132.2539S} Seth, A.~C., Dalcanton, 
J.~J., Hodge, P.~W., \& Debattista, V.~P.\ 2006, \aj, 132, 2539 


\bibitem[Seth et al.(2008a)]{2008ApJ...687..997S} Seth, A.~C., Blum, R.~D., 
Bastian, N., Caldwell, N., \& Debattista, V.~P.\ 2008, \apj, 687, 997 

\bibitem[Seth et al.(2008b)]{2008ApJ...678..116S} Seth, A., Ag{\"u}eros, M., 
Lee, D., \& Basu-Zych, A.\ 2008, \apj, 678, 116 

\bibitem[Shen et al.(2003)]{2003MNRAS.343..978S} Shen, S., Mo, H.~J., 
White, S.~D.~M., et al.\ 2003, \mnras, 343, 978 

\bibitem[Sch{\"o}del et 
al.(2014)]{2014AA...566A..47S} Sch{\"o}del, R., Feldmeier, A., Kunneriath, D., et al.\ 2014, \aap, 566, AA47 



\bibitem[Swinbank et al.(2015)]{2015arXiv150505148S} Swinbank, M., Dye, S., 
Nightgale, J., et al.\ 2015, arXiv:1505.05148 


\bibitem[Taffoni et al.(2003)]{2003MNRAS.341..434T} Taffoni, G., Mayer, L., 
Colpi, M., \& Governato, F.\ 2003, \mnras, 341, 434 

\bibitem[Tremaine et al.(2002)]{2002ApJ...574..740T} Tremaine, S., 
Gebhardt, K., Bender, R., et al.\ 2002, \apj, 574, 740 


\bibitem[Tremaine et al.(1975)]{1975ApJ...196..407T} Tremaine, S.~D., 
Ostriker, J.~P., \& Spitzer, L., Jr.\ 1975, \apj, 196, 407 

\bibitem[Turner et al.(2012)]{2012ApJS..203....5T} Turner, M.~L., 
C{\^o}t{\'e}, P., Ferrarese, L., et al.\ 2012, \apjs, 203, 5

\bibitem[Umemura(2001)]{2001ApJ...560L..29U} Umemura, M.\ 2001, \apjl, 560, 
L29 

\bibitem[van den Bergh~(1986)]{VdB}van den Bergh, S., AJ, 1986, 91, 271

\bibitem[van Meter et al.(2010)]{2010ApJ...719.1427V} van Meter, J.~R., 
Miller, M.~C., Baker, J.~G., Boggs, W.~D., 
\& Kelly, B.~J.\ 2010, \apj, 719, 1427 


\bibitem[Vasiliev et al.(2014)]{lastPc2} Vasiliev, E., 
Antonini, F., \& Merritt, D.\ 2014, \apj, 785, 163 


\bibitem[Vasiliev(2014)]{2014arXiv1411.1762V} Vasiliev, E.\ 2014, 
arXiv:1411.1762 

\bibitem[Vasiliev et al.(2015)]{2015arXiv150505480V} Vasiliev, E., 
Antonini, F., \& Merritt, D.\ 2015, arXiv:1505.05480 

\bibitem[Volonteri et al.(2008)]{2008MNRAS.383.1079V} Volonteri, M., 
Lodato, G., \& Natarajan, P.\ 2008, \mnras, 383, 1079 

\bibitem[Walcher et al.(2005)]{2005ApJ...618..237W} Walcher, C.~J., van der
Marel, R.~P., McLaughlin, D., et al.\ 2005, \apj, 618, 237

\bibitem[Walcher et al.(2006)]{2006ApJ...649..692W} Walcher, C.~J., 
B{\"o}ker, T., Charlot, S., et al.\ 2006, \apj, 649, 692 

\bibitem[Wehner 
\& Harris(2006)]{2006ApJ...644L..17W} Wehner, E.~H., \& Harris, W.~E.\ 2006, \apjl, 644, L17 


\bibitem[Yusef-Zadeh et al.(2012)]{2012ApJ...744...24Y} Yusef-Zadeh, F., 
Bushouse, H., \& Wardle, M.\ 2012, \apj, 744, 24 

\bibitem[Yu(2002)]{yu} Yu, Q.\ 2002, \mnras, 331, 935 

\end{thebibliography}
  \end{document}